\documentclass[fleqn,usenatbib]{mnras}

\usepackage[T1]{fontenc}
\usepackage{newtxtext,newtxmath}

\usepackage{graphicx}
\usepackage{amsmath}
\usepackage{booktabs}
\usepackage{multirow}
\usepackage{array}
\usepackage{subcaption}
\usepackage{float}
\usepackage{xcolor}
\usepackage{orcidlink}

\usepackage{tikz}
\usetikzlibrary{shapes,arrows}

\DeclareRobustCommand{\VAN}[3]{#2}
\let\VANthebibliography\thebibliography
\def\thebibliography{\DeclareRobustCommand{\VAN}[3]{##3}\VANthebibliography}

\title[A New Methodology for Classifying Eclipsing Binaries]
{A New Methodology for Classifying Eclipsing Binaries 
with Kepler Data and Deep Learning}

\author[Mondal et al.]{
Mousam Mondal$^{1}$\orcidlink{0009-0009-1238-5327}\thanks{E-mail: mousamphysics137@gmail.com},
Patricia Cruz$^{2}$\orcidlink{0000-0003-1793-200X},
Hugh R. A. Jones$^{1}$\orcidlink{0000-0003-0433-3665},
M. C. G\'alvez-Ortiz$^{2}$,
and John F. Aguilar$^{3,4}$\orcidlink{0000-0002-4752-2784}
\\
$^{1}$Centre for Astrophysics Research, University of Hertfordshire, College Lane, Hatfield, Hertfordshire AL10 9AB, UK.\\
$^{2}$Centro de Astrobiolog\'{\i}a (CAB), CSIC-INTA, Camino Bajo del Castillo s/n, E-28692, Villanueva de la Ca\~{n}ada, Madrid, Spain.\\
$^{3}$Departamento de Matem\'aticas, Universidad Militar Nueva Granada, kilometro 2 v\'ia Cajic\'a--Zipaquir\'a, c\'odigo postal 110111, Colombia.\\
$^{4}$PhD Programme in Astrophysics, Doctoral School, Universidad Aut\'onoma de Madrid, Ciudad Universitaria de Cantoblanco, 28049 Madrid, Spain.
}

\date{Accepted for publication in RAS Techniques and Instruments}

\pubyear{2026}

\begin{document}
\label{firstpage}
\pagerange{\pageref{firstpage}--\pageref{lastpage}}
\maketitle


\begin{abstract}
We present a new method for the automated classification of eclipsing binaries, into contact, detached, and semi-detached types using {\sl Kepler} data. Phase-folded light curves are generated and chi-square vs. box size plots are constructed by comparing flux values to the median flux, revealing distinct class patterns. These patterns were first modelled using a polynomial damped sinusoidal function, whose period served as classification feature, achieving an overall accuracy of 86.5 percent. To capture more features and enhance accuracy, we trained a convolutional neural network, which improved the total accuracy to 90 percent, including 47 percent for the challenging semi-detached systems. However, several binaries displayed irregular chi-square signatures. To mitigate this, we incorporated simulated light curves generated with the PHOEBE modelling code, achieving 99 per cent accuracy in distinguishing contact and detached binaries. The resulting chi-square morphologies show a strong correlation with orbital period, and a subset of systems exhibit quarterly variability in their light curves and chi-square trends. We designate these as Temporally Varying systems. By measuring the normalized spread of the chi-square period across quarters, we define a statistical threshold that separates these systems from stable binaries. We reported four Temporally Varying systems not previously noted in the literature with magnetic activity that requires further investigation. Furthermore, cooler stars, namely late-F, G, K, and M types, display systematically higher variability than hotter stars. Cross-matching with catalogues of magnetically active stars indicates that stellar flares and starspots are the most likely causes of this enhanced variability.
\end{abstract}


\section{Introduction}

The advancement of space technology and modern telescopes has led to an abundance of high-precision astronomical data. Visual inspection remains a reliable approach for eclipsing binaries (EB) classification, capturing subtle morphological features in photometric light curves (LCs) that are difficult to encode algorithmically. However, its manual nature makes it impractical due to the  rapidly growing volume of survey data. This created increasing opportunities for automated methods and artificial intelligence algorithms to identify patterns, anomalies, and correlations in a fast and efficient way.
Among all the astronomical targets, EBs offer great potential for pattern recognition and classification based on their LCs. The analysis of LCs is crucial for the derivation of physical parameters, such as the relative radii of their components, as well as their 
orbital properties, which can serve as test-beds to improve atmospheric models, formation, and evolution theories 
\citep[e.g.][]{Andersen1991,Torres2010,Frost_2024}.

EBs can be classified according to the shape of their LCs or their physical properties. Based on physical configuration, three main morphological EB categories are defined by the degree to which the component stars fill their Roche lobes \citep{kallrath2009eclipsing}. Detached systems have both components within their respective Roche lobes. In semi-detached systems, only one component fills this boundary. Contact and over-contact binaries occur when both components exceed this limit and share a common envelope.

In the past years, many researchers have employed automated methods to classify EBs using machine learning models trained and validated on large datasets of LCs. For instance, \citet{refId0} developed an automated classification approach using a Bayesian ensemble of neural networks trained on data from the {\sl Hipparcos} satellite. Their method classified both pulsating and EBs based solely on LC morphology and further subdivided EBs into four geometrically meaningful classes. 

Using a collection of long-term time series, \citet{Matijevic} classified EBs in the second {\sl Kepler} catalog \citep{Slawson11} using a nonlinear dimensionality reduction tool known as Locally Linear Embedding (LLE) \citep[e.g.][]{roweis2000nonlinear,de2002pattern}. Using this method, LC data in higher dimensions were projected into lower two-dimensional (2D) space. A classification parameter, $c$, was included to provide the fit between 2-dimensional LC and spline function. They did not report a direct classification accuracy, but instead introduced the $c$ parameter, whose ranges can be mapped to detached, semi-detached, and contact categories.

\citet{2021A&C....3600488C} used deep learning techniques for EB classification.
Their model classified semi-detached EBs as detached systems and achieved an accuracy of 98 percent, as it was primarily trained on detached and overcontact EBs. When semi-detached EBs were excluded, the accuracy improved to 100 percent. More recently, \citet{2023MNRAS.520..828D} employed a supervised machine learning model with a compound decision tree (CDT) to classify EBs into the same classification categories, utilizing VISTA (Visible and Infrared Survey Telescope for Astronomy) in the Vía Láctea (VVV, \citealt{MINNITI2010433,2012A&A...537A.107S,2014Msngr.155...29H}) and the VISTA Variables in the Vía Láctea eXtended Survey (VVVX, \citealt{2018A&A...616A..26M}) survey. The study identified 35 features, such as shape and period. The CDT model reliability was approximately 80 percent when compared to visually performed classifications. In their recent work, \citet{2025AJ....169..202D} developed a neural-network–accelerated PHOEBE framework combined with MCMC to model semi-detached eclipsing binaries from TESS light curves, identifying 327 systems in which the less massive component fills its Roche lobe and three systems in which the more massive component does so. The neural network replaced PHOEBE within the MCMC loop, providing nearly two orders of magnitude faster light-curve generation.

In this work, we propose a different approach, an indirect classification method, based on the LC normalisation technique outlined in \citet{2022MNRAS.515.1416C}, to classify systems into contact EBs (CEBs), semi-detached EBs (SDEBs), and detached EBs (DEBs). For simplicity, we grouped CEBs and over-contact systems together under CEBs, maintaining a classification structure with three primary categories. 
This paper is organized as follows. The process of sample selection is described in Section~\ref{sec:des}, while the LC normalization procedure is explained in Section~\ref{sec:nor}. Section~\ref{sec:mor} provides detailed information about the classification scheme, including the use of two different methodologies. Section~\ref{sec:dis} contains the results and discussion, the summary and our conclusions are presented in Section~\ref{sec:con}.


\section{Description of the sample}\label{sec:des}

To develop our automated classification scheme, a dataset of well-sampled LCs with thousands of epochs each was needed. Therefore, we adopted the complete sample of 2920 EBs from the {\sl Kepler} Eclipsing Binary Catalog\footnote{\url{https://keplerEBs.villanova.edu}} (KEBC, \citealt{2016AJ....151...68K}). In the {\sl Kepler} mission, observations are divided into seventeen observing quarters, each corresponding to an approximately 93-day observing segment defined by 90$^\circ$ spacecraft rolls required to maintain spacecraft orientation \citep{Morris2020KeplerHandbook}. To collect accurate LC data for each quarter for all targets, we utilized the Mikulski Archive for Space Telescopes (MAST)\footnote{\url{https://archive.stsci.edu}}. 

From this sample, we executed a query on the MAST {\sl Kepler} Search portal, setting the condition flag to EB as the target type, and searched for long cadence data within a 0.02 arcmin radius. We specifically selected targets with long-cadence (1765.5 s) LCs to maximize our sample size, as short-cadence (58.85 s) data were available for only a limited number of targets. Subsequently, we downloaded quarter-wise fits files for all selected targets to facilitate further analysis.

\begin{figure}
  \centering
  \begin{subfigure}{0.5\textwidth}
    \includegraphics[width=\textwidth]{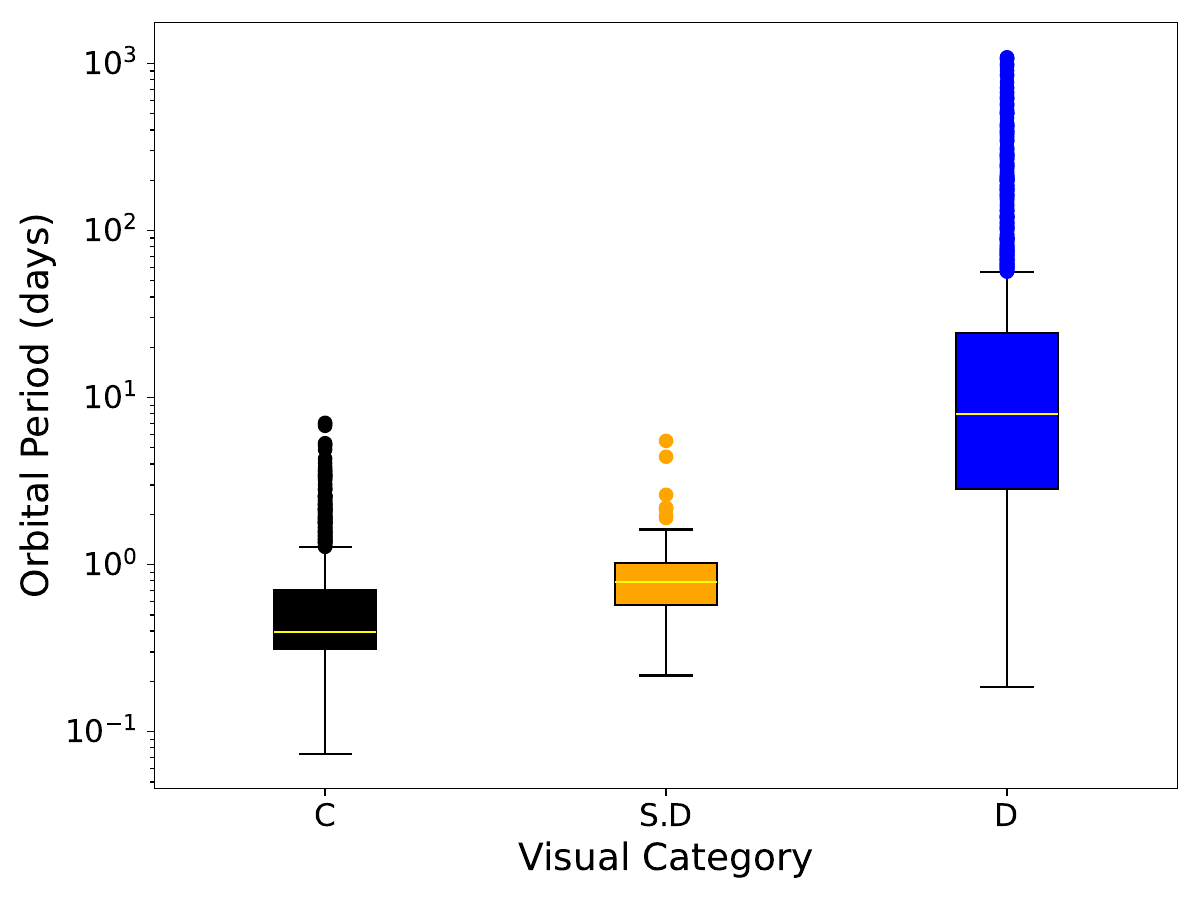}
  \end{subfigure}
  \caption{Box plots illustrated the $P_{\rm orb}$ values of C (black), SD (orange), and D (blue) EBs. Each box represented the interquartile range (IQR), which encompassed 50 percent of the data. The horizontal yellow line within each box indicated the median $P_{\rm orb}$. The whiskers extended to 1.5 times the IQR, and the dots above the whiskers represented outliers — data points that fell outside this range, highlighting extreme values in each category. In the figure, CEBs had 98 outliers (11.88 percent), SDEBs had 12 outliers, and DEBs had 269 outliers (14.04 percent). The error bars, represented by the whiskers, indicated the spread of the data within 1.5 times the IQR.}
  \label{fig:f_2x}
\end{figure}

After cross-matching the KEBC with photometric data from the MAST, we selected 2865 EB systems with long-exposure data suitable for our automated classification scheme. The orbital periods ($P_{\rm orb}$) of these targets, taken from the KEBC, ranged from 0.073 to 1087.3 days (Fig.~\ref{fig:f_2x}). We then examined the $P_{\rm orb}$ distribution for each visually classified EB class to determine the period range of CEBs, SDEBs, and DEBs. The accuracy of our automated classification depends on the visual inspection of the 2865 LCs, where we identified EBs into three morphological categories: CEBs, SDEBs, and DEBs, as described in detail in Section~\ref{sec:mor}. Based on this cross-match, we found that CEBs had $P_{\rm orb}$ ranging from 0.073 to 7.056 days, DEBs from 0.184 to 1087.3 days, and SDEBs from 0.216 to 5.499 days. In this work, CEBs, over-contact systems and ellipsoidal variables were treated as a single CEB category, preserving a three-class structure. We did not treat ellipsoidal variables as a separate class in this work; therefore, any systems showing ellipsoidal-type variability, including the few long-period objects classified as CEBs, are retained within the CEB category.

\section{Normalization procedure}\label{sec:nor}

The automated classification scheme was developed as a by-product of the LC normalization procedure adopted by \citet{2022MNRAS.515.1416C}. In this work, we generated normalized phase-folded LCs for all objects in our sample using the same methodology, as only around 1000 objects were analyzed in \citet{2022MNRAS.515.1416C}. For consistency, we also adopted the $P_{\rm orb}$ values given by the KEBC for all 2865 objects in our sample.
To prepare the data, we first conducted a median-based normalization using the pre-search data conditioning (PDC) flux, where the PDC flux values were divided by the respective median for each quarter; the PDC flux was produced by the {\sl Kepler} pipeline to remove common instrumental trends, such as pointing drifts and thermal effects, present in the raw Simple Aperture Photometry (SAP) flux, while preserving the dominant astrophysical variability \citep{2016ksci.rept....2V,10.1111/j.1365-2966.2012.20644.x}.\\~\\
Normalisation was carried out quarter by quarter using the sliding-median method of \citet{2022MNRAS.515.1416C}. For each quarter, we constructed a local median baseline by moving a window across the LC. In the classification plots, we used the half-window length as the box size and explored box sizes from 10 to 100 epochs, corresponding to total sliding windows of 21 to 201 epochs. For each trial window size, an error-weighted chi-square value was calculated by comparing the observed flux with the corresponding sliding-median baseline, following the procedure described by \citet{2022MNRAS.515.1416C}. The optimal window size was then selected using the same criterion: the chi-square values were scanned in order of increasing window size, and the first value at which the integer part of the chi-square stopped increasing monotonically was adopted, where the integer part was used to suppress noise-level fluctuations that would otherwise cause a premature break in the scan. This selected the smallest window that could follow the dominant long-term trend while avoiding excessive smoothing of the signal.

The sliding-median baseline obtained with the selected window size was then smoothed using a cubic spline. The original LC was normalised by dividing it by this spline-smoothed trend, and the normalised quarterly LCs were then combined to produce the final normalised LC for each target.

The chi-square--box-size relation showed how the LC responded to filtering on progressively longer time-scales. The {\sl Kepler} long-cadence sampling interval used in this work was 1765.5~s, corresponding to approximately 29.43~min. In terms of the plotted box-size scale, a box size of 10 epochs corresponded to about 4.9~h, while a box size of 100 epochs corresponded to about 49.0~h, or roughly 2~d.

\begin{figure*}
  \centering
  \begin{subfigure}{0.29\textwidth}
    \includegraphics[width=\textwidth]{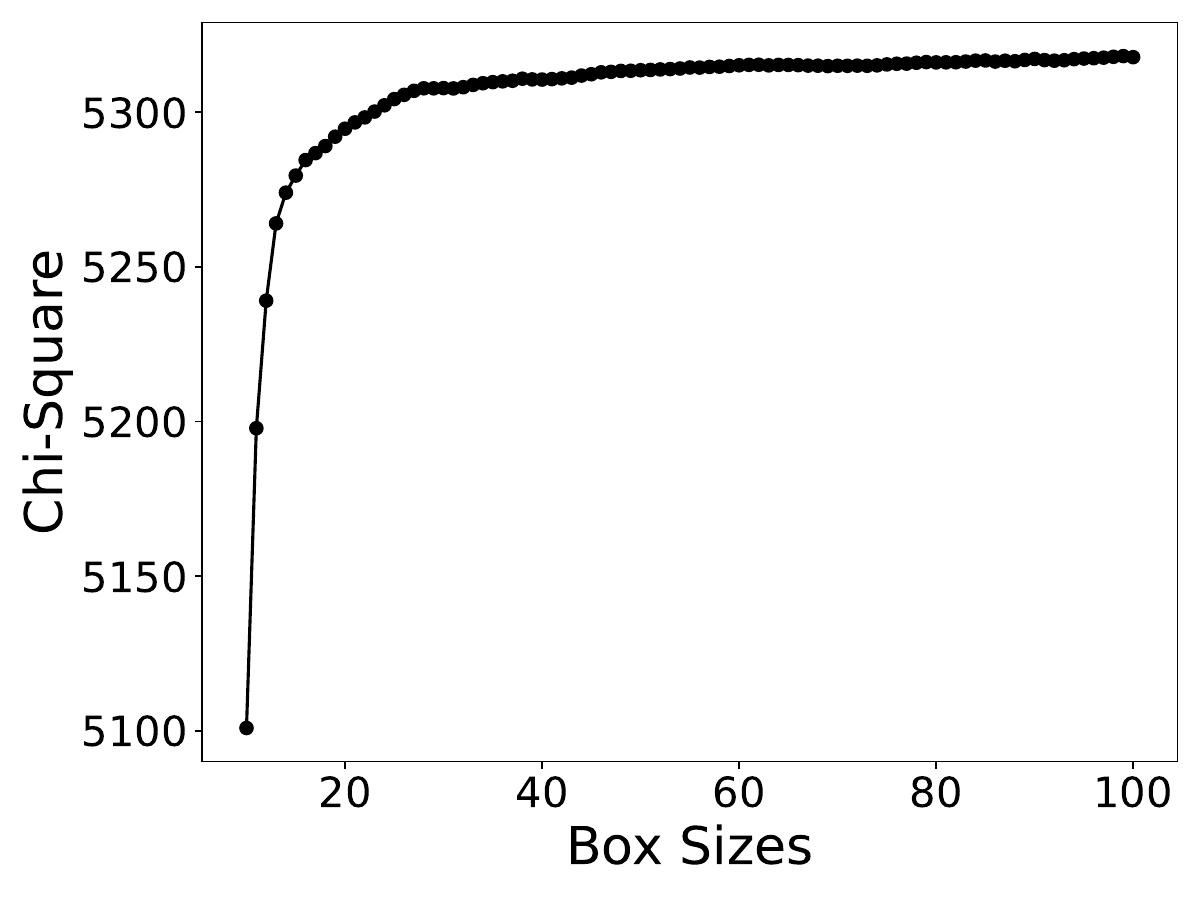}
    \caption{}
  \end{subfigure}
  \begin{subfigure}{0.29\textwidth}
    \includegraphics[width=\textwidth]{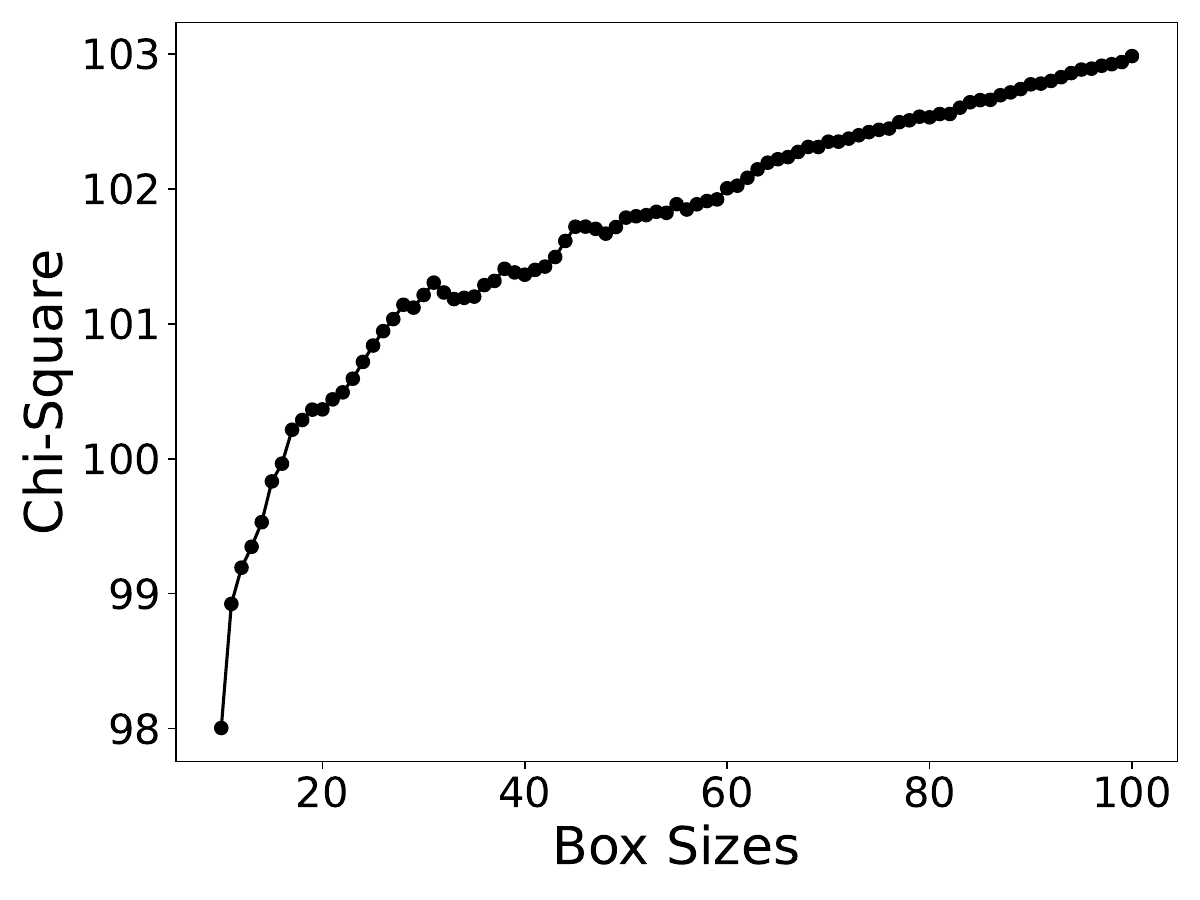}
    \caption{}
  \end{subfigure}
  \begin{subfigure}{0.29\textwidth}
    \includegraphics[width=\textwidth]{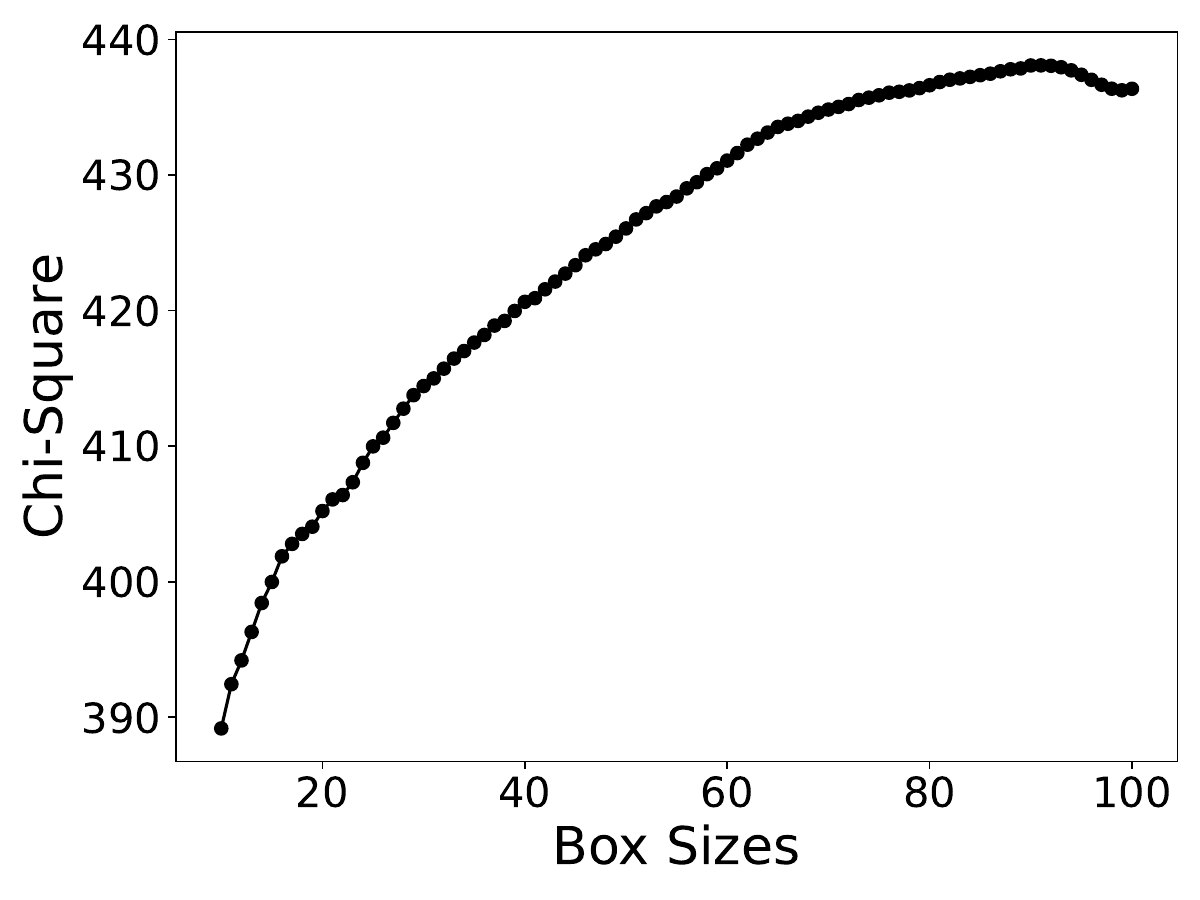}
    \caption{}
  \end{subfigure}
  \begin{subfigure}{0.29\textwidth}
    \includegraphics[width=\textwidth]{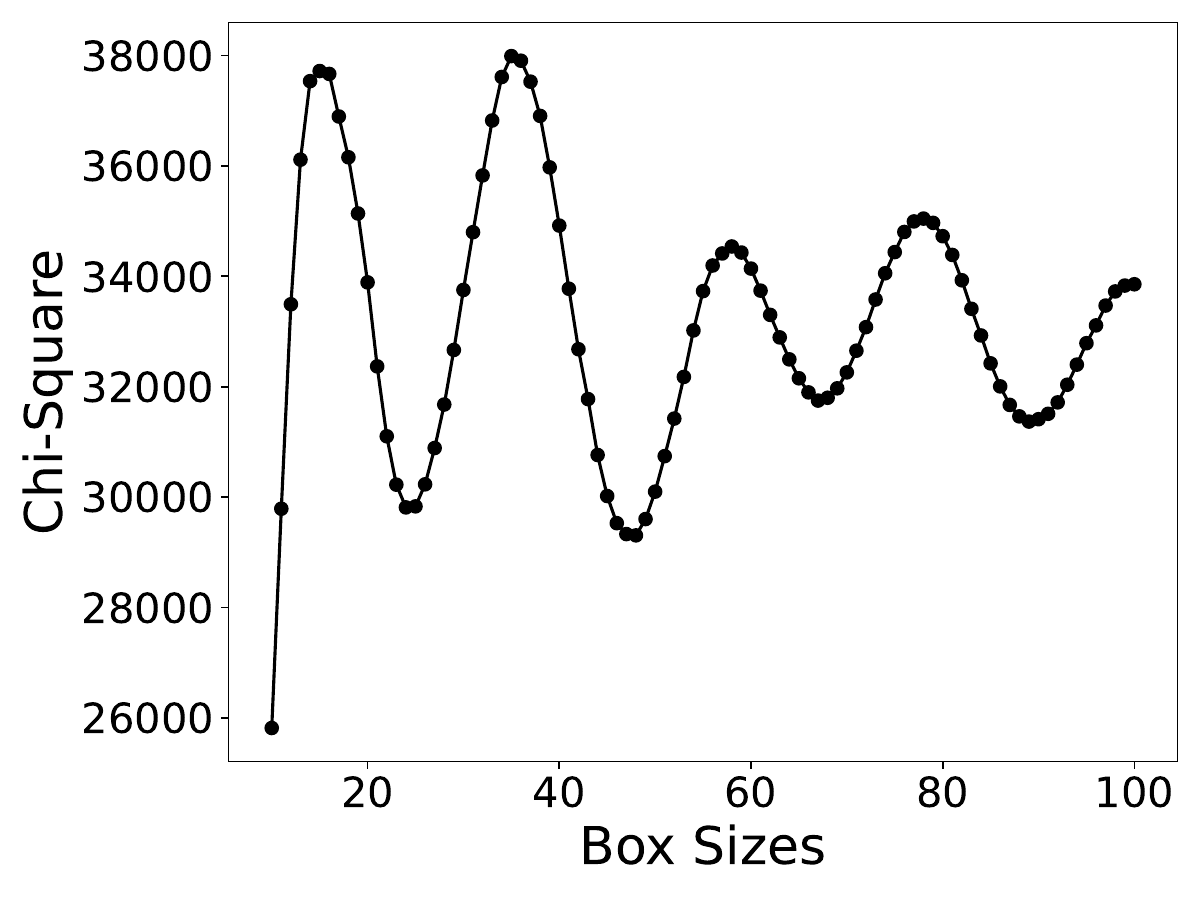}
    \caption{}
  \end{subfigure}
  \begin{subfigure}{0.29\textwidth}
    \includegraphics[width=\textwidth]{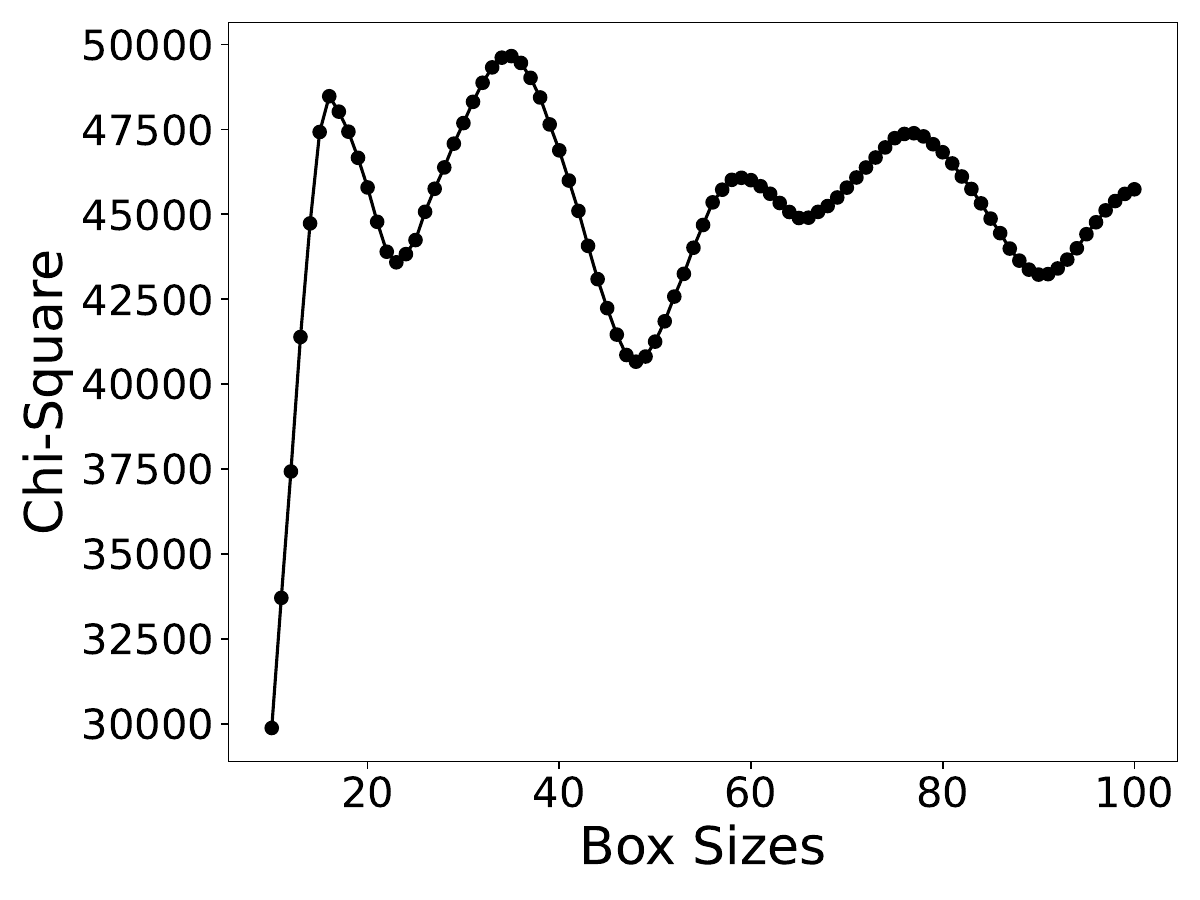}
    \caption{}
  \end{subfigure}
  \begin{subfigure}{0.29\textwidth}
    \includegraphics[width=\textwidth]{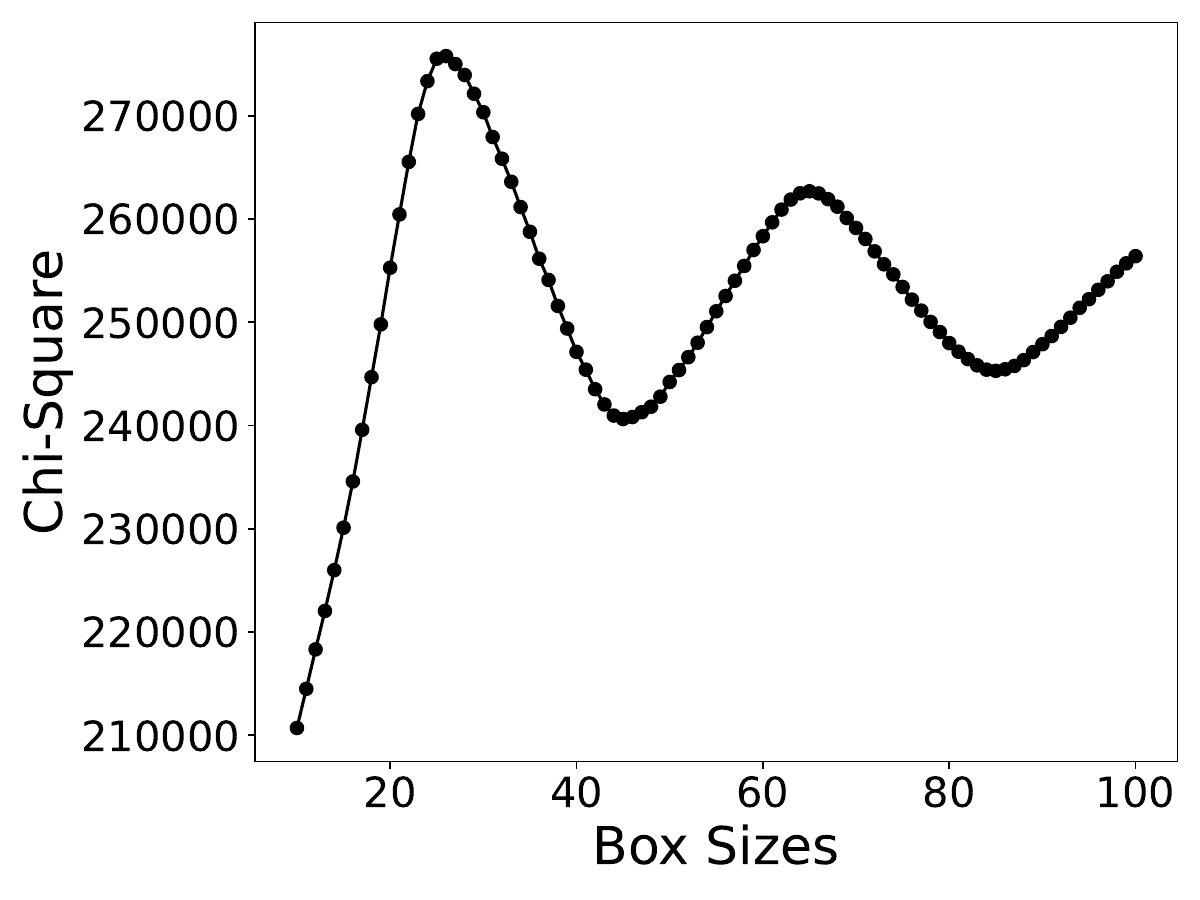}
    \caption{}
  \end{subfigure}
    \begin{subfigure}{0.29\textwidth}
    \includegraphics[width=\textwidth]{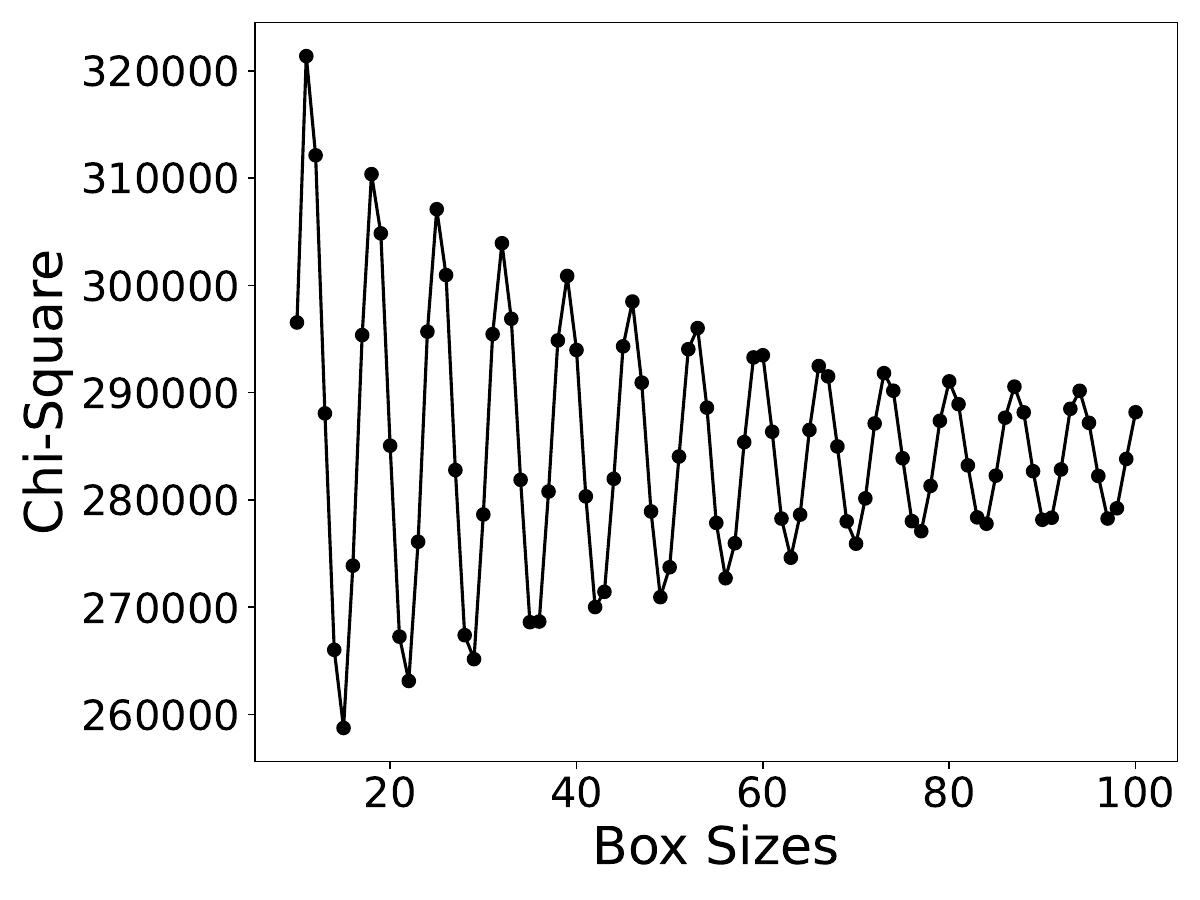}
    \caption{}
  \end{subfigure}
  \begin{subfigure}{0.29\textwidth}
    \includegraphics[width=\textwidth]{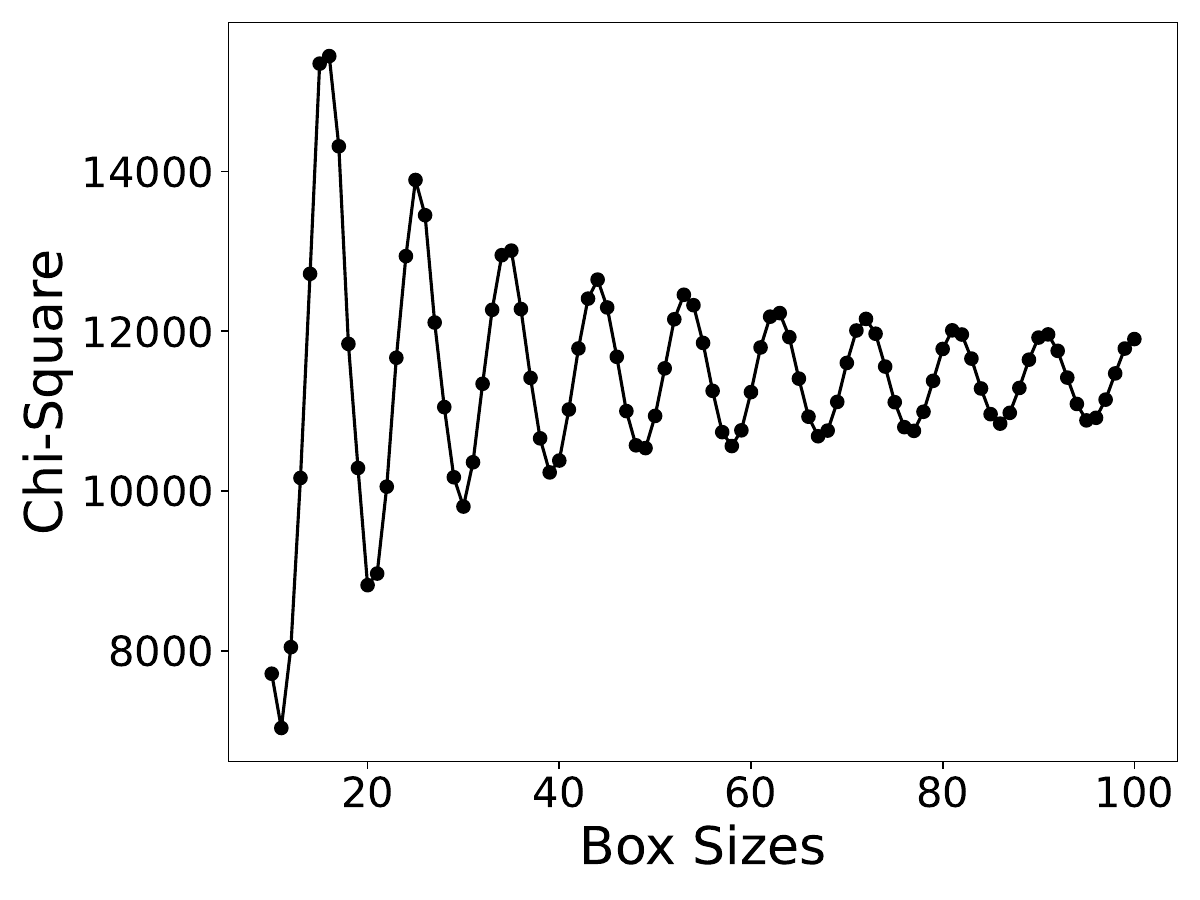}
    \caption{}
  \end{subfigure}
   \begin{subfigure}{0.29\textwidth}
    \includegraphics[width=\textwidth]{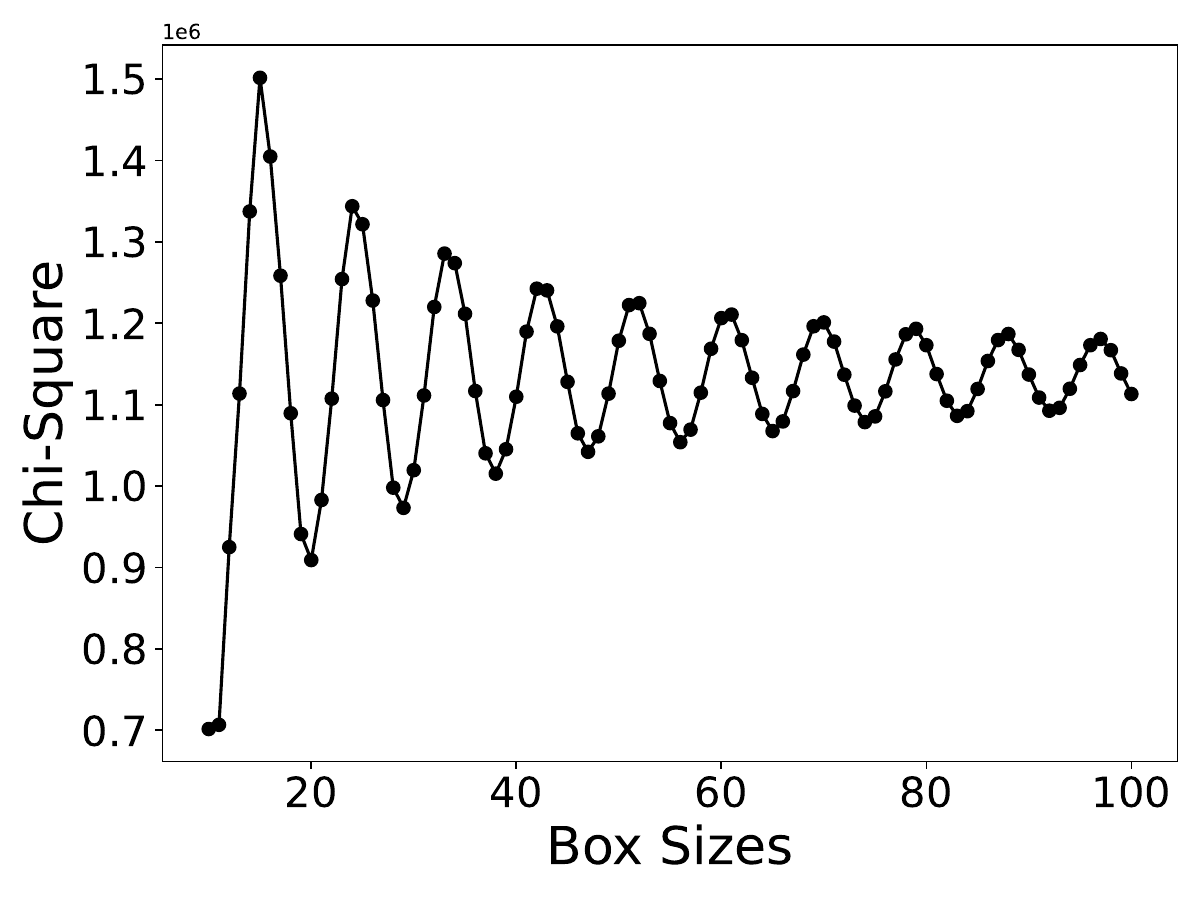}
    \caption{}
  \end{subfigure}
  \caption{The chi-square vs. box size plots for nine different EBs are shown. The top row (a, b, c panels) displays three different DEBs with {\sl Kepler} IDs (KIC): 1026032, 1873513, and 2166200, respectively. The middle row (d, e, f panels) displays three different SDEBs with KIC: 2577756, 5560831, and 5962514, respectively. Lastly, the bottom row (g, h, i panels) displayed three CEBs systems with KIC: 1433410, 2012362, and 2159783, respectively.}
  \label{fig:f_003}
\end{figure*}

The chi-square--box-size curve was used in two ways in this work. First, it helped us choose the normalisation window for each quarter by identifying the smallest box size that could follow the dominant instrumental trend. Second, by visually inspecting the chi-square plots together with the phase-folded LCs, we found that systems in the same morphological class generally showed similar chi-square trends. DEBs usually produced an increasing, polynomial-like or exponential-like curve (Fig.~\ref{fig:f_003}, panels a--c), whereas CEBs showed a damped sinusoidal pattern (Fig.~\ref{fig:f_003}, panels g--i). The SDEB systems showed less distinct behaviour; most displayed a distorted sinusoidal or damped pattern between the CEB and DEB cases (Fig.~\ref{fig:f_003}, panels d--f). These class-dependent differences made the chi-square--box-size curves useful diagnostic features for separating the three EB morphological classes.

For a given target, the chi-square curves were generally similar across different observing quarters, indicating that this diagnostic was stable over the {\sl Kepler} baseline. Since the number of data points varied from quarter to quarter, we used the quarter with the largest number of epochs for the subsequent classification analysis.

To test whether the selected range was robust, we also calculated chi-square--box-size curves up to 1000 epochs for a representative subset of systems. For CEBs, the damped sinusoidal structure remained visible up to about 180--200 epochs. Beyond this range, the curves became almost flat. This flattening occurred because larger windows covered multiple orbital cycles, so the sliding median averaged over all orbital phases and approached a nearly constant baseline. DEBs showed a simpler behaviour. Their chi-square values increased at small box sizes, approximately below 25 epochs, and then remained nearly constant for larger windows. For long-period DEBs, this early plateau is expected to be related to the eclipse duration rather than the full $P_{\rm orb}$, because individual eclipses are narrow compared with $P_{\rm orb}$. Once the window became wider than the characteristic eclipse width, the sliding-median response saturated. This was also seen in our tests up to 1000 epochs, corresponding to approximately 20.4~days for the {\sl Kepler} long-cadence sampling used here, where the DEB chi-square behaviour showed no significant change beyond approximately 25 epochs. Since the main aim of this work was morphological classification rather than optimal detrending of individual systems, we restricted the main analysis to box sizes of 10--100 epochs. This range captured the relevant class-dependent structure while remaining applicable to the full sample.


To verify the consistency of the above-described behaviour, we analyzed the original LCs, using PDC flux values available in the {\sl Kepler} \texttt{FITS} files, as well as the normalized LCs. The analysis was performed for both, individual quarter data and the complete LCs which included all quarters. In most cases, we noticed similar trends, also suggesting a chi-square plot dependency on the $P_{\rm orb}$ of the EBs. The major discrepancies were observed in the chi-square plots of some SDEBs and short $P_{\rm orb}$ DEBs, even for different quarters of observations for a given object, which we discussed further in Section~\ref{sec:dis}.


\section{Morphological Classification}\label{sec:mor}

Our classification techniques comprised two processes: the curve fitting approach, described in Section~\ref{sec:mor1}, and the deep learning approach, described in Section~\ref{sec:ai}.

During LCs visual inspection, we identified EBs whose LCs did not fall under the three standard morphological classes (C, SD, D). These were systems that did not show any primary or secondary eclipses visible in the phase-folded LC with the given $P_{\rm orb}$ (e.g., KIC 9512958). Earlier authors \citet{Slawson11} and \citet{Matijevic} mentioned these systems in the {\sl Kepler} EB catalogs and termed them as uncertain categories (UNC) \citet{Slawson11} included UNC as a morphological category. Also, the KEBC \citep{2016AJ....151...68K} catalog included various types of EB LC, which do not satisfy our three classes: heartbeat stars (e.g., KIC 5944240), UNC classes (e.g., KIC 5823121), pulsation EBs (e.g., KIC 7515679), and eclipse timing variations. We classified  all these ambiguous cases under the  UNC category and they were excluded from our analysis.

After visually classifying the LCs, we found that out of 2865 objects, there were 832 CEBs, 1820 DEBs, 154 SDEBs, and 59 UNC systems. This brought our total visually classified sample size to 2806, which contained only the three morphological classes. This classification was used for comparison with the automated classification result.


\subsection{{Classification using the PDS function}}\label{sec:mor1}

Our first approach was based on analyzing the shape of the chi-square plots and finding the best fitting function. Since polynomial or damping-like trends shown in the plots of DEBs and CEBs dominated our chi-square plots (95 percent of the sample), with SD systems showing a similar trend or something in between, we adopted the following polynomially damped sinusoidal (PDS) function.

\begin{equation}
y = A \cdot x^n \cdot \exp(-B x) \cdot \sin(\omega x + \phi) + C,
\label{eq:eq_1}
\end{equation}

\noindent
where \( A \) is the amplitude of the function, \( \omega \) is the angular frequency, and \( \phi \) is the phase angle. The \( B \) reflected the decay rate in the exponential decline of the oscillatory component with respect to the box size \( x \) in our chi-squared plots. The parameter \( C \) is a constant vertical offset parameter that shifts the fitted curve along the y-axis and accounts for any residual baseline level in the normalized chi-square values. Additionally, \( x^n \) polynomial modulation term, where \( n \) is the polynomial exponent.

\begin{figure}
  \centering
  \begin{subfigure}{0.5\textwidth}
    \includegraphics[width=\textwidth]{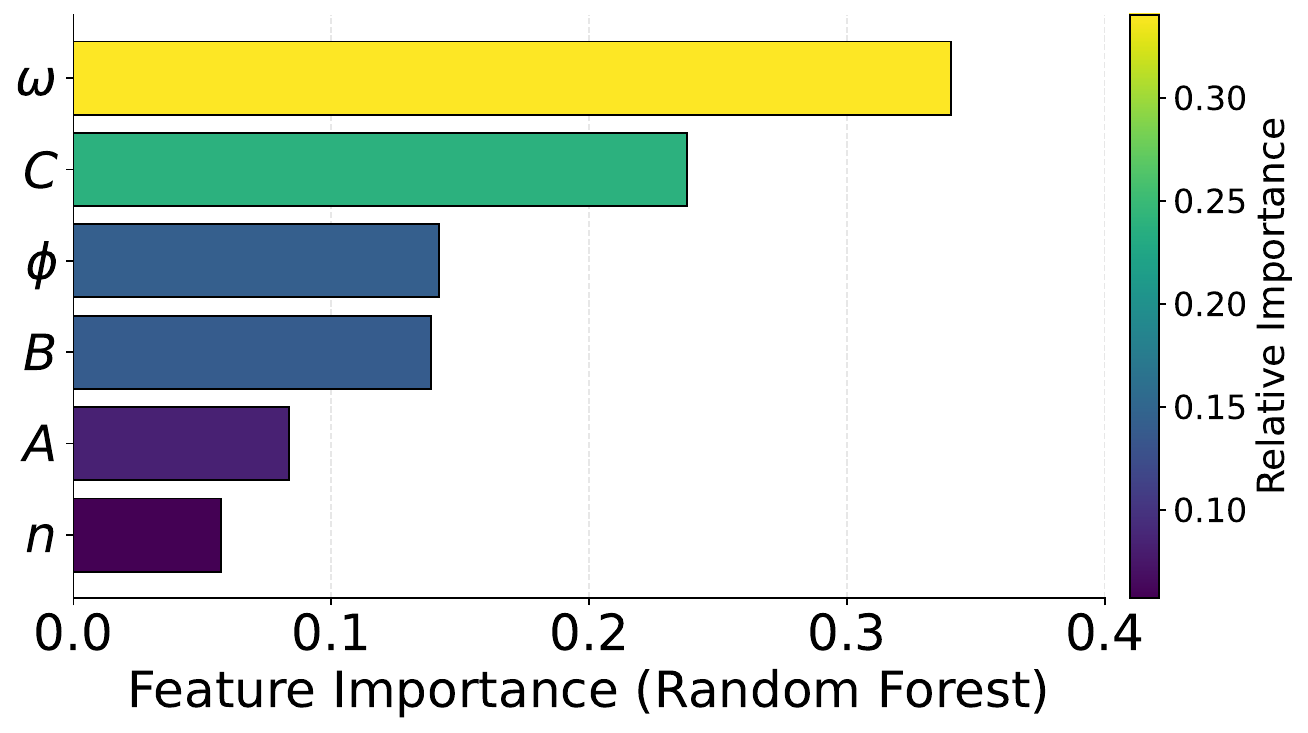}
  \end{subfigure}
  \caption{Random Forest feature-importance ranking of the six free parameters of the PDS function for distinguishing CEBs, DEBs, and SDEBs. The vertical axis lists the six fit parameters: $\omega$---the angular frequency; $C$---the constant vertical offset parameter of the fit; $\phi$---the phase angle of the sinusoidal component; $B$---the exponential damping coefficient; $A$---the overall amplitude of the function; and $n$---the exponent of the polynomial term. The horizontal axis gives the normalized feature importance of each parameter, while the colour scale represents the corresponding relative importance, with brighter colours indicating a higher contribution to the Random Forest classifier.}
  \label{fig:rf_3}
\end{figure}

We utilized Python's \texttt{Scipy curve fit} \citep[e.g.][]{2020SciPy-NMeth} code for optimizing parameters and fitting functions to data. We tried fitting the chi-square values directly without the offsets, scaling, and alternative non-linear optimization strategies; however, these approaches did not yield accurate recovery of the oscillatory PDS component across the full sample and were unable to reproduce the sinusoidal behaviour. To simplify the fitting procedure, we normalised the data, which scaled the data values between $[0, 1]$ and then the PDS function was fitted to the normalised data, with the baseline offset absorbed by the constant offset parameter in equation~(\ref{eq:eq_1}), which accounts for any residual vertical shift during fitting. As outlined in equation~(\ref{eq:eq_1}), the PDS function encompassed multiple parameters crucial for accurately capturing the system's behaviour. \\~\\
To examine which PDS parameters most effectively distinguished the three EB classes, we trained a Random Forest classifier \citep{breiman2001random} using the 2806 visually classified systems in our sample. The six fitted PDS parameters were used as input features. The classifier achieved a 5-fold cross-validation accuracy of $87.6 \pm 2.8$ percent, and the corresponding feature-importance ranking was shown in Fig.~\ref{fig:rf_3}. Among the parameters, $\omega$ emerged as the strongest discriminator with an importance of $0.340$, while the constant vertical offset parameter ranked second at $0.238$. These normalized scores summed to unity and represented each parameter's fractional share of the classifier's total discriminative power across the three EB classes, with a score of $1$ indicating a feature that did all the work alone and $0$ indicating a feature that contributed nothing. Constant offset parameter was mainly associated with separating D systems from the other two classes. This was expected, since D systems often produced more polynomial-like chi-square plots, whereas C and SDEBs showed more continuously varying morphologies and therefore overlapped more strongly in baseline level. Since the broad separation between D and the other two systems was already largely captured by $P_{\rm PDS}$, we used the coefficient of determination (\( R^2 \)) as a complementary diagnostic of the PDS fit quality. In this context, $R^{2}$ acted as a shape indicator: high values corresponded to the well-behaved chi-square plots typically seen in C and D systems, while lower values highlighted the distorted plots more commonly associated with SD systems.

The PDS period $P_{\rm PDS}$, used as our principal physical descriptor throughout this work, was obtained directly from the fitted $\omega$ as $P_{\rm PDS} = 2\pi/\omega$. Since $\omega$ determined the spacing of extrema in box size, this period estimate was independent of the amplitude, baseline, or normalization applied to the chi-square plots.

\begin{figure*}
  \centering
  \begin{subfigure}{0.32\textwidth}
    \includegraphics[width=\textwidth]{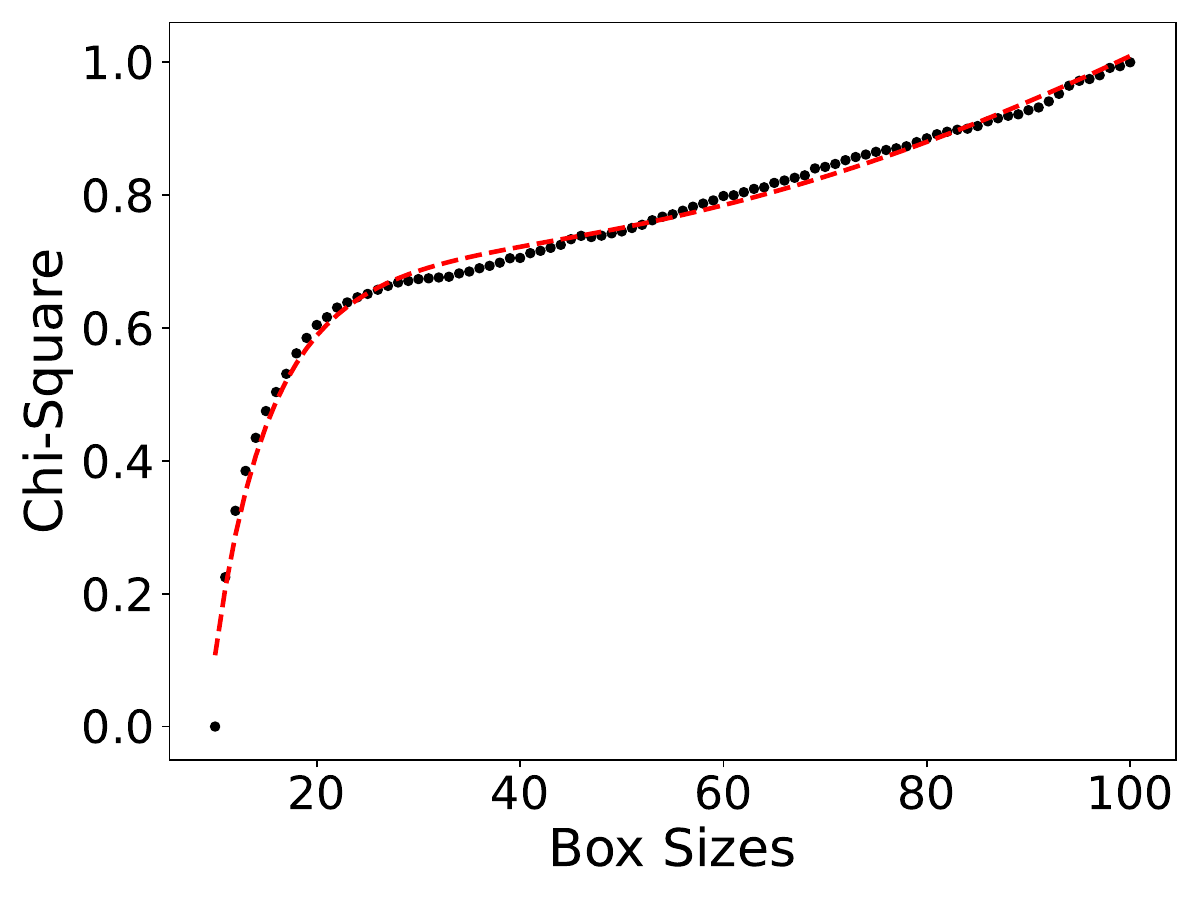}
    \caption{}
  \end{subfigure}
  \begin{subfigure}{0.32\textwidth}
    \includegraphics[width=\textwidth]{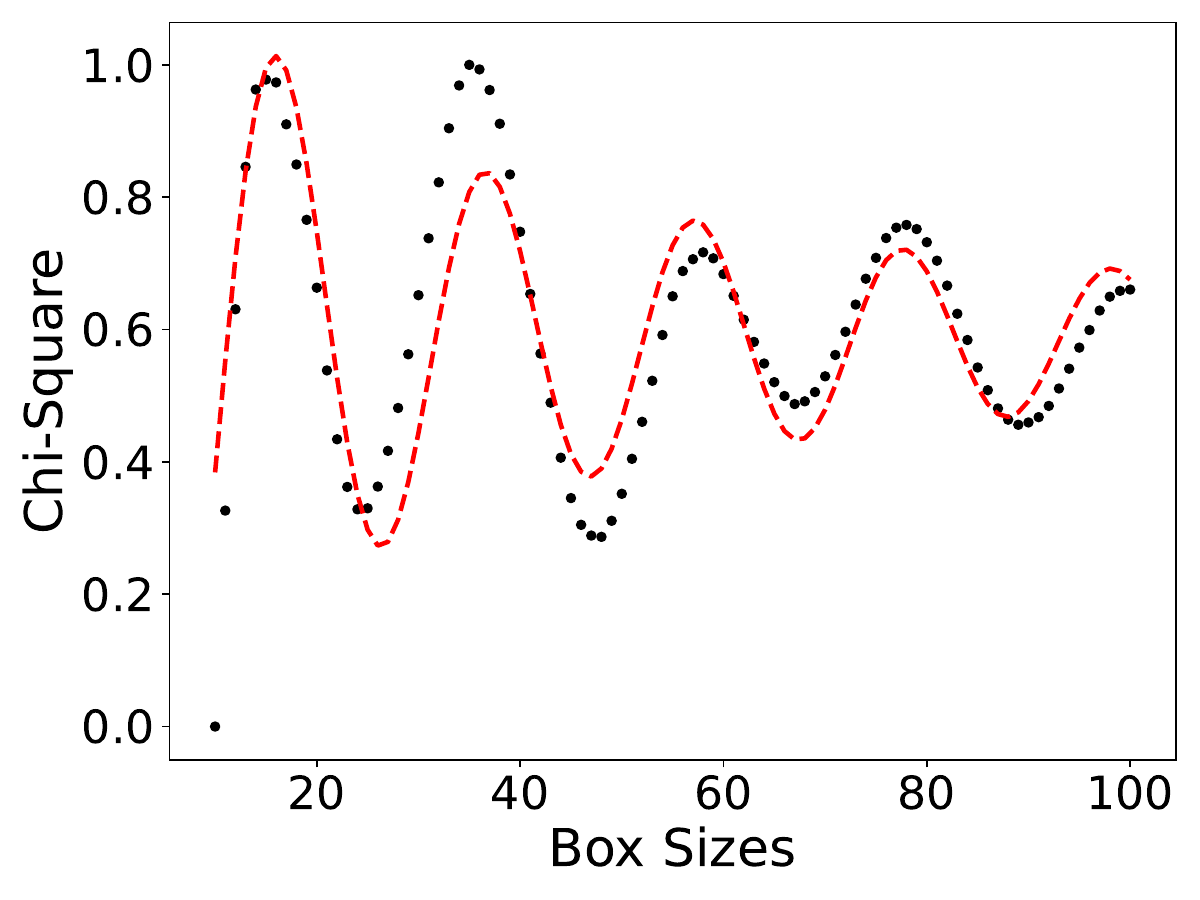}
    \caption{}
  \end{subfigure}
  \begin{subfigure}{0.32\textwidth}
    \includegraphics[width=\textwidth]{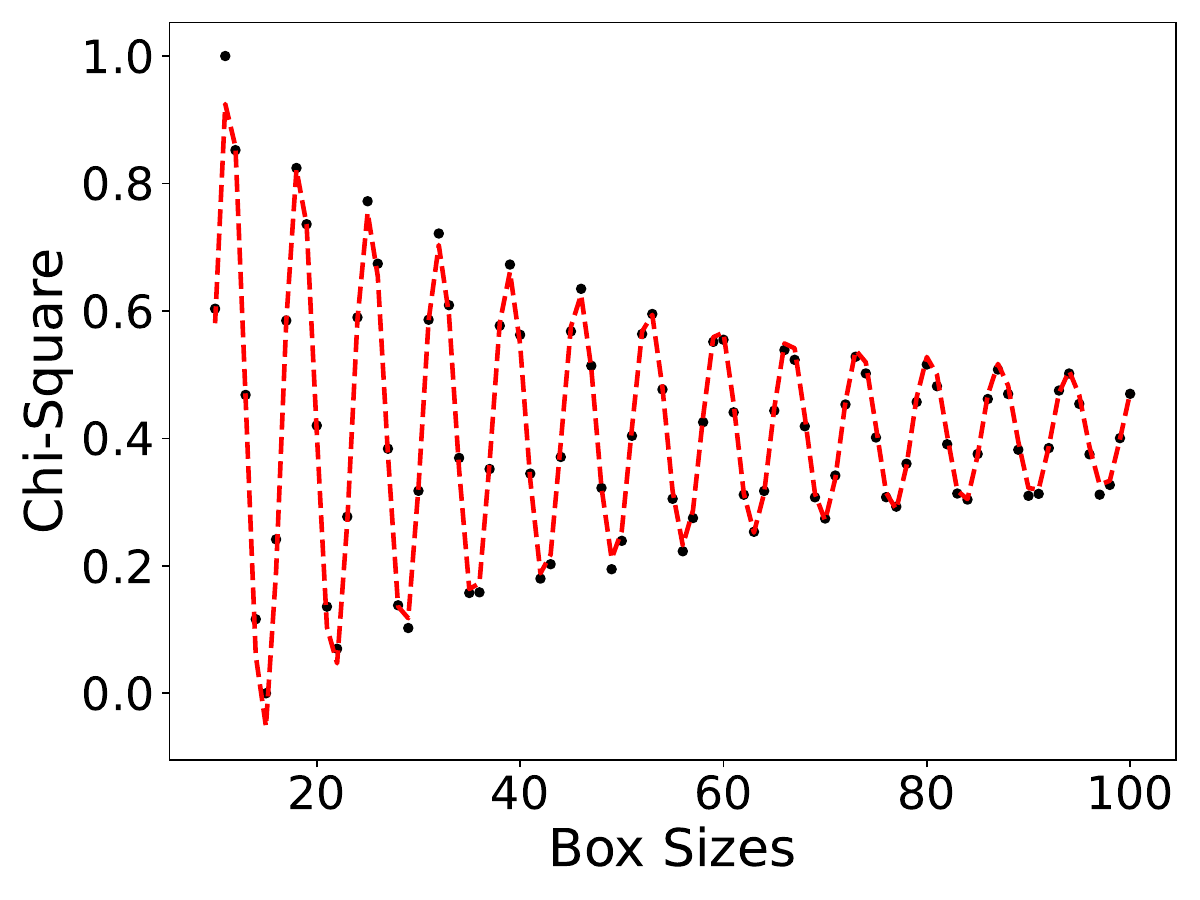}
    \caption{}
  \end{subfigure}
  \caption{Chi-square plot examples of three EB classes: (a) DEB (KIC~8299947); (b) SDEB (KIC~2577756); and (c) CEB (KIC~1433410). The chi-square values are Min–Max normalized to the range [0, 1], and the red curve shows the PDS function fitting}  
  \label{fig:f_01}
\end{figure*}

After selecting the initial fitting values and identifying $P_{\rm PDS}$ as the primary classification feature, our code proceeded with the fitting using the curve-fit method. We then adopted $R^2$ as an additional fit-quality metric to evaluate how well the PDS function described each chi-square curve. We used $R^2$ as a scale-independent indicator of how well the PDS function captured the shape of the normalized chi-square versus box-size plots; prior to fitting, the chi-square curves were rescaled to the range [0,1] to enable consistent comparison across targets. The $R^2$ scores for both DEBs and CEBs were consistently greater than 0.9, indicating good fits, as shown in Fig.~\ref{fig:f_01}~(a) and Fig.~\ref{fig:f_01}~(c), respectively. On the other hand, SDEBs showed comparatively poorer fits than the other two classes, as illustrated in Fig.~\ref{fig:f_01}~(b), where most SDEBs had $R^2$ values below 0.9, often by a substantial margin. Accordingly, while $P_{\rm PDS}$ effectively separated CEBs and DEBs, $R^2$ was adopted as a complementary diagnostic to identify SD systems, since the PDS model provided a comparatively poor description of their chi-square plots. The combined use of $P_{\rm PDS}$ and $R^2$ therefore enabled a more consistent classification across all three EB classes.

\begin{figure*}
  \centering
  \begin{subfigure}{0.39\textwidth}
    \includegraphics[width=\textwidth]{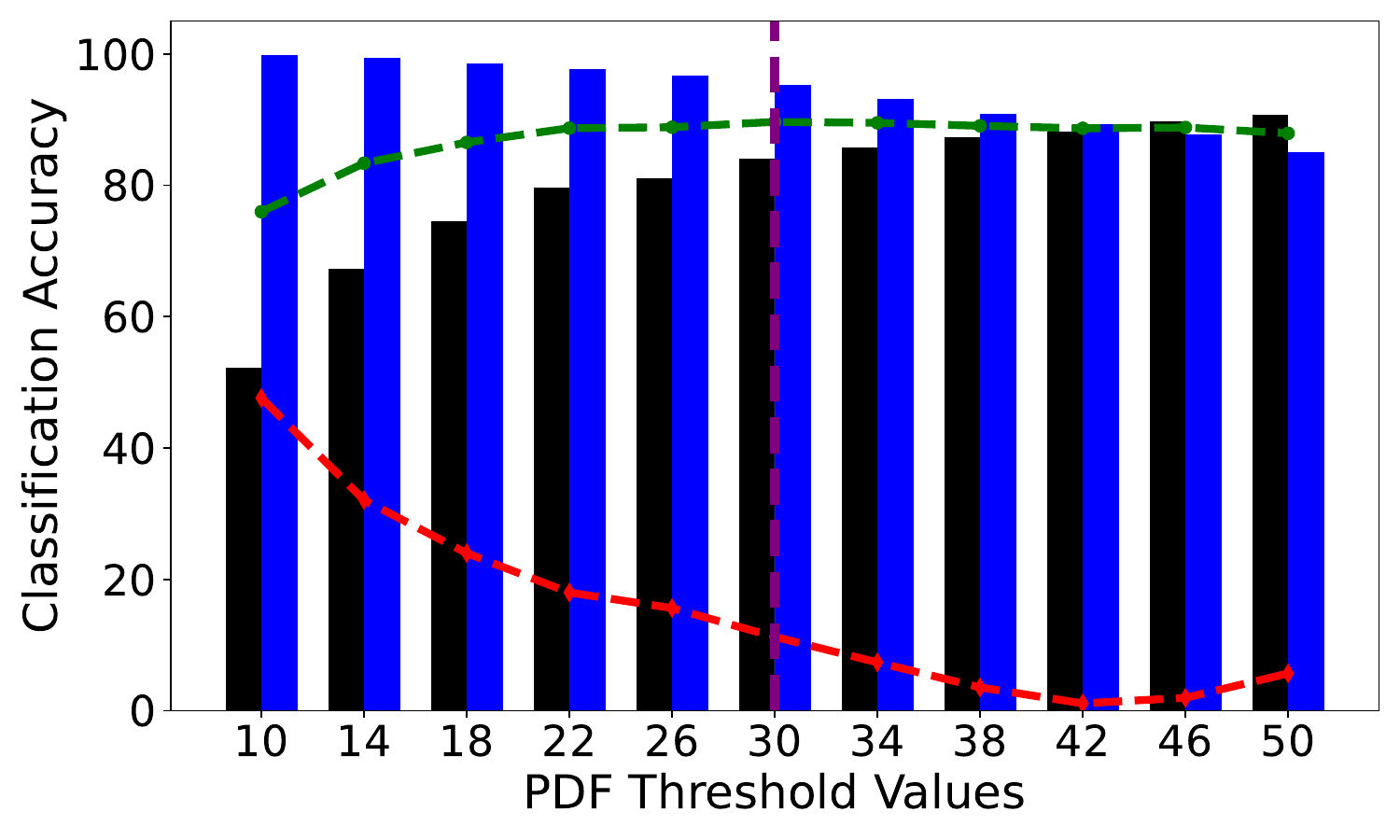}
    \caption{}
  \end{subfigure}
  \begin{subfigure}{0.39\textwidth}
    \includegraphics[width=\textwidth]{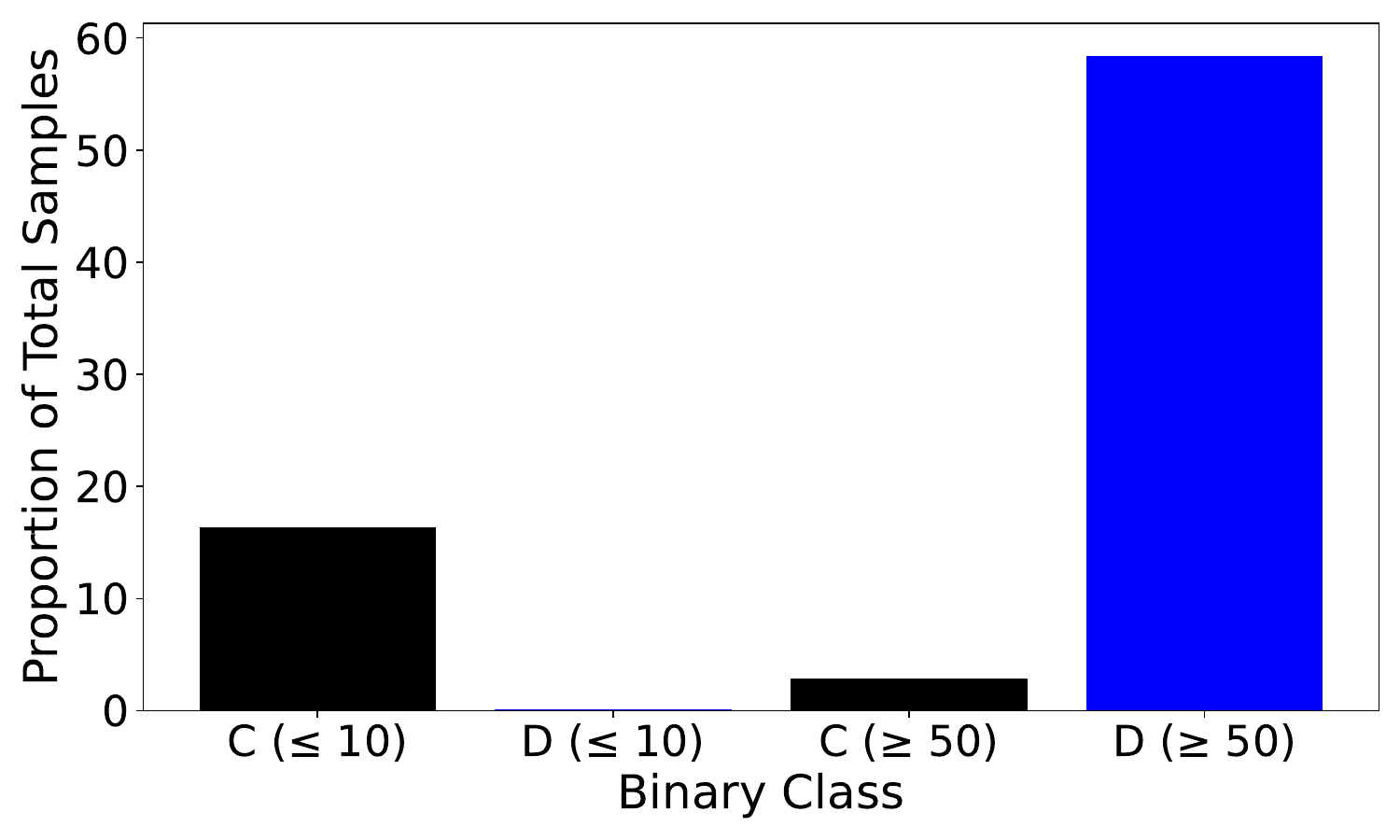}
    \caption{}
  \end{subfigure}
  \caption{(a) Plot shows how  different $P_{\rm PDS}$ values can separate visually identified CEBs (black) and DEBs (blue). The green dashed line represented the average classification accuracy for both CEBs and DEBs, the red dashed line indicated the accuracy difference between CEBs and DEBs, and the purple vertical line marked the threshold value of 30, which was the best $P_{\rm PDS}$ value for classification. (b) The x-axis in the second plot represented CEBs (black) and DEBs (blue) under two distinct $P_{\rm PDS}$ thresholds: $\leq 10$ and $\geq 50$. The y-axis illustrated the percentage of CEBs and DEBs in the entire sample, highlighting how their distribution varied across these thresholds. This comparison emphasized how each $P_{\rm PDS}$ range captured specific EB types effectively.}  \label{fig:f_0XX}
\end{figure*}

\begin{table*}
    \centering
    \caption {Different \( P_{\rm PDS} \) values related to  visually classified C and D systems and their ratios, along with the average ratio and ratio difference.}
    \label{tab:threshold_analysis}
    \begin{tabular}{ccccccccc}
        \hline
        $P_{\rm PDS}$ value & C Below & D Above & C Ratio (\%) & D Ratio (\%) & Average Ratio $(\overline{R}$, \%) & Ratio Difference (\%) \\
        \hline
        10 & 434 & 1816 & 52.16 & 99.78 & 75.97 & 47.62 \\
        14 & 560 & 1809 & 67.31 & 99.40 & 83.35 & 32.09 \\
        18 & 620 & 1793 & 74.52 & 98.52 & 86.52 & 24.00 \\
        22 & 663 & 1778 & 79.69 & 97.69 & 88.69 & 18.00 \\
        26 & 674 & 1759 & 81.01 & 96.65 & 88.83 & 15.64 \\
        30 & 699 & 1734 & 84.01 & 95.27 & 89.64 & 11.26 \\
        34 & 714 & 1696 & 85.82 & 93.19 & 89.50 & 7.37 \\
        38 & 726 & 1653 & 87.26 & 90.82 & 89.04 & 3.56 \\
        42 & 733 & 1624 & 88.10 & 89.23 & 88.67 & 1.13 \\
        46 & 747 & 1598 & 89.78 & 87.80 & 88.79 & 1.98 \\
        50 & 755 & 1548 & 90.75 & 85.05 & 87.90 & 5.69 \\
        \hline
    \end{tabular}
\end{table*}

Initially, we conducted our analysis on a small sample of 100 EBs. We found that CEBs generally exhibited shorter \( P_{\rm PDS} \) values compared to DEBs. 
To select the best \( P_{\rm PDS} \) value for classification, we first calculated how many systems visually classified as DEBs or CEBs fell under certain threshold values of \( P_{\rm PDS} \). In Table~\ref{tab:threshold_analysis}, we showed different \( P_{\rm PDS} \) values and, corresponding to each value, the percentage of visually identified CEBs that fell below the threshold, the percentage of visually identified DEBs that exceeded the threshold, as well as their average ratio percentage and ratio difference percentage.  
Our selection criteria focused solely on maximizing overall accuracy without favouring either class. Because DEBs and CEBs accounted for approximately 95 percent of our sample, we based the criteria on the classification performance of these two classes, iterating the defined thresholds as follows. In Fig.~\ref{fig:f_0XX}~(a), we presented the classification accuracy of CEBs (in black) and DEBs (in blue) for different $P_{\rm PDS}$ values using an initial combined sample of 100 systems. The accuracy was based on visual versus automated comparisons, with the automated classifications using different $P_{\rm PDS}$ thresholds shown in Fig.~\ref{fig:f_0XX}~(a) and the \( R^2 \) criteria listed in Table ~\ref{tab:bigdata1}. 
The green line in this plot represented the total average accuracy of CEBs and DEBs, and the red line represented the difference between C and D accuracy. 
We wanted to choose the optimum \( P_{\rm PDS} \) value for classification by comparing \( P_{\rm PDS} \) values from 10 to 50. 
We wanted to select a threshold \( P_{\rm PDS} \) value that was not biased toward any class. For example, if we selected a threshold \( P_{\rm PDS} \) value of 18, it would have given a CEB accuracy of 74 percent but a DEB accuracy of 98.5 percent. However, as DEBs dominated the sample, the overall accuracy would have increased, but we aimed to select a \( P_{\rm PDS} \) value that represented both classes equivalently with the highest accuracy. This is why the \( P_{\rm PDS} \) value of 30 was selected as the optimal choice. At this threshold, The overall misclassification rate, defined as $100 - \bar{R}$, where $\bar{R}$ represents the average classification ratio of CEBs and DEBs, reached its lowest value of 10.36 per cent at $P_{\rm PDS} = 30$, corresponding to an average classification ratio of 89.64 percent. This confirmed that $P_{\rm PDS} = 30$ provided the best balance between the two dominant classes. Furthermore, the CEBs accuracy reached 84.01 percent, while DEBs accuracy remained high at 95.27 percent. This combination provided a reasonable balance, ensuring that neither class was disproportionately prioritized, as shown in Table~\ref{tab:threshold_analysis}. The \( P_{\rm PDS} \) value of 30 thus represented the best compromise between high overall accuracy and fairness in class representation. The purple vertical line depicted \( P_{\rm PDS} \) value 30 in Fig.~\ref{fig:f_0XX}~(a). Along with this, in Fig.~\ref{fig:f_0XX}~(b), we also showed that when \( P_{\rm PDS} \) value was 10 or less, EBs were primarily CEBs, with only 0.15 percent of samples being DEBs. When \( P_{\rm PDS} \) value was 50 or higher, EBs were primarily DEBs, with only 2.9 percent of samples being CEBs.

\begin{table}
    \centering
    \caption{Adopted criteria in our PDS-fitting method.}
    \label{tab:bigdata1}
    \begin{tabular}{lc}
        \hline
        Classification & Fitting Criteria \\
        \hline
        Contact & \(( R^2 \geq 0.9\ \text{and}\ P_{\rm PDS} < 30) \text{ or } P_{\rm PDS} \leq 10\) \\
        Detached & \(( R^2 \geq 0.9\ \text{and}\ P_{\rm PDS} \geq 30) \text{ or } P_{\rm PDS} \geq 50\) \\
        Semi-Detached & \((R^2 < 0.9)\) \\
        \hline
    \end{tabular}
\end{table}

Finally, we used a combination of both $R^2$ and $P_{\rm PDS}$ to distinguish between CEB and DEB systems. The final classification criteria were presented in Table ~\ref{tab:bigdata1}. 
Compared with the visual inspection, out of 832 CEBs, 680 were correctly classified by the PDS function, yielding an accuracy of 81.7 percent. Out of 1,820 DEBs, 1,714 were classified as D, yielding an accuracy of 94.1 percent. Finally, out of 154 SDEBs, only 34 were classified as such, yielding a poor accuracy of 22.1 percent. Had we prioritized SDEB accuracy, our overall performance would have suffered significantly; therefore, we chose a method that was biased toward the two dominant classes (CEBs and DEBs) and incorporated the $R^2$ criterion to recover some SDEBs. Considering the whole sample, the total accuracy obtained with the PDS-function fitting was 86.5 percent. 

Table~\ref{tab:DD2}  presented a portion of our classification results, showing first 10 systems from each morphological class (C, SD, D) in ascending order based on their {\sl Kepler} ID. 
The complete version of this table, listing all 2,865 targets, is available online.

Analyzing the misclassified cases, we found some EBs with clearly different class assignments when comparing the visual classification of the phase-folded LC and the chi-squared plot fitting procedure, as well as targets with curves that could not be clearly discriminated. Since the PDS function was an attempt to explain the chi-squared plot behaviour and, in general, it fitted the SDEBs poorly, the results were not ideal and presented lower accuracy than initially expected. Furthermore, the $P_{\rm PDS}$ values are directly correlated with the binary $P_{\rm orb}$, which ultimately represents a separation of periods rather than a purely morphological classification feature, as discussed further in Section~\ref{sec:dis}. Therefore, we adopted a deep-learning technique to address the classification problem.

\begin{table*}
    \centering
    \caption{The table presents the classification of the first 10 CEBs, DEBs, and SDEBs that are common across all three methods and are sorted by their {\sl Kepler} ID in ascending order. The {\sl Kepler} ID column represents the unique Kepler identification number, while the Visual column indicates the visual classification of those targets. The $R^2$ column denotes the goodness-of-fit of the PDS function, and the $P_{\rm PDS}$ column lists the periods derived from the PDS method. The PDS Classification column refers to the classification output from the PDS fitting method. The Prediction of CNN1 column shows the classification probability from a 1D Convolutional Neural Network trained on {\sl Kepler} all-quarter light curve data, and CNN1 represents the corresponding classification output. Similarly, Prediction of CNN2 provides the classification probability from a 1D-CNN model trained on both {\sl Kepler} data and synthetic data generated using PHOEBE, with the final classification output shown in the CNN2 column. The full version of this table is available online.}
    \label{tab:DD2}
    \begin{tabular}{lcccccccc}
        \hline
        {\sl Kepler} ID & Visual Classification & \( R^2 \) & $P_{\rm PDS}$ & PDS Classification & Prediction of CNN1 & CNN1 & Prediction of CNN2 & CNN2 \\
        \hline
KIC 002437038 & C & 0.988 & 6.489 & C & 0.957 & C & 0.915 & C \\
KIC 002571439 & C & 0.987 & 7.787 & C & 0.969 & C & 0.879 & C \\
KIC 002715417 & C & 0.996 & 5.725 & C & 0.907 & C & 0.924 & C \\
KIC 002717141 & C & 0.998 & 9.560 & C & 0.889 & C & 0.758 & C \\
KIC 002858322 & C & 0.816 & 10.643 & S.D & 0.526 & S.D & 0.643 & C \\
KIC 003127873 & C & 0.996 & 16.302 & C & 0.737 & C & 0.529 & C \\
KIC 003221207 & C & 0.997 & 11.489 & C & 0.717 & C & 0.943 & C \\
KIC 003232823 & C & 0.992 & 10.264 & C & 0.943 & C & 0.773 & C \\
KIC 003353233 & C & 0.988 & 7.664 & C & 0.969 & C & 0.840 & C \\
KIC 003437800 & C & 0.995 & 8.796 & C & 0.918 & C & 0.878 & C \\
KIC 001026957 & D & 0.998 & 211.866 & D & 0.956 & D & 0.792 & D \\
KIC 001571511 & D & 0.993 & 41.430 & D & 0.946 & D & 0.872 & D \\
KIC 002438490 & D & 0.998 & 122.812 & D & 0.962 & D & 0.796 & D \\
KIC 002576692 & D & 0.995 & 56.430 & D & 0.931 & D & 0.631 & D \\
KIC 002719873 & D & 0.952 & 78.341 & D & 0.945 & D & 0.941 & D \\
KIC 002860788 & D & 0.894 & 1960.258 & D & 0.964 & D & 0.934 & D \\
KIC 002987433 & D & 0.961 & 13695137.660 & D & 0.946 & D & 0.852 & D \\
KIC 003102024 & D & 0.960 & 140.488 & D & 0.930 & D & 0.922 & D \\
KIC 003122985 & D & 0.995 & 77.939 & D & 0.730 & D & 0.640 & D \\
KIC 003230227 & D & 0.993 & 16355.969 & D & 0.869 & D & 0.896 & D \\
KIC 002305372 & S.D & 0.906 & 32.274 & D & 0.887 & D & 0.932 & D \\
KIC 002577756 & S.D & 0.743 & 20.478 & S.D & 0.670 & S.D & 0.425 & C \\
KIC 003730067 & S.D & 0.991 & 7.114 & C & 0.961 & C & 0.868 & C \\
KIC 005077994 & S.D & 0.990 & 16.814 & C & 0.719 & S.D & 0.417 & C \\
KIC 005962514 & S.D & 0.988 & 37.378 & D & 0.590 & C & 0.631 & D \\
KIC 006046061 & S.D & 0.116 & 25941846.164 & D & 0.969 & C & 0.832 & C \\
KIC 006220497 & S.D & 0.987 & 31.425 & D & 0.717 & S.D & 0.912 & D \\
KIC 006387887 & S.D & 0.995 & 10.510 & C & 0.929 & C & 0.776 & C \\
KIC 006443392 & S.D & 0.931 & 18.629 & C & 0.776 & S.D & 0.539 & C \\
KIC 006516874 & S.D & 0.505 & 29450204.538 & D & 0.860 & S.D & 0.549 & S.D \\
        \hline
    \end{tabular}
    \vspace{0.9cm}
\end{table*}


\subsection{Classification using a deep-learning method}\label{sec:ai}

We employed the convolutional neural network (CNN) approach to perform more in-depth pattern recognition on chi-square plots and to enhance the previous classification accuracy across three classes. We utilized a two-step approach. First, we normalized the chi-square data to bring it to the same scale, which helped in analyzing the meaningful structure of the plots rather than their varying magnitudes. For this purpose, we utilized a Min--Max normalization technique, which scaled the data values between $[0, 1]$. This normalization was particularly suitable for the adopted one-dimensional CNN (1D-CNN) model, as neural networks generally performed better when input data was standardized to a fixed range, ensuring faster convergence and improved training stability \citep{pedregosa2011scikit}.

Before selecting the final architecture, we evaluated multiple pattern-recognition approaches, including Fourier-based features, wavelet-transformed representations, and higher-dimensional CNN variants (e.g., 2D and 3D CNNs). While these methods captured certain global characteristics of the sequences, they did not provide consistent improvements in validation accuracy and often increased model complexity or training instability.

Given that the Chi-square curves represent structured one-dimensional sequences, a 1D CNN provides a natural inductive bias by learning local morphological patterns directly from the ordered data. 1D-CNNs have been widely and successfully applied in astronomical time-series analysis, including light-curve classification \citep{2021A&C....3600488C} and solar flare prediction \citep{2025ApJS..280...52L}, where they effectively capture local temporal features while maintaining computational efficiency. Motivated by these successful 1D-CNN applications, we adopted a compact 1D-CNN with two Conv1D layers per branch and a lightweight three-branch ensemble, which provided stable convergence and strong performance without resorting to deeper architectures.

The model followed a three-branch 1D-CNN design implemented through two custom functions. The first function (\texttt{model\_architect}) built a single-scale 1D-CNN with five convolutional layers (kernel size of 8, $\tanh$ activation), followed by a global max-pooling layer that captured the most informative local patterns across the input sequence. The pooled output was passed through a Dense layer with 50 units and then a dropout layer with a rate of 10 percent to reduce overfitting. The second function (\texttt{core\_model}) instantiated three parallel branches with identical architecture, each processing the same input sequence. The resulting feature representations were concatenated into a 150-dimensional vector and passed to a dense softmax classifier to predict the probabilities of the three classes (C, SD, D). Training used the Adam optimizer \citep{kingma2014adam} together with a focal loss function $\gamma = 2, \alpha = 0.25$ \citep{8417976}, which reduced the impact of easily classified examples and focused on harder ones. To prevent overfitting, early stopping with a patience of 20 epochs was applied. The model was trained with a batch size of 64, which provided a balance between efficiency and stable gradient updates, and a maximum of 100 epochs to ensure convergence without excessive training. For clarity and reproducibility, the complete CNN architecture, including all layers and hyperparameters, is summarized in Table ~\ref{tab:cnn_architecture}. The above complete CNN algorithm is available on GitHub\footnote{\url{https://github.com/mousam100}}.
\begin{table*}
    \centering
    \caption{Summary of the CNN architecture and training configuration used for morphological classification of EBs.}
    \label{tab:cnn_architecture}
    \begin{tabular}{p{4cm} p{11cm}}
        \hline
        \textbf{Architecture component} & \textbf{Architecture details} \\
        \hline

        Input representation &
        Each sample is a chi-square-boxsize time-series stored as a sequence of length 91 with two channels (box size and chi-square value), giving an input shape of $(91,2)$. \\

        \texttt{model\_architect} (single branch) &
        Returns a Keras sub-model that maps an input sequence of shape $(91,2)$ to a 50-dimensional embedding: \\
        & Conv1D (50 filters, kernel size = 8, $\tanh$, same padding) $\rightarrow$
        Conv1D ($\times$4, 50 filters, kernel size = 8, $\tanh$, same padding) $\rightarrow$
        GlobalMaxPooling1D $\rightarrow$
        Dense (50, $\tanh$) $\rightarrow$
        Dropout (0.1). \\

        \texttt{core\_model} (multi-branch assembly) &
        Instantiates three parallel branches by calling \texttt{model\_architect} three times and applying each branch to an input of shape $(91,2)$. The three branch outputs are concatenated and passed to a softmax classifier. \\

        Branch configuration &
        Three branches with identical layer structure and hyperparameters. In the implemented code, branches are instantiated separately (i.e., each branch has its own learnable weights). \\

        Merge &
        Concatenation of three embeddings ($3 \times 50 = 150$). \\

        Classifier &
        Dense (3, softmax). \\

        Loss function &
        Focal loss ($\gamma = 2$, $\alpha = 0.25$). \\

        Optimizer &
        Adam. \\

        Training setup &
        Batch size = 64; maximum epochs = 100; early stopping (patience = 20, monitored on validation loss, best weights restored). \\
        \hline
    \end{tabular}
\end{table*}

\subsubsection{CNN with Kepler Data}\label{sec:ai1} 

CNN required a large amount of data to train effectively. However, in our {\sl Kepler} sample, we only had 2806 EB systems. Therefore, we used the chi-square plots of all the quarters from all systems to increase our dataset. Chi-square plots for most targets exhibited similar patterns across quarters. Here, we used a labelled train-test split, where the quarters with the highest number of epochs, the same analyzed in Section~\ref{sec:mor1}, were included in the testing set only and were randomly selected. For training, we used chi-square plots from the remaining sample; we have also seen that some chi-square plots show contradiction to their majority shape, and we have taken out these targets from the training set and only selected those matching Fig.~\ref{fig:f_003} and which were consistent across all quarters. 

To maintain the randomness of the split and avoid any biased results, a Python-based random sampling algorithm was applied. After excluding 9 per cent of the EB sample whose chi-square plots showed morphologies inconsistent with their assigned classes, the final dataset consisted of 48 per cent of the objects for training and 43 per cent for testing. These EBs were excluded to help the model learn representative patterns of the majority classes, and we discussed these systems in more detail in Section~\ref{sec:dis}. The test set contained 1,220 EB systems, including 800 DEBs, 350 CEBs, and 70 SDEBs, and the 48 percent training set contained 21,564 chi-square plots in total: 14,370 plots from 900 DEBs, 6,639 plots from 423 CEBs, and 555 plots from 36 SDEBs. From this training set, we further selected 20 percent (4,313 plots) for validation during model training. Since the chi-square plots were mixed across systems, it was not possible to specify how many individual EBs contributed to the validation subset, but the proportion of chi-square plots remained 20 percent of the training set.

\begin{figure}
  \centering
  \begin{subfigure}{0.39\textwidth}
    \includegraphics[width=\textwidth]{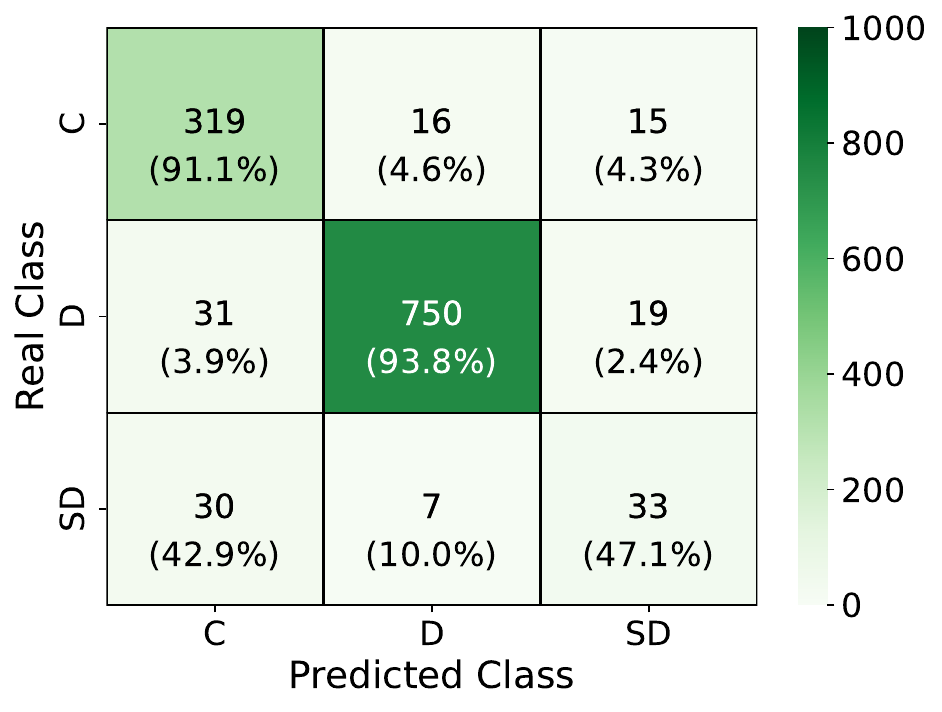}
    \caption{}
  \end{subfigure}
  \begin{subfigure}{0.39\textwidth}
    \includegraphics[width=\textwidth]{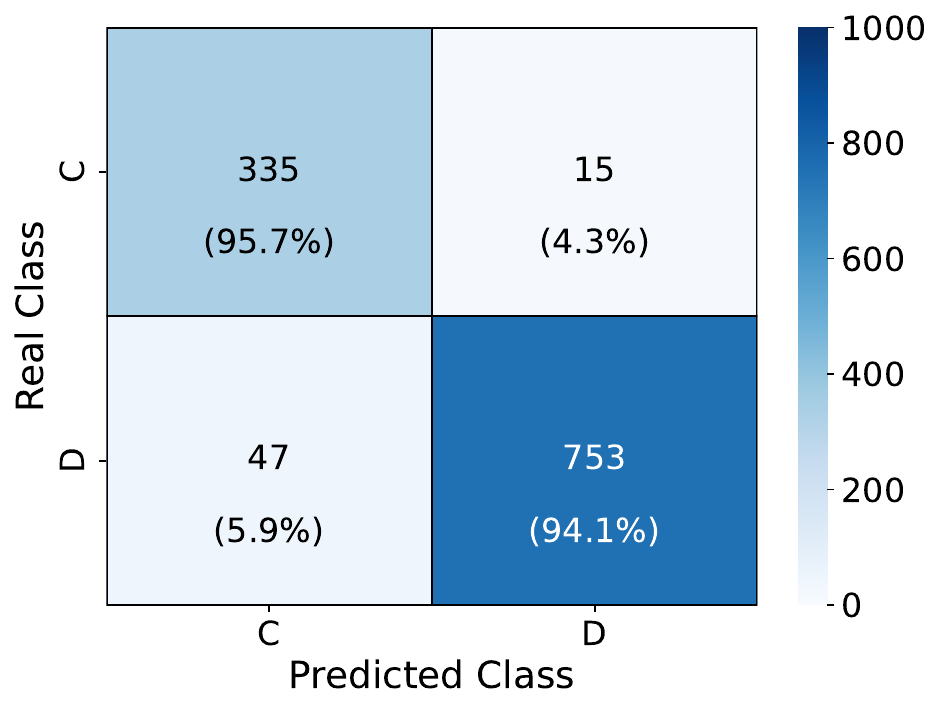}
    \caption{}
  \end{subfigure}
  \caption{The confusion matrix of the 1D-CNN model using chi-square plots from {\it Kepler} data is shown here. (a) Displays the confusion matrix for all three EB classes: CEB, DEB, and SDEB. (b) Displays the confusion matrix for the binary classification involving only the two major classes: CEB and DEB.}
  \label{fig:f_M2}
\end{figure}

Our model achieved an overall accuracy of 90.3 percent when classifying into three categories. Among the 800 DEBs, 750 were correctly classified, giving an accuracy of 93.8 percent. For the 350 CEBs, 319 were accurately identified, resulting in an accuracy of 91.1 percent. Lastly, from the 70 SDEBs, 33 were correctly detected, yielding an accuracy of 47 percent. These results were  depicted in the confusion matrix in Fig.~\ref{fig:f_M2}(a). 

Some of the SD chi-square plots had significant similarities with those from C and D systems. Therefore, in our second experiment, we classified only CEBs and DEBs, excluding the SDEBs, to assess the model’s accuracy without the uncertainty introduced by SD systems. We used the same labeled train–test and validation sample proportion method as in the previous experiment and kept the same DEB and CEB sample sizes, excluding only the SDEBs. A total of 14,370 chi-square plots from 900 DEBs and 6,639 plots from 423 CEBs were used for training. In the test data, out of 350 CEBs, 335 were correctly classified, which resulted in an accuracy of 95.7 percent. Similarly, out of 800 DEBs, 753 were correctly classified, which gave an accuracy of 94.1 percent. This brought the overall classification accuracy to 95 percent. The confusion matrix for this classification was shown in Fig.~\ref{fig:f_M2}(b). 

The precision, recall, and F1-score metrics offer valuable insights into the accuracy of a CNN model's classification of the test sample. Precision measures the accuracy of true positive predictions out of all positive predictions, with a high precision score indicating a high likelihood of true positive prediction. Recall, also known as the true positive rate, gauges the precision or true positive rate out of all positive instances. The F1 score balances precision and recall, serving as the harmonic mean of these two metrics. Together, these three scores provide information about the model's performance and its effectiveness in classification tasks.In Table~\ref{tab:M3}, we presented the Precision, Recall, and F1-scores for these two analyses. The columns labeled C (CNN1), D (CNN1), and SD (CNN1) represent the metrics using only the {\sl Kepler} data, where the first three rows of CNN1 correspond to the first analysis -- considering all three morphological classes-- and the last three rows correspond to the analysis with only CEBs and DEBs.

One of the possible reasons why our SDEB classification do not achieve the expected quality parameters, can be attributed to the very limited sample size -- only 154 SDEBs in total, of which just 106 were used for training and testing -- posing a challenge for the model to learn reliable patterns. Unlike CEBs or DEBs, SDEB chi-square plots often lacked clear distinguishing features: many exhibited distorted damping behaviour, as shown in Fig.~\ref{fig:f_003}(d,e), while others resembled CEB-like curves or the examples shown in Fig.~\ref{fig:f_003}(f). Consequently, our SDEB classification rates remained low. To address these issues, we augmented our dataset by combining actual {\it Kepler} light curves with synthetic light curves generated using PHOEBE modelling code \citep{prsa2011phoebe}.

\begin{table*}
	\centering
	\caption{Precision, Recall, and F1-scores for the classification of CEBs, DEBs, and SDEBs using chi-square plots from the {\it Kepler} and {\it Kepler + PHOEBE} datasets. CNN1 represents the experiment conducted with only {\it Kepler} data, while CNN2 represents the experiment conducted with the combined {\it Kepler} and {\it PHOEBE} data. The first three rows show the results for the C, D, and SD classes, and the following three rows show the results for only CEBs and DEBs.}
	\label{tab:M3}
	\begin{tabular}{lcccccc} 
		\hline
		\textbf{Metric} & C (CNN1) & D (CNN1) & SD (CNN1) & C (CNN2) & D (CNN2) & SD (CNN2) \\
		\hline
		Precision & 0.84 & 0.97 & 0.49 & 0.98 & 0.98 & 0.96 \\
		Recall    & 0.91 & 0.94 & 0.47 & 0.99 & 0.99 & 0.77 \\
		F1-score  & 0.87 & 0.95 & 0.48 & 0.98 & 0.99 & 0.86 \\
        \hline
		Precision & 0.88 & 0.98 & ...  & 0.99 & 0.98 & ...  \\
		Recall    & 0.96 & 0.94 & ...  & 0.98 & 0.99 & ...  \\
		F1-score  & 0.92 & 0.96 & ...  & 0.98 & 0.99 & ...  \\
		\hline
	\end{tabular}
\end{table*}

\subsubsection{CNN using {\it Kepler} and PHOEBE data combined}

PHOEBE \citep{prvsa2016physics} is an EB modelling code based on the Wilson--Devinney model \citep{wilson1971realization}. Using the PHOEBE forward-modelling algorithm,\footnote{We used 
PHOEBE version~2.4.15.} we generated synthetic LCs along with three-dimensional visualizations of each system’s surface geometry and orbital configuration (mesh plots) for DEB, SDEB, and CEBs. From these synthetic LCs, we then produced the corresponding chi-square vs box-size plots. To construct the models, we adopted appropriate PHOEBE configurations for D, SD, and CEB systems and varied key physical parameters such as mass ratio, \( P_{\rm orb} \) , effective temperatures of both components, orbital inclination, and orbital separation. To ensure that the simulated chi-square plots were based on physically realistic systems, we used representative systems from the Catalog and Atlas of Eclipsing Binaries \citep[CALEB;][]{bradstreet2004catalog} as templates. The specific CALEB systems used were as follows: for DEBs, V541 Cyg (15.337 days), LV Her (18.435 days), and AI Phe (24.592 days); for SDEBs, DD Mon (0.568 days), XZ CMi (0.579 days), and BO Peg (0.58 days); and for CEBs, CC Com (0.2207 days), V523 Cas (0.2337 days), and BS Vul (0.4759 days). These systems were selected because their \( P_{\rm orb} \) values were consistent with the range observed in our {\it Kepler} sample. From these representative systems, we generated a large number of synthetic LCs by systematically varying the binary parameters across physically plausible ranges. The resulting set of simulated LCs was then used to generate synthetic chi-square plots following the same procedure applied to the {\it Kepler} data. This simulated dataset was subsequently used to train our 1D-CNN model.

\begin{figure*}
  \centering
  \begin{subfigure}{0.3\textwidth}
    \includegraphics[width=\textwidth]{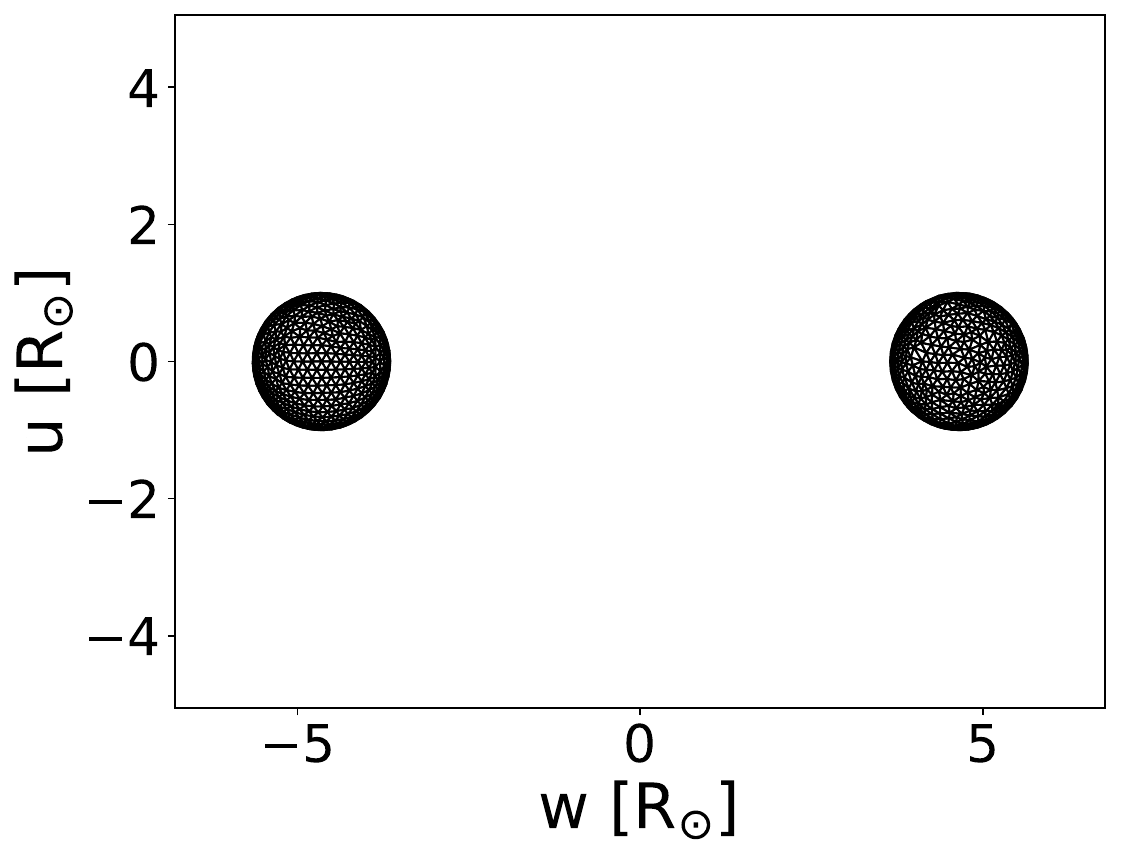}
    \caption{}
  \end{subfigure}
  \begin{subfigure}{0.315\textwidth}
    \includegraphics[width=\textwidth]{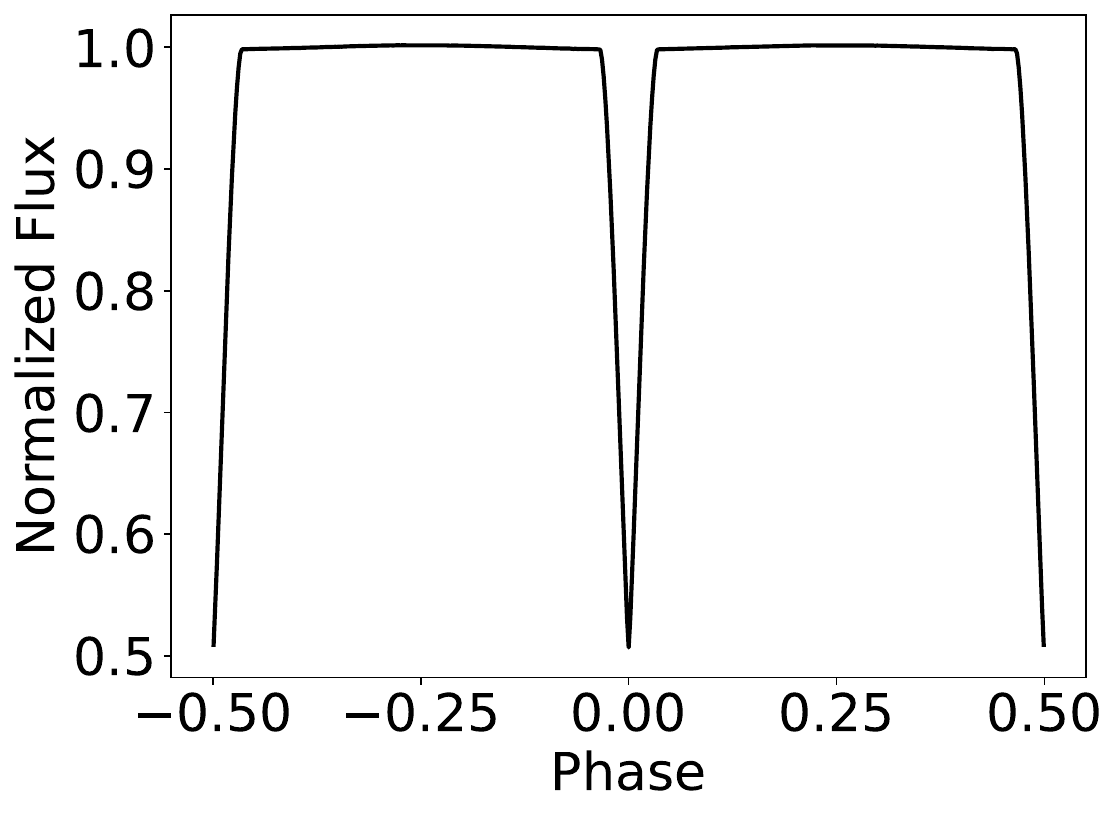}
    \caption{}
  \end{subfigure}
  \begin{subfigure}{0.32\textwidth}
    \includegraphics[width=\textwidth]{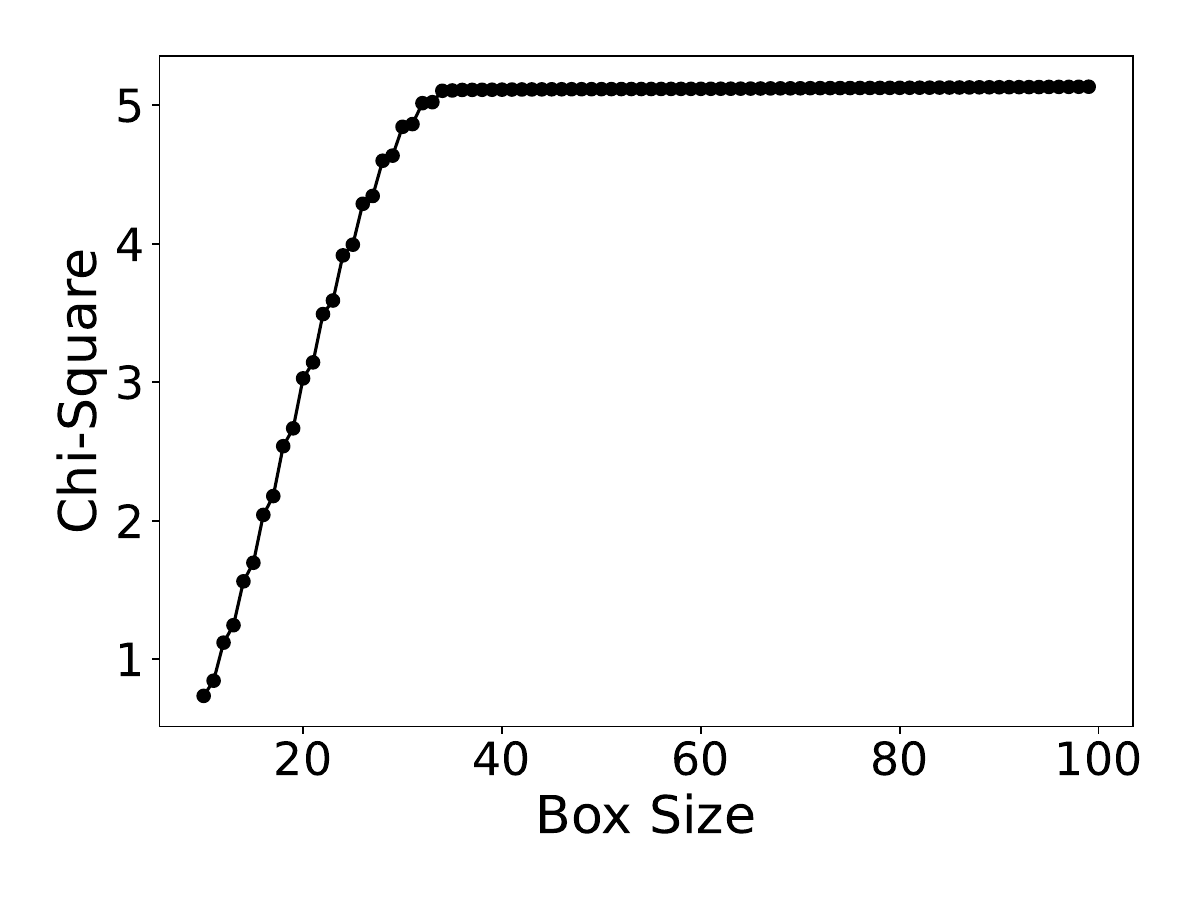}
    \caption{}
  \end{subfigure}
  \begin{subfigure}{0.3\textwidth}
    \includegraphics[width=\textwidth]{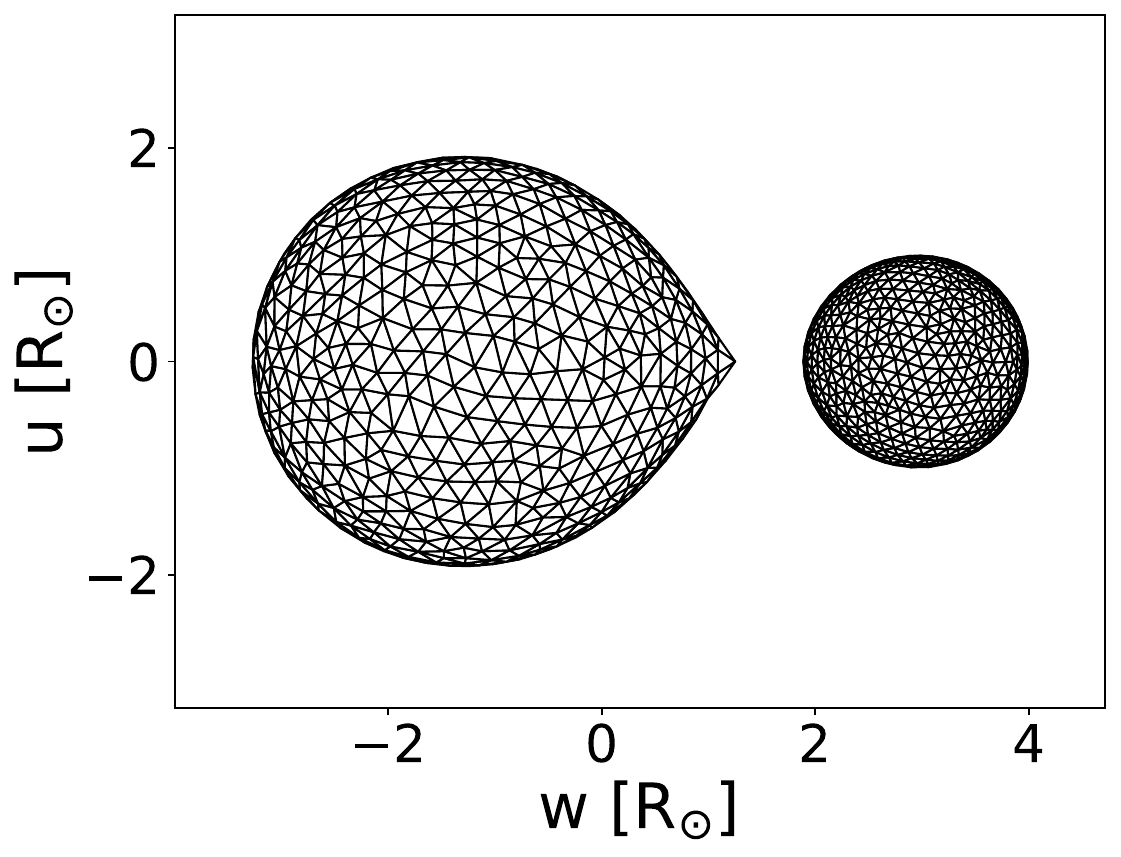}
    \caption{}
  \end{subfigure}
  \begin{subfigure}{0.315\textwidth}
    \includegraphics[width=\textwidth]{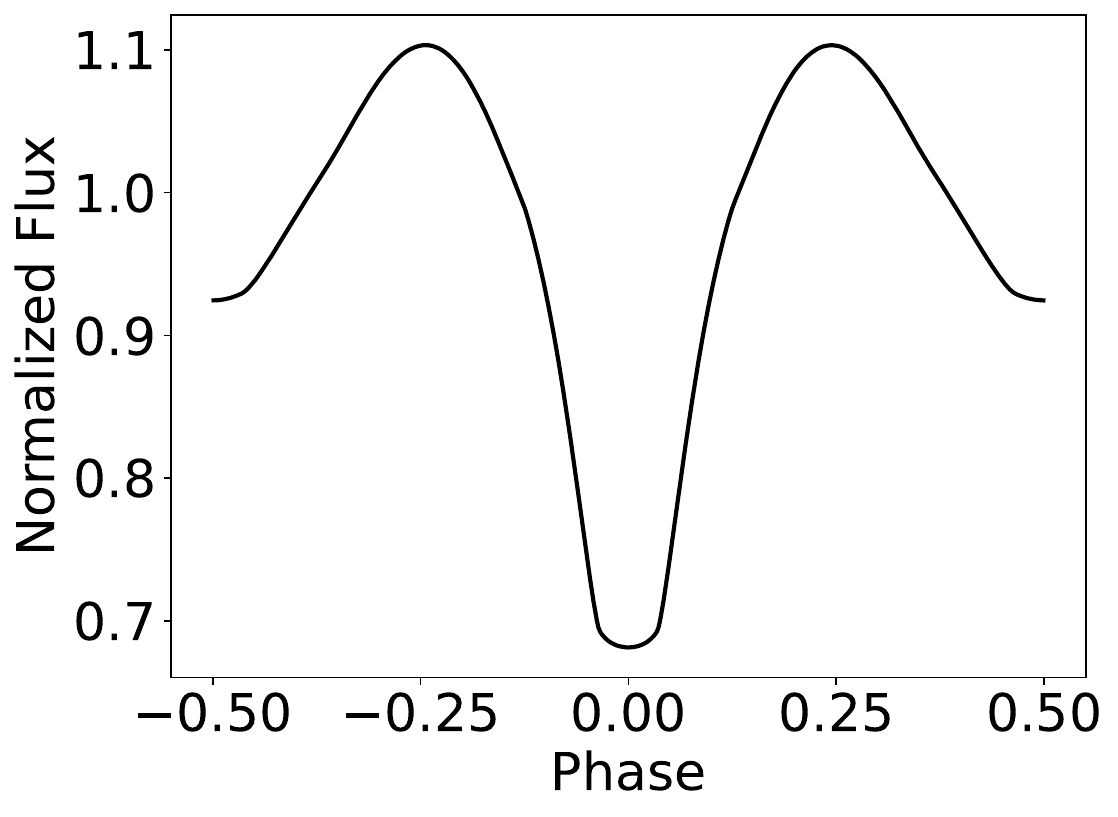}
    \caption{}
  \end{subfigure}
  \begin{subfigure}{0.32\textwidth}
    \includegraphics[width=\textwidth]{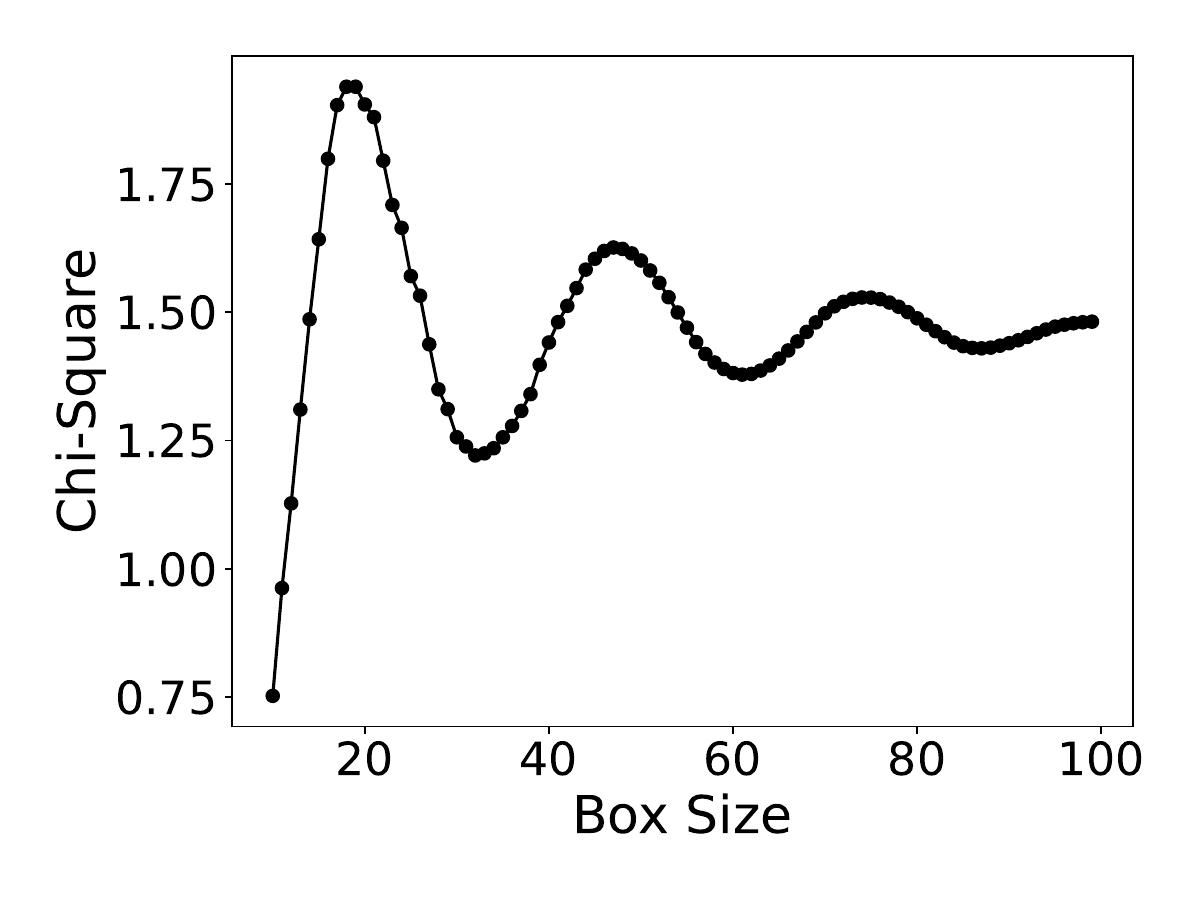}
    \caption{}
  \end{subfigure}
  \begin{subfigure}{0.3\textwidth}
    \includegraphics[width=\textwidth]{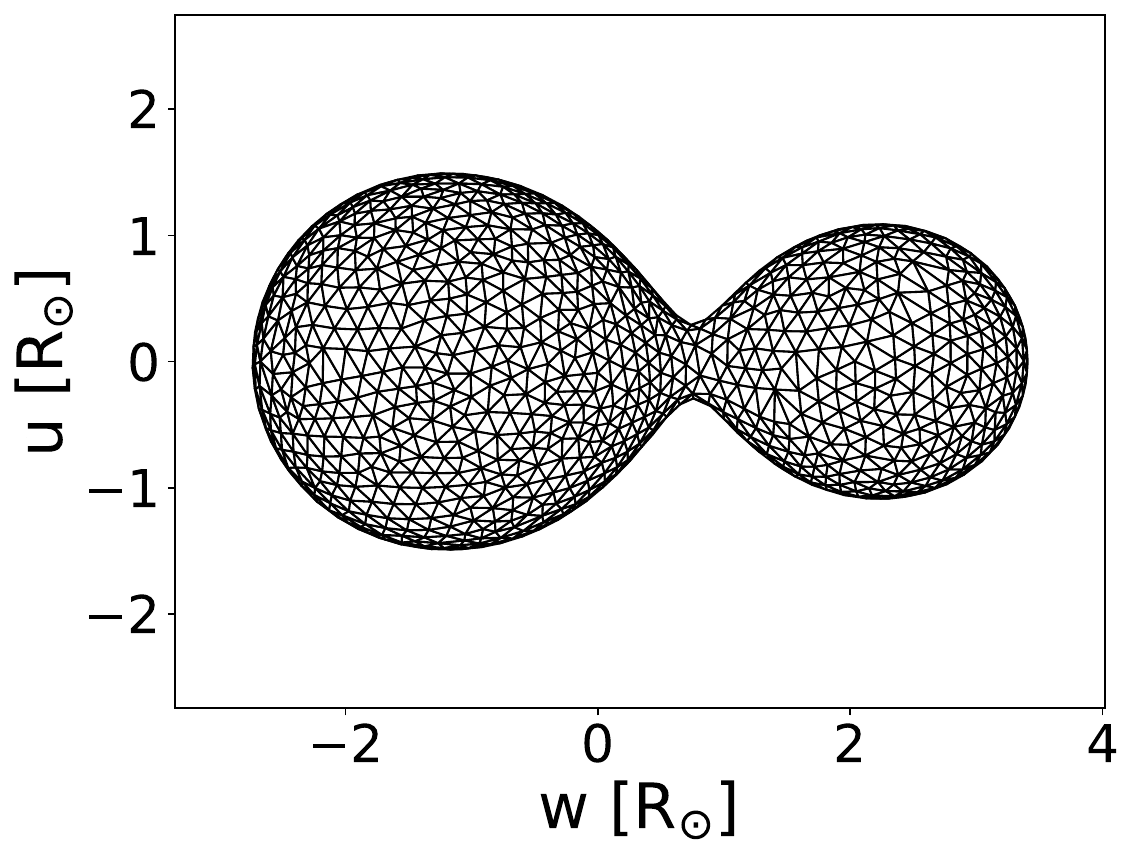}
    \caption{}
  \end{subfigure}
  \begin{subfigure}{0.315\textwidth}
    \includegraphics[width=\textwidth]{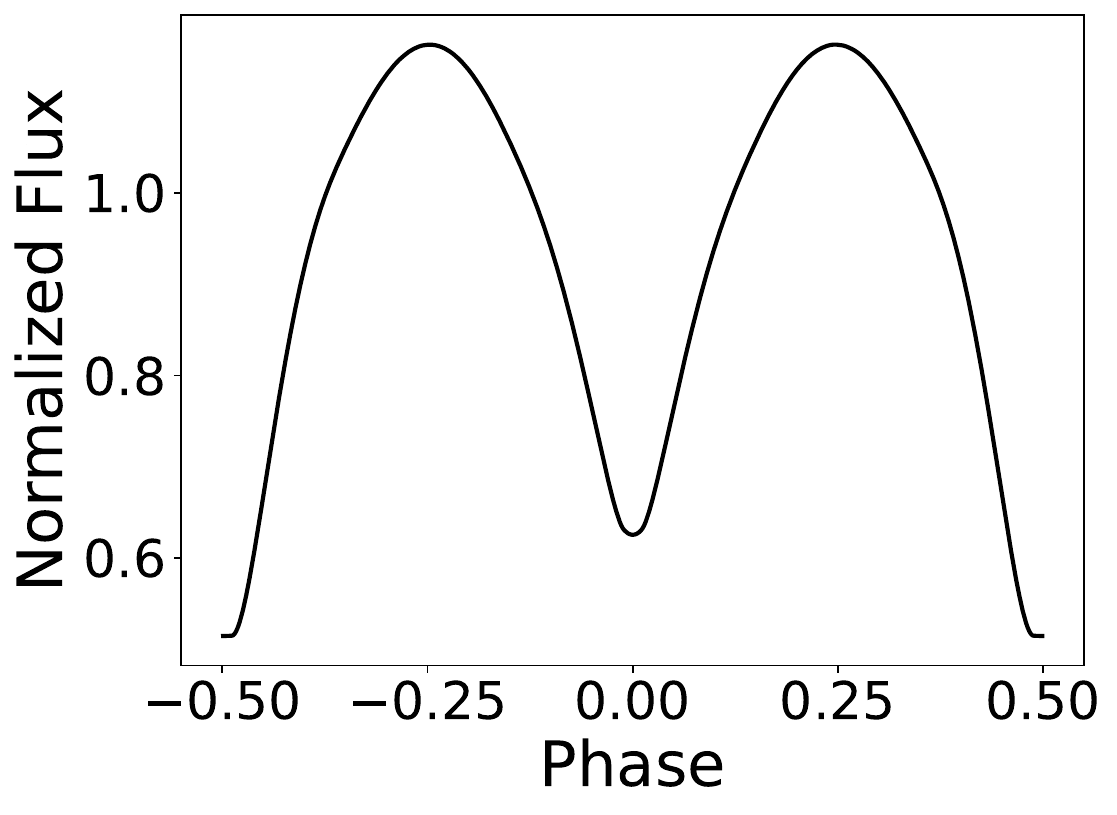}
    \caption{}
  \end{subfigure}
  \begin{subfigure}{0.32\textwidth}
    \includegraphics[width=\textwidth]{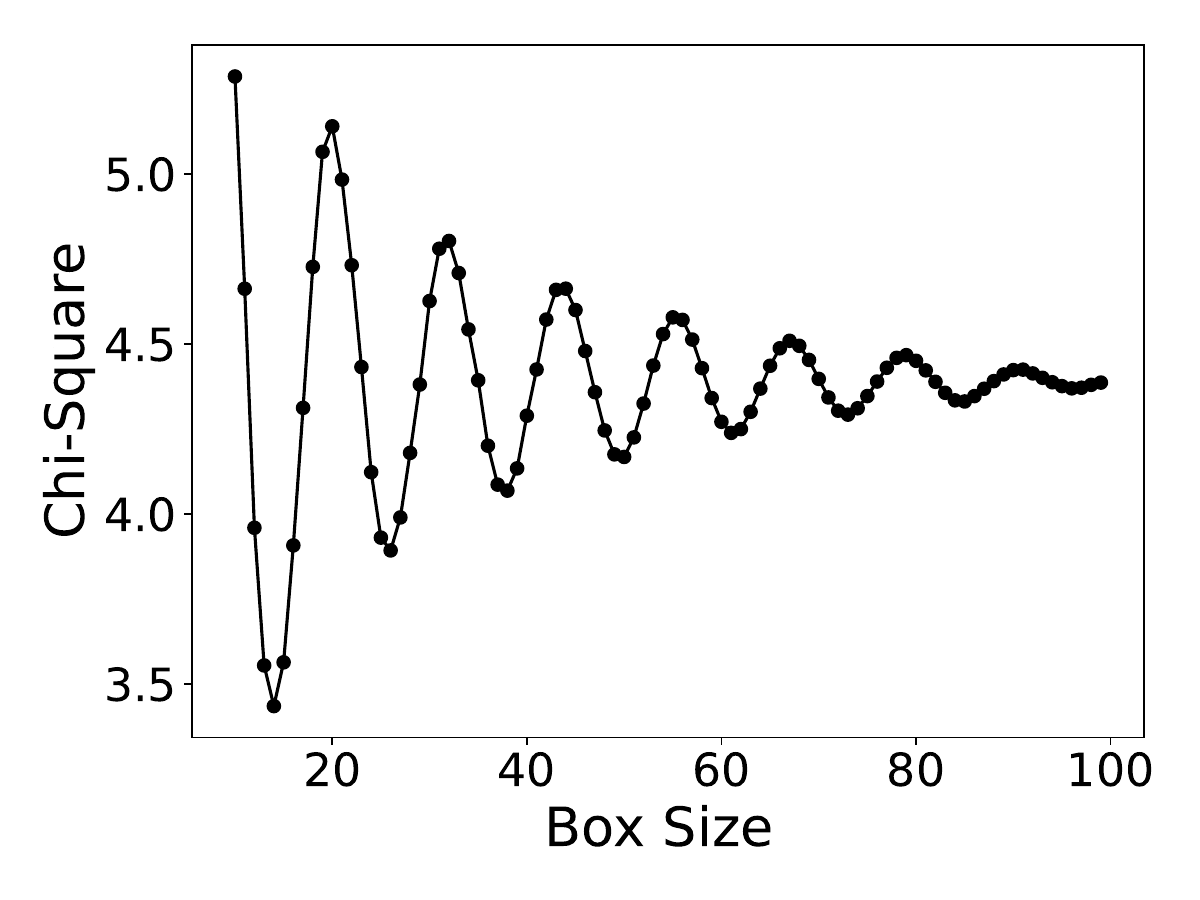}
    \caption{}
  \end{subfigure}
  \caption{PHOEBE mesh, phase-folded light curve, and chi-square plots for D (a, b, c panels), SD (d, e, f panels), and C (g, h, i panels) binaries. All the data were generated from the CALEB catalog.}
  \label{fig:f_00A3}
\end{figure*}

In Fig.~\ref{fig:f_00A3}, examples of mesh plots, phase-folded LCs, and chi-square plots of D, SD, and CEBs are presented, which were generated using PHOEBE modelling code. In this figure, the DEB was based on the orbital parameters of V541 Cyg, the SDEB was based on XZ Pup, and the CEB was based on the V523 Cas system. The chi-square diagram in Fig.~\ref{fig:f_00A3}(c) differs from Fig.~\ref{fig:f_003}(a, b, c). However, it is expected as DEBs exhibited various chi-square shapes and in general they are not as distinct as the damping sinusoidal pattern observed in CEBs. Nevertheless, in all cases, the D chi-square plots displayed an exponentially increasing trend. 
The chi-square plots generated from PHOEBE data, compared to those generated with data from {\sl Kepler}, revealed significant similarities for C and D systems. However, the SD chi-square plots from PHOEBE did not show the same variations seen in observed SDs and matched only a few cases.

We adopted a sample size of 13,872, consisting of both synthetic chi-square plots generated by PHOEBE and those from {\it Kepler} data (only with the highest epochs), which 2806 plots were from observed LCs and the remainder were from simulated data. Among the 6,171 CEBs, 832 were from {\it Kepler} data. 
We adopted 7,077 DEBs and 624 SDEBs, from which 1,820 and 154 were from {\it Kepler} data, respectively. 

We used a 60:40 split for training and testing the data. Here, the data was split using Scikit-learn’s \citep{pedregosa2018scikitlearnmachinelearningpython} built-in train test function, which performs random sampling based on the specified ratio. From the initial 60 percent training data, we further set aside 20 percent for validation, which resulted in 48 percent training and 12 percent validation samples, making the final dataset consist of 48 percent training and 12 percent validation data and 40 percent test data. The reason for using a higher proportion of data for testing was that our sample contained a large amount of synthetic data generated with PHOEBE, while the number of {\sl Kepler} SD sample was very limited. To ensure that some {\sl Kepler} data were included in the test set amid the abundance of synthetic data, we opted for a larger test set than usual.
\begin{figure}
  \centering
  \begin{subfigure}{0.41\textwidth}
    \includegraphics[width=\textwidth]{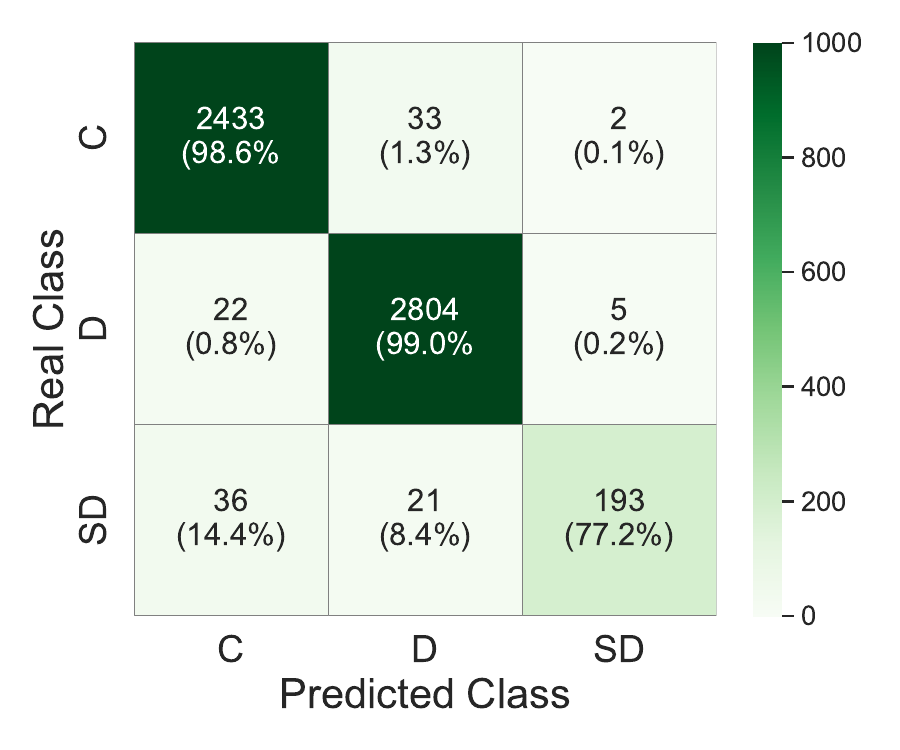}
    \caption{}
  \end{subfigure}
  \begin{subfigure}{0.37\textwidth}
    \includegraphics[width=\textwidth]{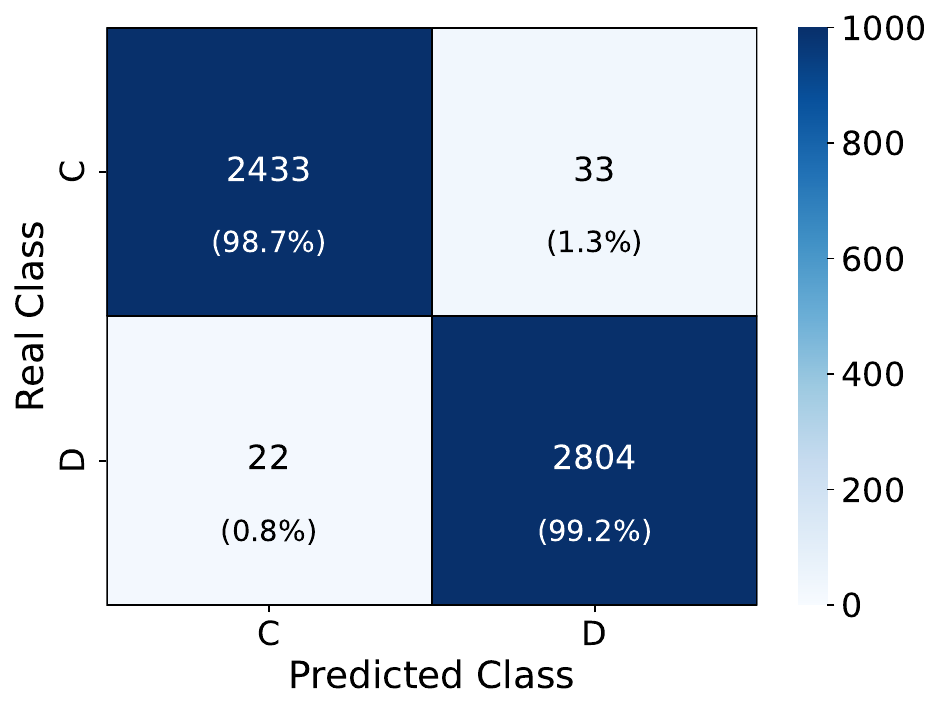}
    \caption{}
  \end{subfigure}
  \caption{The confusion matrix of the 1D-CNN model using chi-square plots from the combined {\it Kepler} and PHOEBE data is shown here. Panel (a) shows the confusion matrix for all three eclipsing binary classes: CEB, DEB, and SDEB. Panel (b) shows the confusion matrix for the binary classification involving only the two major classes: CEBs and DEBs.}
  \label{fig:f_M}
\end{figure}
After testing our 1D-CNN model, we found that it had an increased overall accuracy of 98 percent. Out of 2,831 DEBs in the test set, 2,804 were well classified, resulting in an accuracy of 99 percent. For the 2,468 CEBs in the test sample, 2,433 were well classified, with an accuracy of 98.6 percent. Finally, out of 250 SD targets in the test sample, 193 were well classified, resulting in an accuracy of 77.2 percent, a clear improvement in the identification of this type. The resulting confusion matrix was shown in Fig.~\ref{fig:f_M}(a).

We also performed the same analysis with SDEBs excluded, keeping the train, validation, and test sample ratio the same as before, at 48:12:40. In this case, the model was able to distinguish between C and D classes with 99 percent accuracy. Out of 2,466 CEBs in the test sample, 2,433 were well classified, resulting in an accuracy of 98.7 percent. Additionally, out of 2,826 D instances in the test sample, 2,804 were well classified, resulting in an accuracy of 99.2 percent. The confusion matrix for this analysis was shown in Fig.~\ref{fig:f_M}(b). In Table~\ref{tab:M3}, we presented the Precision, Recall, and F1-scores for these two analyses. The columns labeled C (CNN2), D (CNN2), and SD (CNN2) represented the metrics computed using {\it Kepler} and PHOEBE data combined. The first and last three rows corresponded to CNN2 with three classes and with only CEBs and DEBs, respectively.\\
The classification results from both CNN methods (with and without simulated data) are also presented in Table~\ref{tab:DD2}.  
While the PDS method analyzed all available targets, the two CNN methods had different test sets due to their independent and unbiased sampling strategies. Therefore, only the common targets among all three approaches are displayed to evaluate classification accuracy in Table~\ref{tab:DD2}\footnote{The complete version of this table is available online.}

\section{Discussion}\label{sec:dis}
\subsection{Comparison between classifications}
\begin{figure*}
  \centering
  \begin{subfigure}{0.7\textwidth}
    \includegraphics[width=\textwidth]{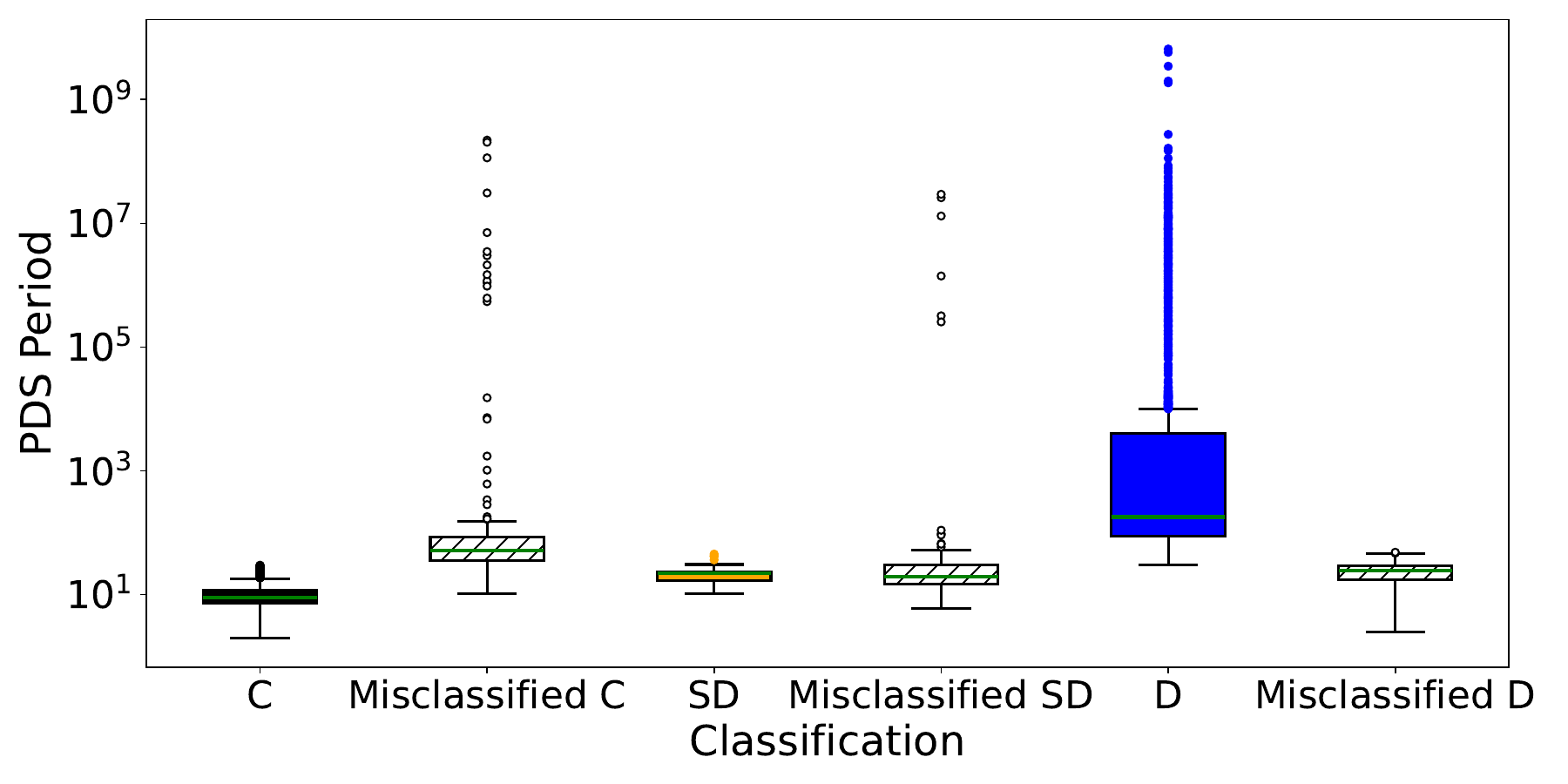}
  \end{subfigure}
  \caption{
  Box plot of $P_{\rm PDS}$ values and the EB morphological classification. The black, yellow, and blue box plots represent the correctly classified C, SD, and D systems, respectively. The misclassified C, SD, and DEBs are shown with a hatch pattern. The green line inside each box indicates their median value (Q2). Dots beyond the whiskers denote outliers.}
  \label{fig:f_cvx}
\end{figure*}

We compared our visual classification with the classification scheme in KEBC, that is based on the $c$ value. 
If the EBs had a $c$ value less than 0.1, they were classified as well D systems. $c$ values between 0.1 and 0.5 indicated primarily DEBs, while SDEBs had $c$ values between 0.5 and 0.7. CEBs had $c$ values between 0.7 and 0.8, and ellipsoidal EBs had $c$ values greater than 0.8 up to 1. We classified both CEBs and ellipsoidal EBs under the CEB category. Out of 2806 common EBs (excluding 59 UNC cases) between the KEBC and our sample, 2502 matched our visual classification, yielding an overall accuracy of 89 percent with respect to the KEBC. 

Additionally, comparing KEBC's classification with our method based on PDS, 2230 EBs matched both classification, representing 79.5 percent of the sample. 

A total of 2201 targets matched across automated, visual and KEBC classification, resulting in an overall agreement of 78.5 percent. The mismatch in classifications was primarily due to the SD and CEB overlapped classes. We observed that some targets, such as KIC 11230837 and KIC 8288741, had $c$ values of 0.61 and 0.64 respectively, which were classified as SD according to the KEBC catalogue, but visually they appeared to be DEBs. Moreover, there was ambiguity in cases where the $c$ value was exactly 0.7, as both SD and CEB classifications overlapped at this threshold.  \\~\\
The PDS function showed limitations in classifying high-$P_{\rm orb}$ CEBs and SDEBs (Fig.~\ref{fig:f_cvx}). Compared to this, the 1D-CNN (Section~\ref{sec:ai1}) improved classification performance, particularly for SDEBs. In the second analysis (CNN2), augmenting with PHOEBE-generated synthetic data yielded 99~percent accuracy for CEB and DEB systems. However, CNN2 reflected bias introduced by the synthetic data when classifying the three classes, which was primarily observed in SDEBs, as the synthetic data did not capture the physical origin of the distorted nature in the chi-square plots. Thus, while CNN2 excelled at separating CEBs and DEBs, CNN1 provided the most reliable overall classification across our different methods.

Earlier studies employing machine learning models classified EBs primarily based on the morphology of their LCs. \citet{2021A&C....3600488C} trained their deep-learning model on 491,425 synthetic LCs generated using the \texttt{ELISa} software and evaluated it on 100 observed LCs from published sources. Their model was trained on over-contact and DEBs, while the evaluation set also contained SDEBs that were treated as DEBs during testing. The model achieved an overall accuracy of 98 percent which increased to 100 percent when the SDEBs were excluded from the evaluation set. In contrast, our work classified {\sl Kepler} EB data into three primary classes (DEBs, SDEBs, and CEBs), achieving an SDEB accuracy of 47 percent, and 99 percent accuracy when classification was performed only on DEBs and CEBs using both synthetic and {\sl Kepler} data combined.

\citet{Ula2023ADL} reported an overall accuracy of 92 percent, with class-wise F1 scores of 0.935 (CEB), 0.940 (DEB), and 0.872 (SDEB). Their dataset combined Kepler, ASAS, and CALEB light curves, using catalog-based morphology labels and a balanced sample across the three classes. In comparison, our chi-square-based CNN achieved an overall accuracy of 90.3 percent. The model produced similar performance for CEBs and DEBs systems, comparable to LC based aproaches, while the SD class showed lower accuracy. The lower SDEB accuracy was influenced by intrinsic morphological complexity and class imbalance, as SD systems formed a much smaller fraction of our dataset. In addition, we analysed the full {\sl Kepler} catalogue which increased the classification difficulty. The aim of this work was not to replace LC-based classification, but to assess whether chi-square morphology could provide class-wise discrimination and uncover variability signatures beyond static LC shapes  (Section~\ref{sec:TVS}).

\subsection{PDS period correlation with orbital period}\label{sec:subsec5.2}

Analyzing the different $P_{\rm PDS}$ values in Table~\ref{tab:DD2}, we observed that those for some DEBs, or systems with DEB-like chi-squared plots, were generally higher than those for CEBs, and in some cases, unusually large values were observed. For example, KIC~2987433 had a $P_{\rm PDS}$ value of \(1.37 \times 10^{7}~\text{box-size steps}\), which was extremely large, yet it was classified as a DEB according to the fitting criteria in Table~\ref{tab:bigdata1}. This occurred due to the non-sinusoidal shape of the chi-squared plot for DEBs, which affected the PDS fitting. To further support this observation, we examined the visually classified SDEB systems such as KIC~6046061 and KIC~6516874 and they also had extremely large $P_{\rm PDS}$ values and were misclassified by the PDS function as DEBs, again due to their non-sinusoidal chi-squared shapes. This confirmed that the extreme values arose from the PDS function’s limitations in accurately fitting such profiles. 
  
The results of our PDS classification scheme using box plots for the correctly assigned C (black), SD (orange) and D (blue) systems are presented in Fig.~\ref{fig:f_cvx}, showing a clear separation between them in terms of $P_{\rm PDS}$. The misclassified EBs (hatch pattern) are also illustrated, displaying relatively higher and lower $P_{\rm PDS}$ values in respect to their classes.
Each box represents the interquartile range (IQR), containing the middle 50 percent of the data from the first quartile (Q1) to the third quartile (Q3). The IQR is defined as $\mathrm{IQR} = Q3 - Q1$, and it provides statistical dispersion by minimizing the influence of extreme values. The green lines inside the boxes indicate the median $P_{\rm PDS}$ values (Q2), while whiskers extend from Q1 to the lowest value within $Q1 - 1.5 \times \mathrm{IQR}$ and from Q3 to the highest value within $Q3 + 1.5 \times \mathrm{IQR}$. These whiskers illustrate the typical range of variability in the $P_{\rm PDS}$ values for each classification, excluding extreme outliers. Data points beyond these whiskers represent outliers, highlighting extreme values within each category. We omitted a few large \( P_{\rm PDS} \) values from Fig.~\ref{fig:f_cvx} for better visualization, and the y-axis \( P_{\rm PDS} \) values were plotted on a logarithmic scale. The systems that presented a good classification had well-separated PDS periods.

The misclassified EBs had \( P_{\rm PDS} \) values outside the expected range for their class, except for SD systems. For CEBs, the median $P_{\rm PDS}$ was 8.88 box-size steps, whereas for misclassified C targets, this value was comparatively higher at 51.04 box-size steps, which was far from the threshold \(
\bigl(R^2 \ge 0.9 \;\text{and}\; P_{\rm PDS} < 30 \;\text{or}\; P_{\rm PDS} \le 10\bigr)
\),
established for the classification (see Table~\ref{tab:bigdata1}). 
For DEBs, the median $P_{\rm PDS}$ was 180.62 box-size steps. However, the misclassified DEBs showed a lower value of only 24.79 box-size steps, also outside our empirically defined threshold \(
\bigl(R^2 \ge 0.9 \;\text{and}\; P_{\rm PDS} \ge 30 \;\text{or}\; P_{\rm PDS} \ge 50\bigr)\). 
The misclassified SDEBs did not present a clear \( P_{\rm PDS} \) value separation like that seen for DEBs or CEBs. In fact, the SD \( P_{\rm PDS} \) values showed an alignment with comparatively larger \( P_{\rm PDS} \) values of C and smaller \( P_{\rm PDS} \) values of DEBs, as seen in Fig.~\ref{fig:f_cvx}, where for the well-classified SD systems the median value was 22.15 box-size steps in comparison to 29.32 box-size steps for misclassified SDs. This could be in part because the classification scheme adopted was primarily based on the \(R^2\) value.

\begin{figure*}
  \centering
  \begin{subfigure}{0.32\textwidth}
    \includegraphics[width=\textwidth]{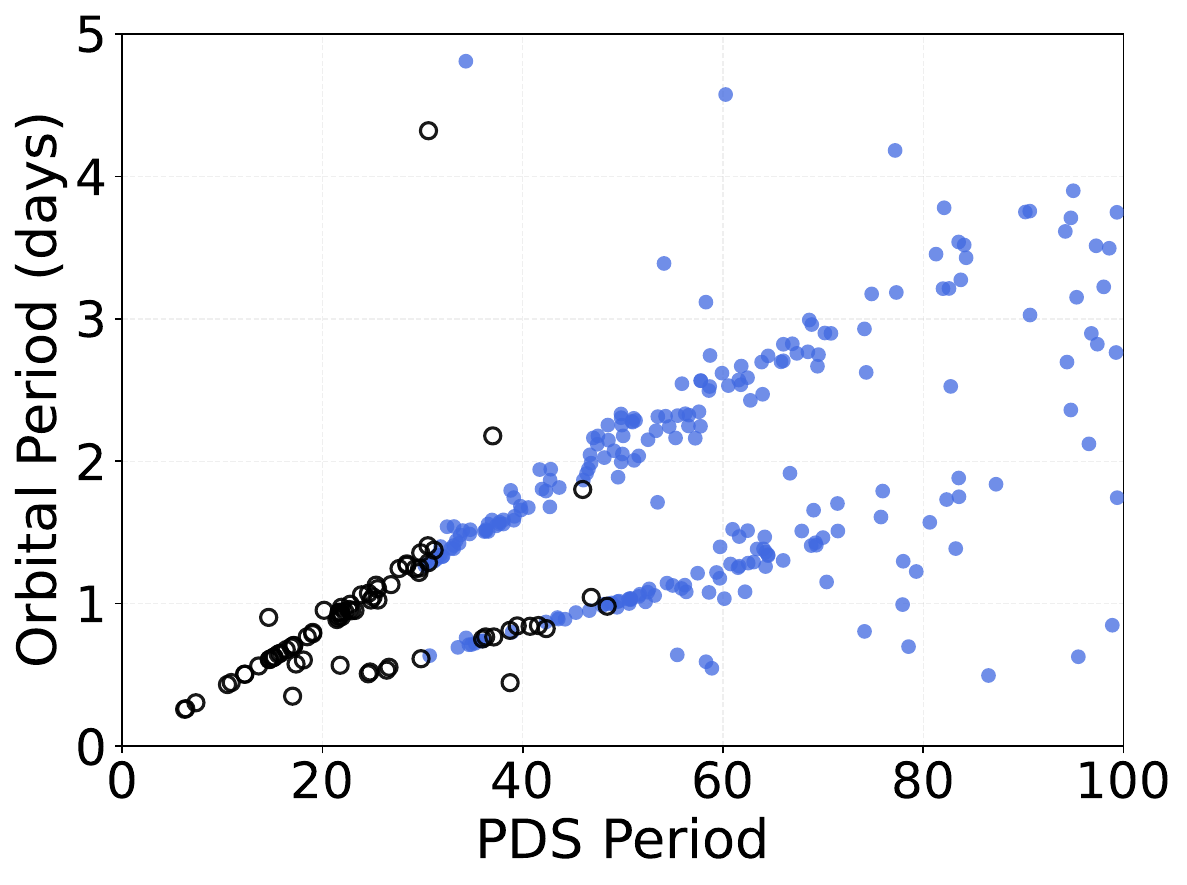}
    \caption{}
  \end{subfigure}
  \begin{subfigure}{0.32\textwidth}
    \includegraphics[width=\textwidth]{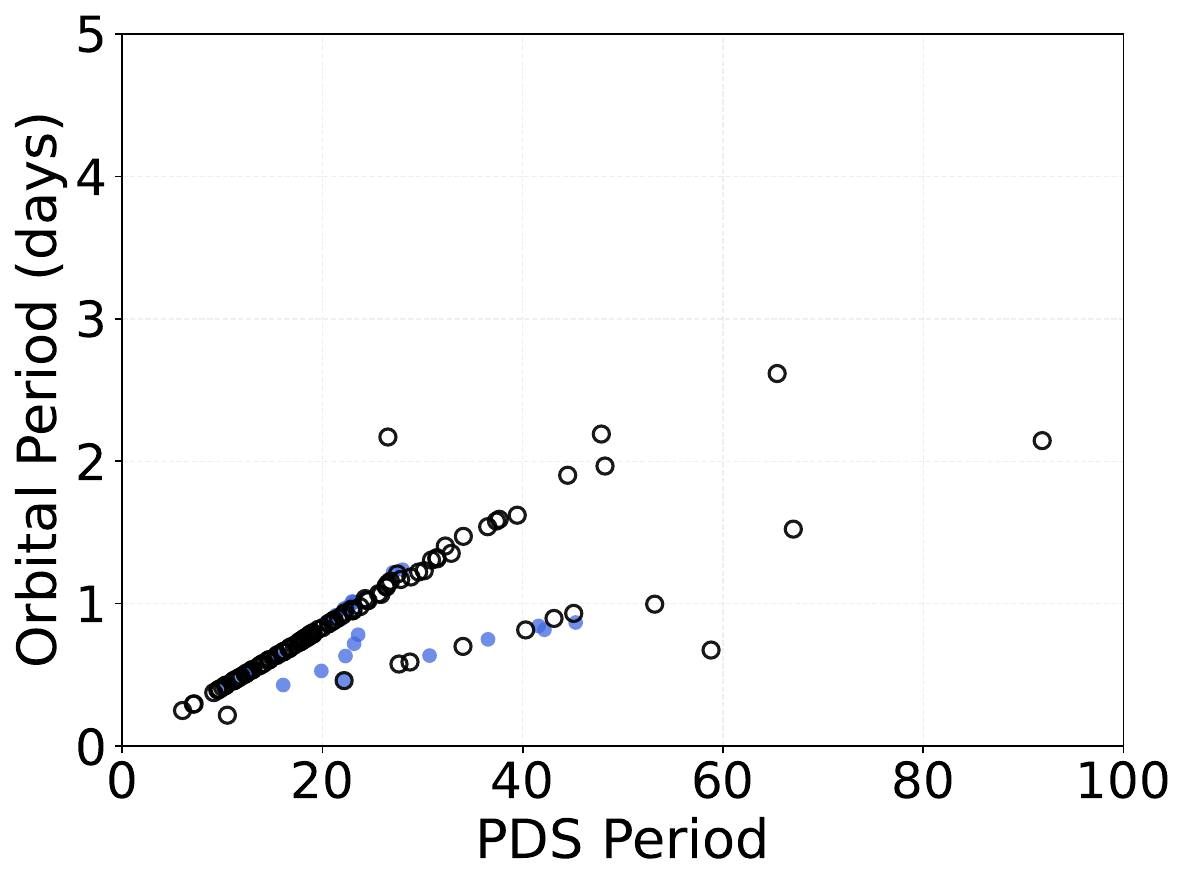}
    \caption{}
  \end{subfigure}
  \begin{subfigure}{0.32\textwidth}
    \includegraphics[width=\textwidth]{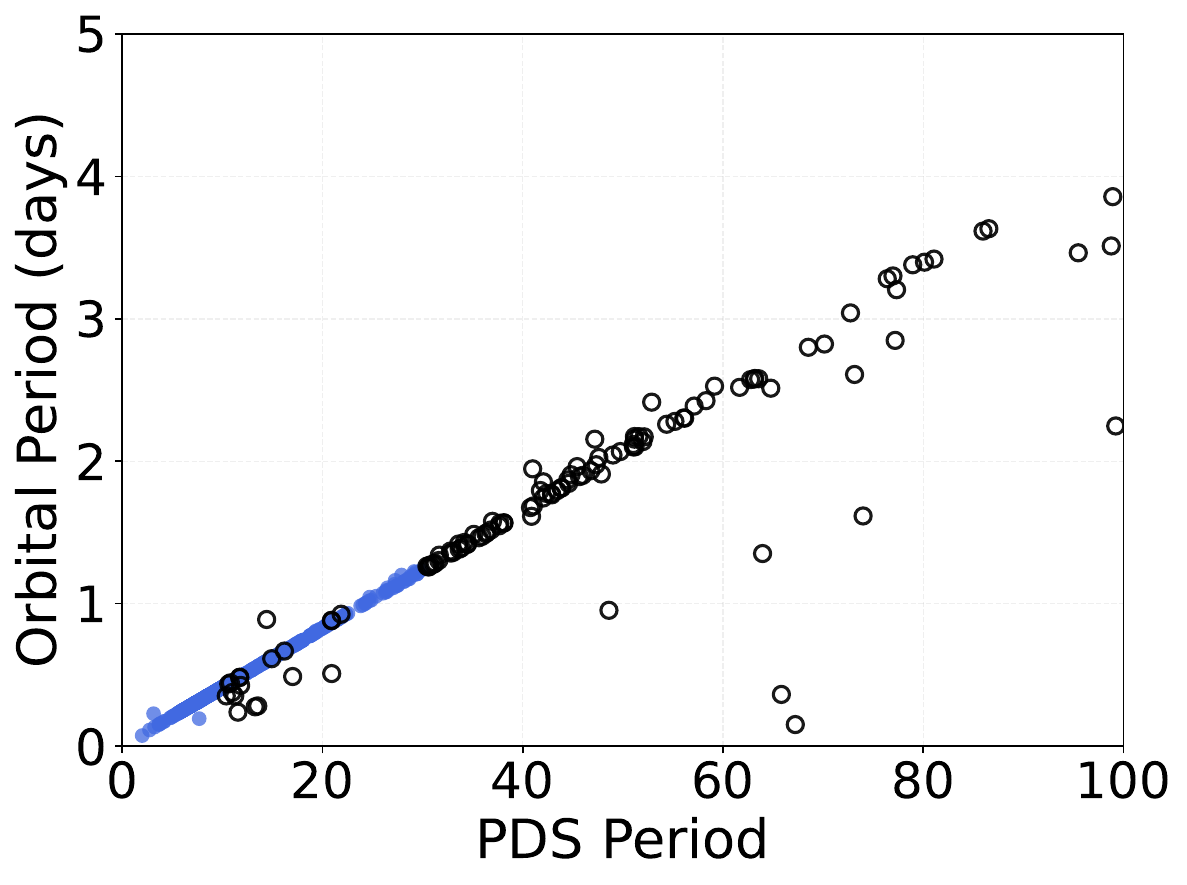}
    \caption{}
  \end{subfigure}
  \caption{Distribution of EB targets in the parameter space 
defined by $P_{\rm PDS}$ and $P_{\rm orb}$. Panels show 
the three EB classes separately: (a)~DEBs, (b)~SDEBs, 
and (c)~CEBs. Targets for which the visual classification 
agrees with the PDS classification are shown as solid blue 
markers, while targets where the two classifications 
disagree are represented by open black circle markers.}  
  \label{fig:f_201x1}
\end{figure*}

In Fig.~\ref{fig:f_201x1}, we explored the correlation between $P_{\rm orb}$ and the corresponding $P_{\rm PDS}$ values for different EB morphological classes. 
The three panels show the distributions separately 
for (a)~DEBs, (b)~SDEBs, and (c)~CEBs. In each 
panel, correctly classified systems are shown as 
solid blue markers, while misclassified systems 
are represented by open black circle markers. We considered systems with $P_{\rm PDS}$ values up to 100 box-size steps and $P_{\rm orb}$ less than 5 days. 
This range included most CEBs and SDEBs as well as short-period DEBs, which allowed us to better identify trends and examine the relationship between the orbital period and the corresponding $P_{\rm PDS}$ values across the different EB classes.

The majority of EB systems in our sample with $P_{\rm orb}$ of less than 5 days seemed to follow two different tracks: one where most of the CEBs lay and another where we found only D and SD systems. Most misclassified CEBs tended to have larger $P_{\rm orb}$, while some systems with smaller $P_{\rm orb}$ did not follow the main C track, as shown in Fig.~\ref{fig:f_201x1}~(c).
These targets may have had ambiguous visual classifications, as they did not show clear LC behaviour, which could have led to a misclassification. Thus, the $P_{\rm PDS}$ and $P_{\rm orb}$ comparison plot helped to detected such cases. 
From Fig.~\ref{fig:f_201x1}~(a), we observed that misclassified DEBs generally had shorter $P_{\rm orb}$ values, which could be a result of the previously defined criteria shown in Table~\ref{tab:bigdata1}. However, misclassified SDEBs may have chi-square plots either like C or D systems, and the majority of them is often aligned with the C track, as shown in Fig.~\ref{fig:f_201x1}~(b). By examining the period comparison plot, we could identify these ambiguous systems, which contributed to the drop in accuracy of our classification method.


We developed an automated technique to identify the trends shown in Fig.~\ref{fig:f_201x1}. The algorithm first calculated the ratio 
$P_{\rm PDS}/P_{\rm orb}$ for each data point. It then applied the unsupervised machine learning model, k-means clustering \citep{mcqueen1967smc}, on these ratios to divide the data into two distinct trend groups. For both clusters, the code performed independent linear regressions to determine the best-fit equations. 10 percent of the outliers were removed within each cluster to ensure that the slopes were not influenced by them.

We identified that 1051 EBs were used by our automated technique to determine the first slope (solid red line in Fig.~\ref{fig:DVS}), and the equation of this slope is 
\begin{equation}
P_{\rm orb} = 0.0415 P_{\rm PDS} - 0.001
\label{eq:orbital_period} . 
\end{equation}
The Pearson correlation coefficient for the first slope was 0.99, where the majority of the sample, 727 out of 1051, were CEBs. The code used 148 EBs to determine the second slope (dashed red line in Fig.~\ref{fig:DVS}), where the majority of the sample consisted of DEBs and SDEBs (121 DEBs and 21 SDEBs). This second slope is of the form
\begin{equation}
P_{\rm orb} = 0.02 P_{\rm PDS} + 0.006 ,
\label{eq:manual_slope_sd}
\end{equation}
with a Pearson correlation coefficient of 0.76. 

We note that the slopes in equations~(\ref{eq:orbital_period}) and~(\ref{eq:manual_slope_sd}) are specific to the {\sl Kepler} long-cadence sampling interval of 29.43~min (1765.5~s). Since $P_{\rm PDS}$ is measured in box-size steps, each corresponding to one long-cadence epoch, these relations would yield different slopes if applied to data with a different 
cadence, such as {\sl TESS} or {\sl Kepler} short-cadence observations.

\subsection{Temporally varying systems}\label{sec:TVS}
Considering the \textit{Kepler} data, the majority of the sample showed consistent chi-square plots across different quarters, and consequently similar phase-folded LCs. However, a subset displayed clear quarter-to-quarter changes in the  chi-square plots across, followed by corresponding changes in their LCs. We refer to these objects as temporally varying (TV) systems.
\begin{figure*}
  \centering
  \begin{subfigure}{0.35\textwidth}
    \includegraphics[width=\textwidth]{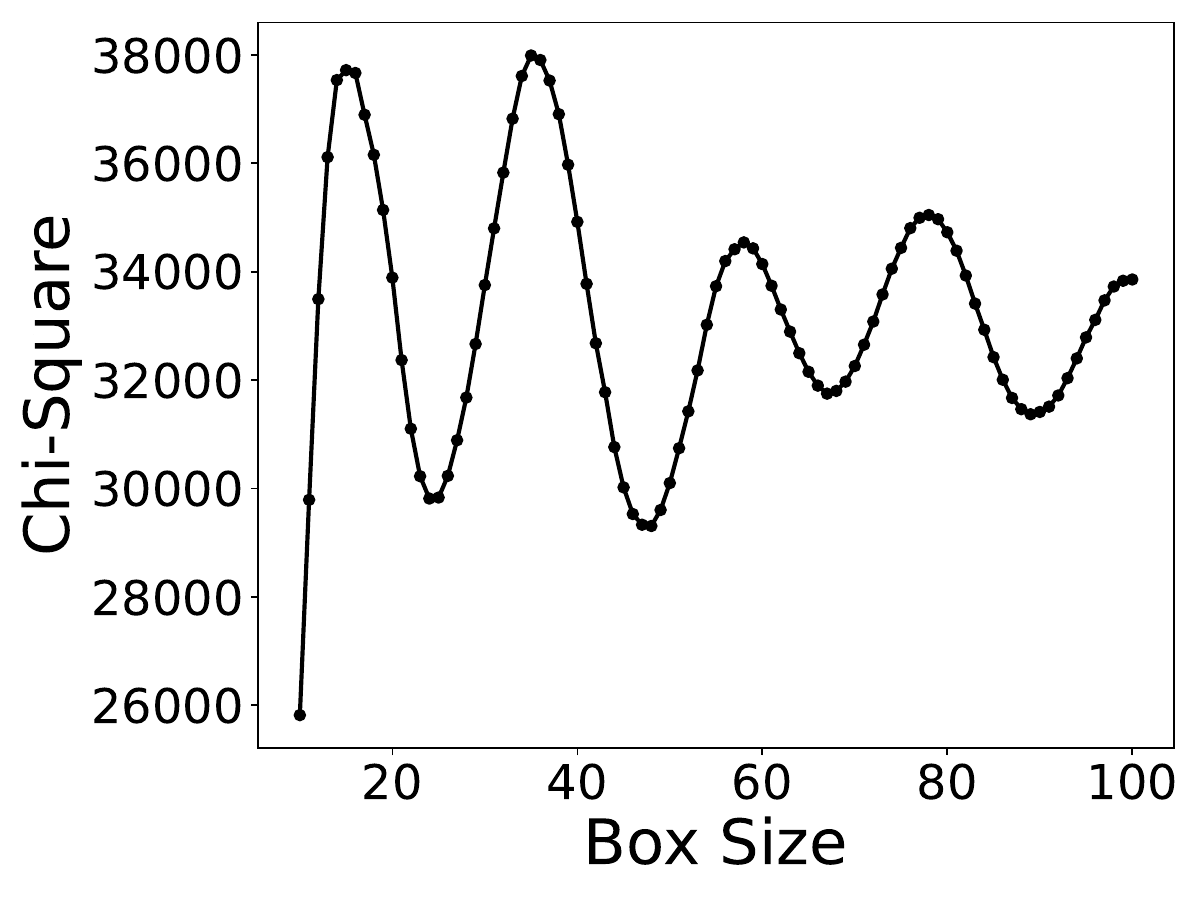}
    \caption{}
  \end{subfigure}
  \begin{subfigure}{0.35\textwidth}
    \includegraphics[width=\textwidth]{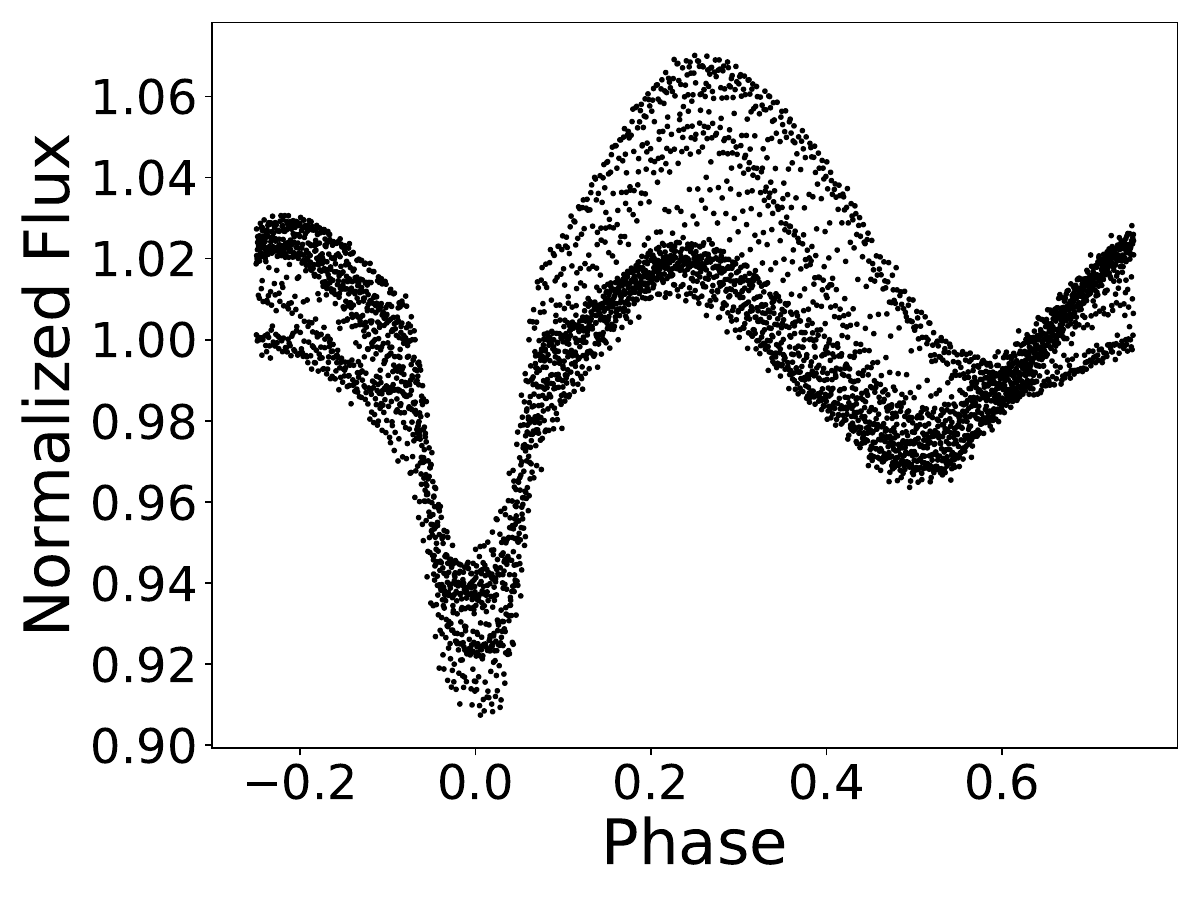}
    \caption{}
  \end{subfigure}
  \begin{subfigure}{0.35\textwidth}
    \includegraphics[width=\textwidth]{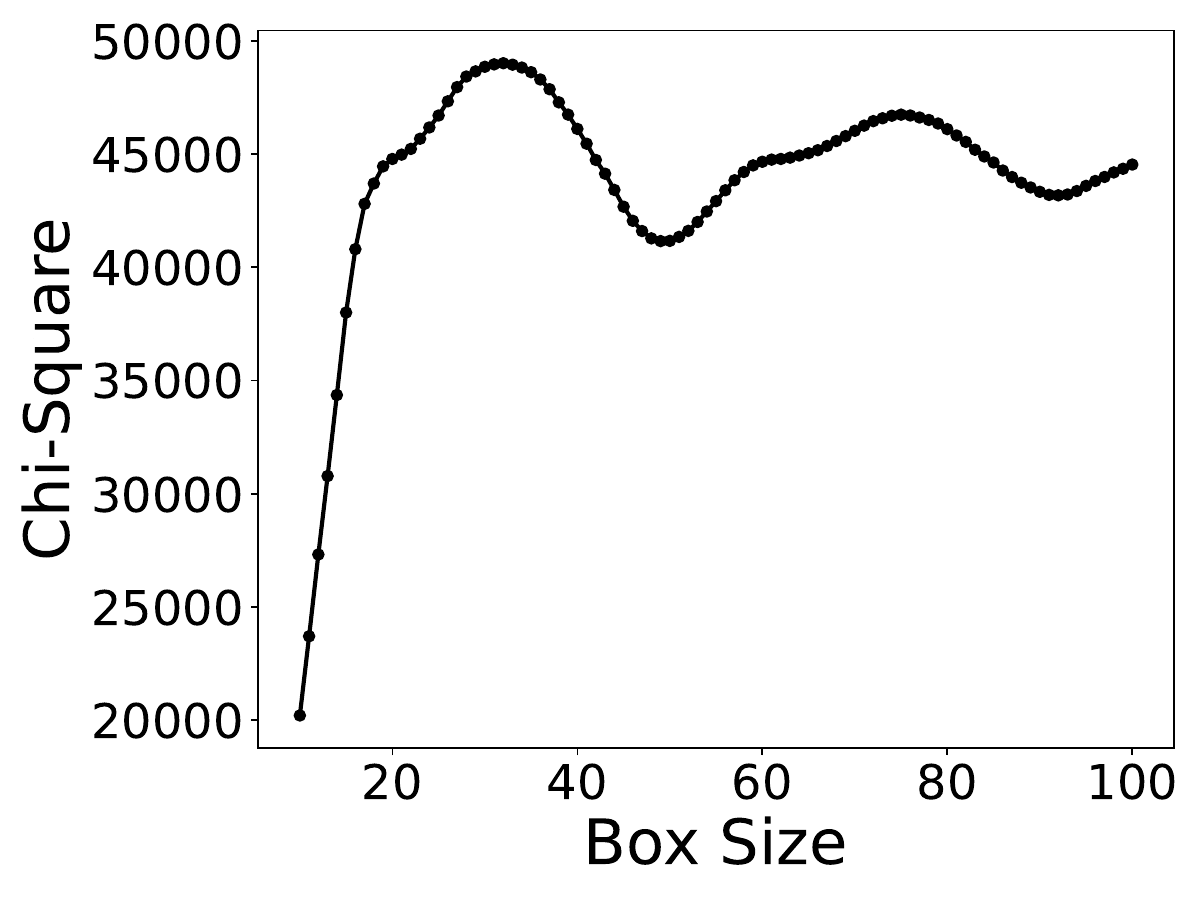}
    \caption{}
  \end{subfigure}
  \begin{subfigure}{0.35\textwidth}
    \includegraphics[width=\textwidth]{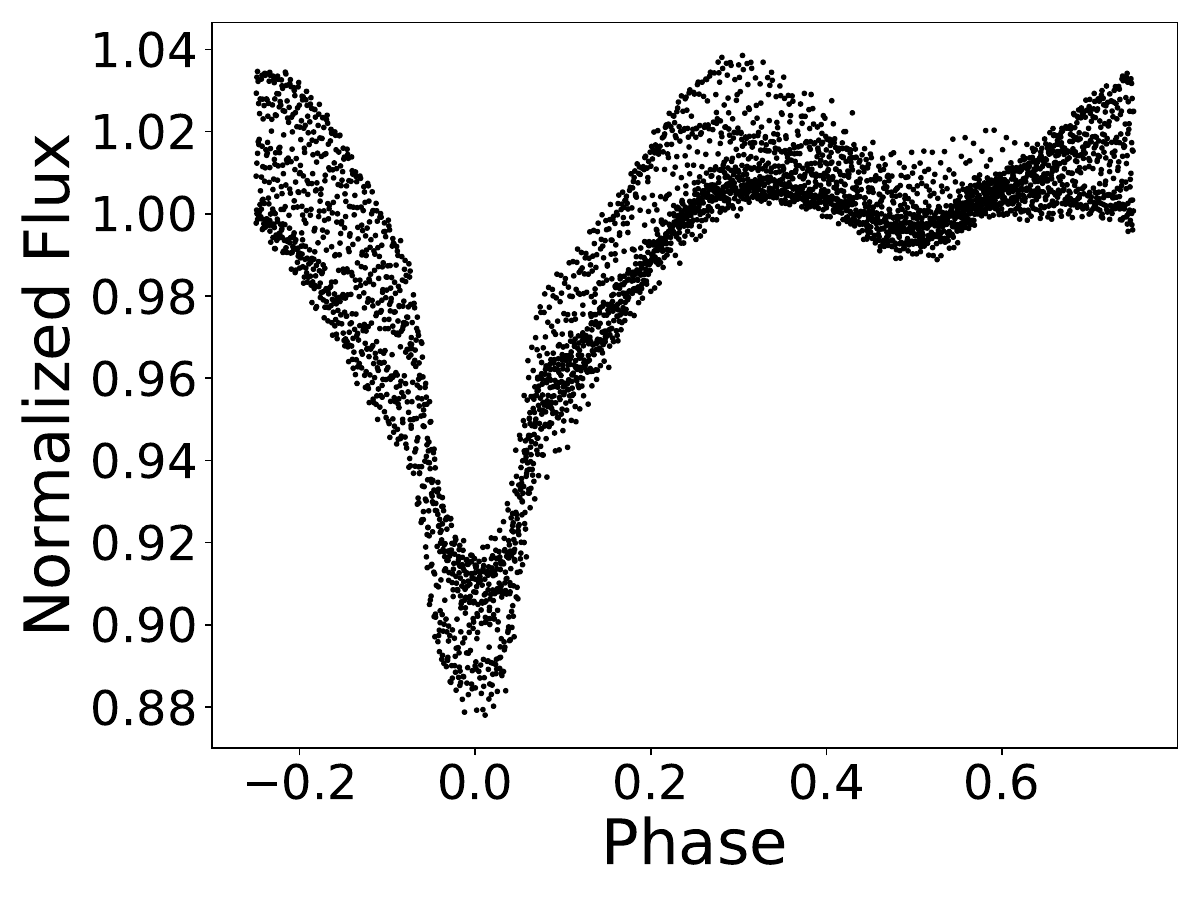}
    \caption{}
  \end{subfigure}
  \begin{subfigure}{0.35\textwidth}
    \includegraphics[width=\textwidth]{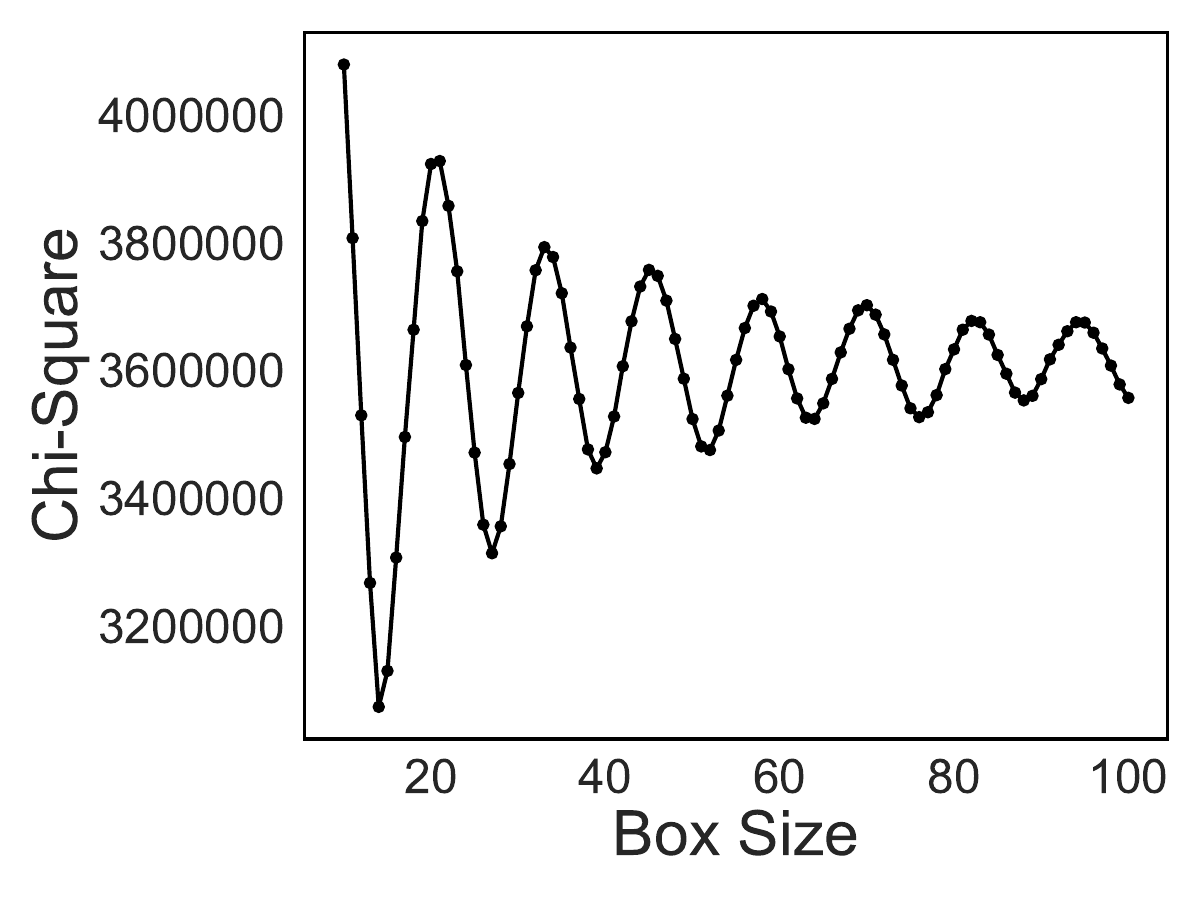} 
    \caption{}
  \end{subfigure}
  \begin{subfigure}{0.35\textwidth}
    \includegraphics[width=\textwidth]{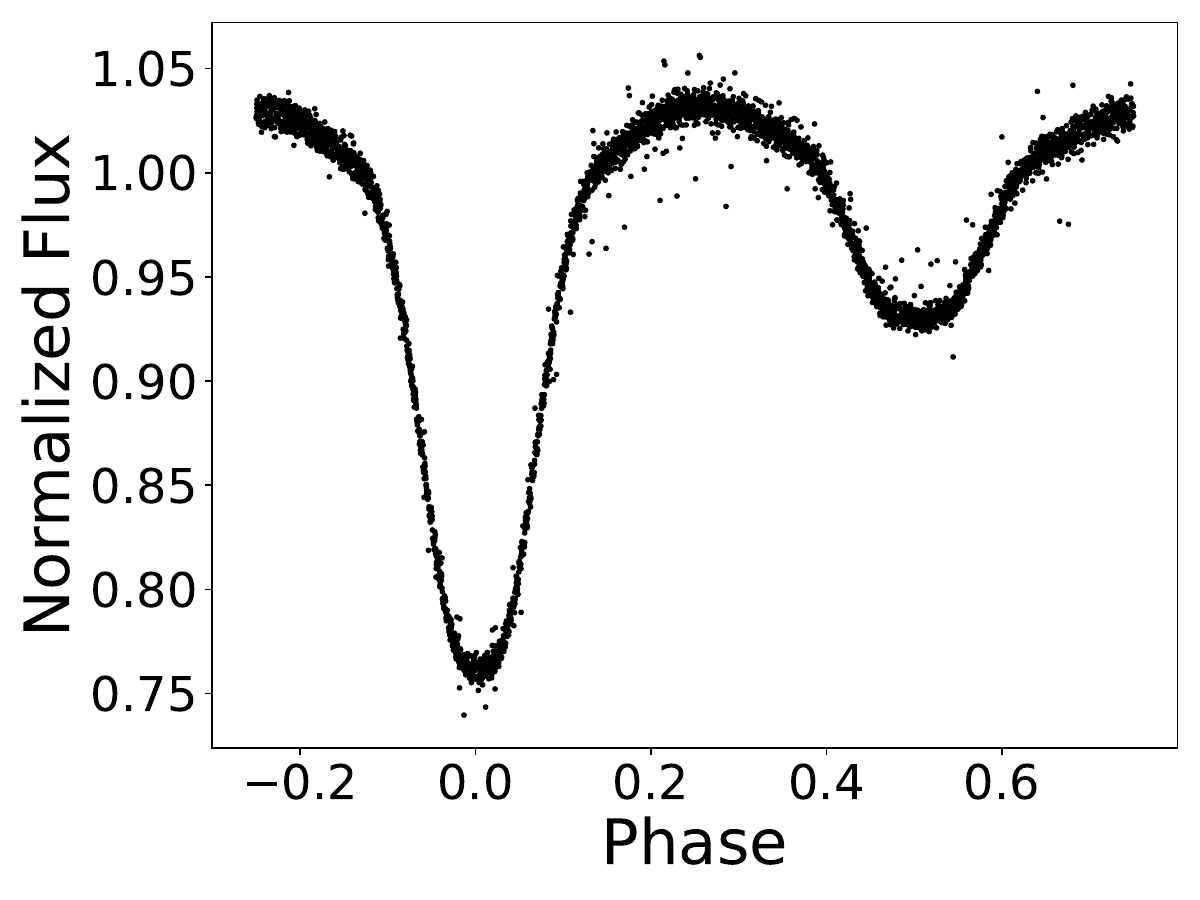} 
    \caption{}
  \end{subfigure}
  \begin{subfigure}{0.35\textwidth}
    \includegraphics[width=\textwidth]{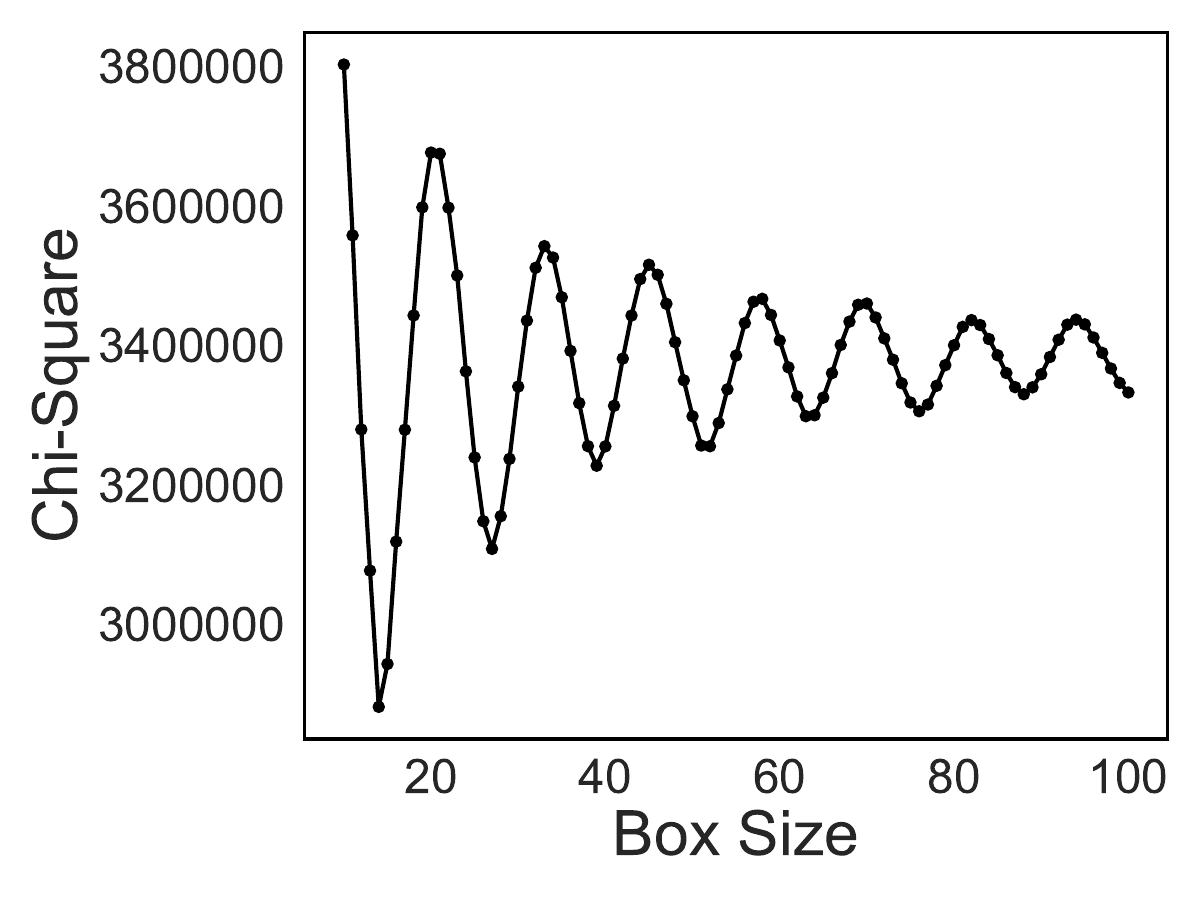} 
    \caption{}
  \end{subfigure}
  \begin{subfigure}{0.35\textwidth}
    \includegraphics[width=\textwidth]{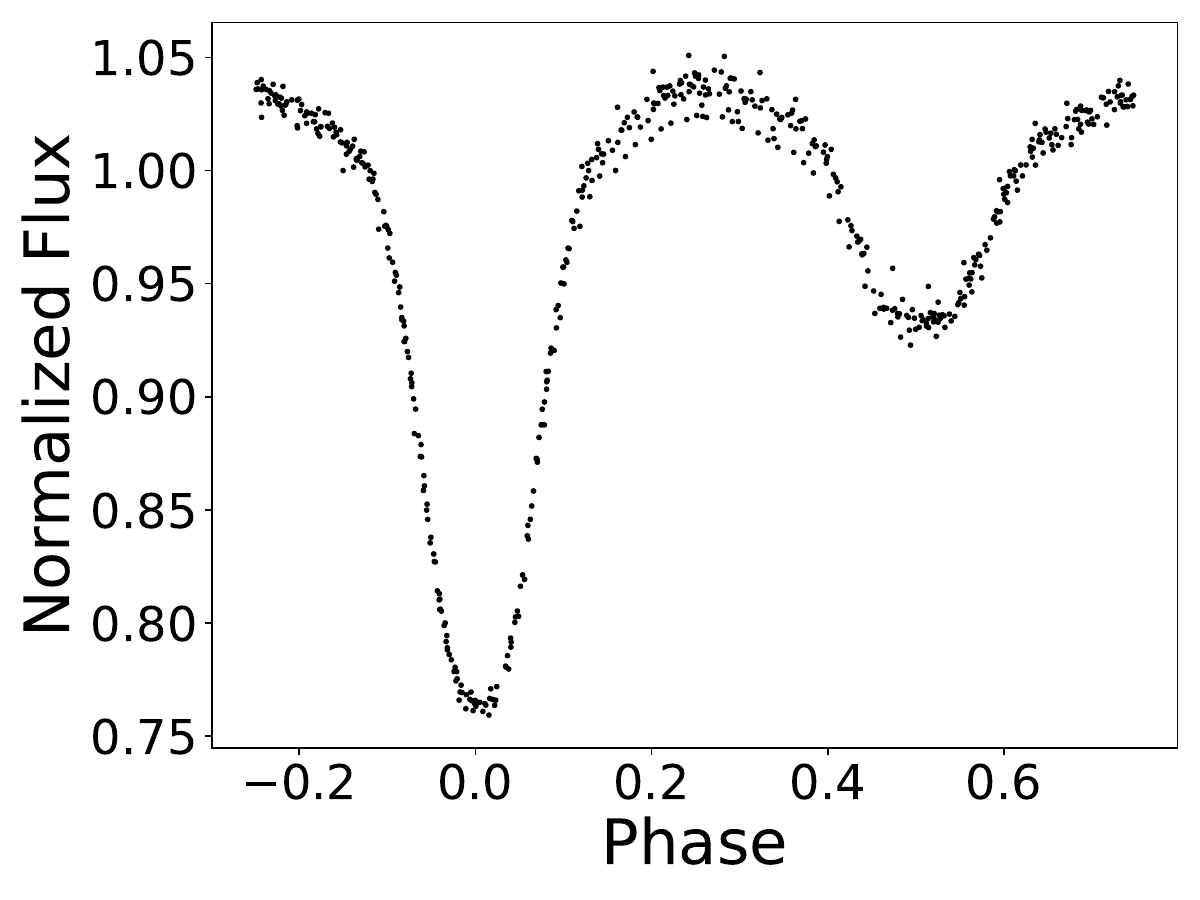}
    \caption{}
  \end{subfigure}

  \caption{A varying SD system, KIC~2577756, and a stable SD system, KIC~10920314, are shown here. 
Panel~(a) illustrates the chi-square plot of quarter~9 for KIC~2577756, and Panel~(b) shows its phase-folded LC for the same quarter. Panel~(c) presents the chi-square plot of quarter~14 for the same system, while Panel~(d) shows the corresponding phase-folded LC. Panel~(e) displays the chi-square plot of quarter~9 for the stable SD system KIC~10920314, and Panel~(f) shows its LC for the same quarter. Panel~(g) illustrates the chi-square plot of quarter~15 for KIC~10920314, and panel~(h) shows the corresponding LC.}
  \label{fig:12}
\end{figure*}

\begin{figure}
  \centering
  \begin{subfigure}{0.5\textwidth}
    \includegraphics[width=\textwidth]{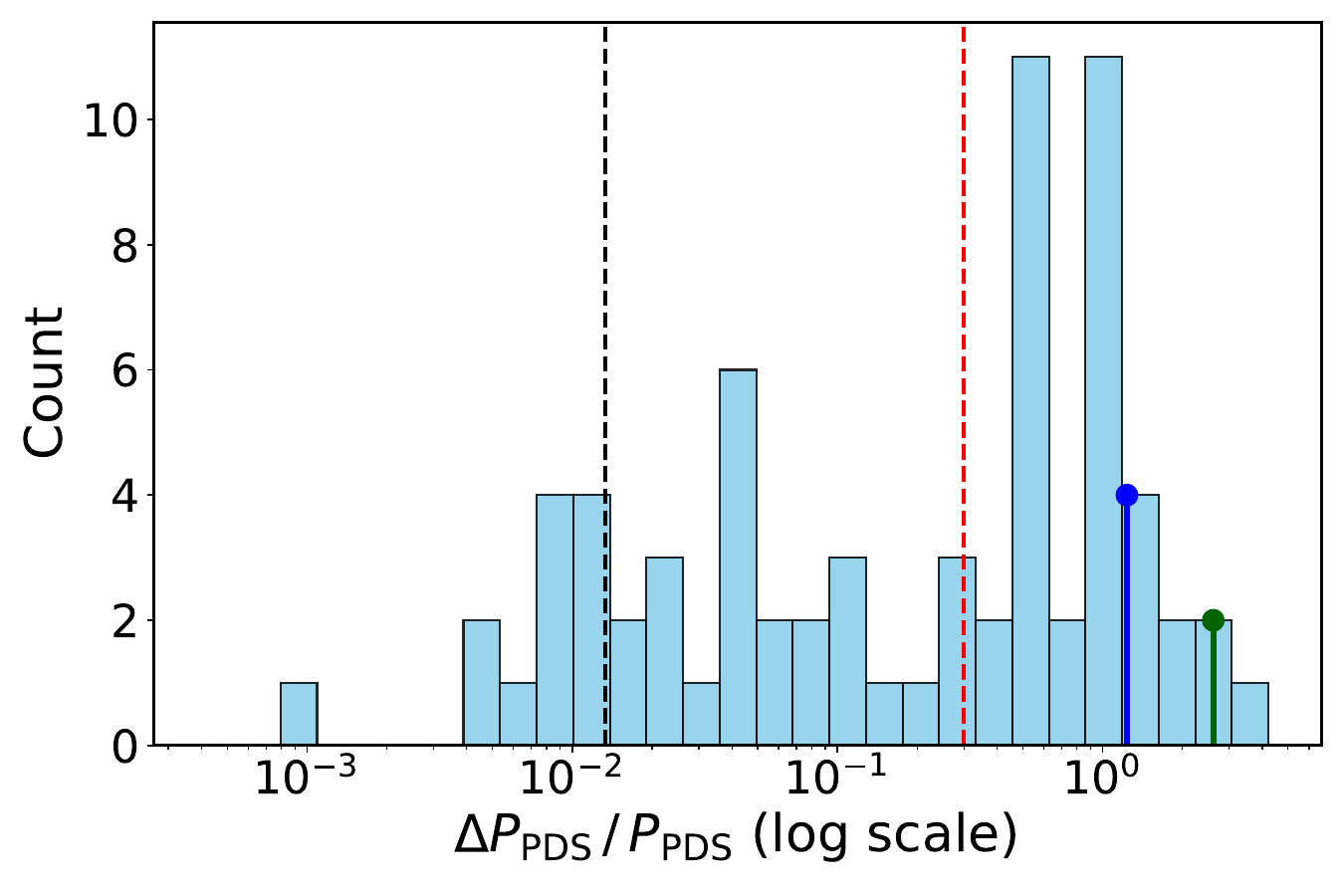}
  \end{subfigure}
  \caption{Distribution of $\Delta P_{\rm PDS}/P_{\rm PDS}$ for the test subset. 
The x-axis represents $\Delta P_{\rm PDS}/P_{\rm PDS}$, shown with logarithmically spaced bins. 
The red dashed line marks the median of the distribution, while the black dashed line indicates the median$-$MAD threshold. 
The solid blue and green vertical lines correspond to KIC~2577756 ($\Delta P_{\rm PDS}/P_{\rm PDS}=1.234$) and KIC~7284688 ($\Delta P_{\rm PDS}/P_{\rm PDS}=2.612$), respectively.}
  \label{fig:f_WBULL}
\end{figure}

Fig.~\ref{fig:12} illustrated two representative SDEBs: a TV system (KIC~2577756) and a stable system (KIC~10920314). For KIC~2577756, both the chi-square plots and phase-folded LC changed between quarters, indicating a temporal evolution in the LC shape. In contrast, KIC~10920314 showed consistent chi-square structure and an essentially unchanged phase-folded LC across quarters. Similar quarter-scale LC variations in {\it Kepler} EBs were well known and were commonly associated with magnetic activity (e.g.\ evolving starspots), flares, and O'Connell-effect variability \citep{knote2022characteristics}.\\
To identify TV candidates, we measured the $P_{\rm PDS}$ independently for all available quarters for each EB. Since TV systems exhibited changing chi-square morphologies across quarters, their fitted $P_{\rm PDS}$ values tended to vary more strongly than in stable systems. We therefore defined $\Delta P_{\rm PDS}$ as the difference between the maximum and minimum $P_{\rm PDS}$ values measured across all available quarters, and adopted the normalized spread $\Delta P_{\rm PDS}/P_{\rm PDS}$ as a proxy for temporal variability. To avoid cases where the PDS fit became unreliable, we discarded targets with $P_{\rm PDS} > 100$ box-size steps, and we restricted the analysis to systems where the PDS fit was well defined (primarily CEBs, SDEBs, and short-period DEBs). We also excluded long-period DEBs, where polynomial-like chi-square trends could dominate and affect the reliability of the derived $P_{\rm PDS}$ values; specifically, we adopted $P_{\rm orb} \leq 4.2$~days, consistent with $P_{\rm PDS}=100$ in equation~(\ref{eq:orbital_period}).\\
To calibrate a practical threshold for TV selection, we cross-matched our sample with the catalogue of TV systems compiled by \citet{knote2022characteristics}, identifying 72 systems in common (hereafter the test subset). We computed $\Delta P_{\rm PDS}/P_{\rm PDS}$ for this subset and examined its distribution (Fig.~\ref{fig:f_WBULL}). The resulting distribution was strongly right-skewed, as confirmed by a Weibull-based skewness analysis \citep{weibull1951statistical}, and we therefore adopted a Median--MAD criterion to define a robust threshold. This yielded a threshold value of 0.0133, and systems with $\Delta P_{\rm PDS}/P_{\rm PDS}$ above this value were classified as TV candidates.

\begin{figure*}
  \centering
  \begin{subfigure}{0.6\textwidth}
    \includegraphics[width=\textwidth]{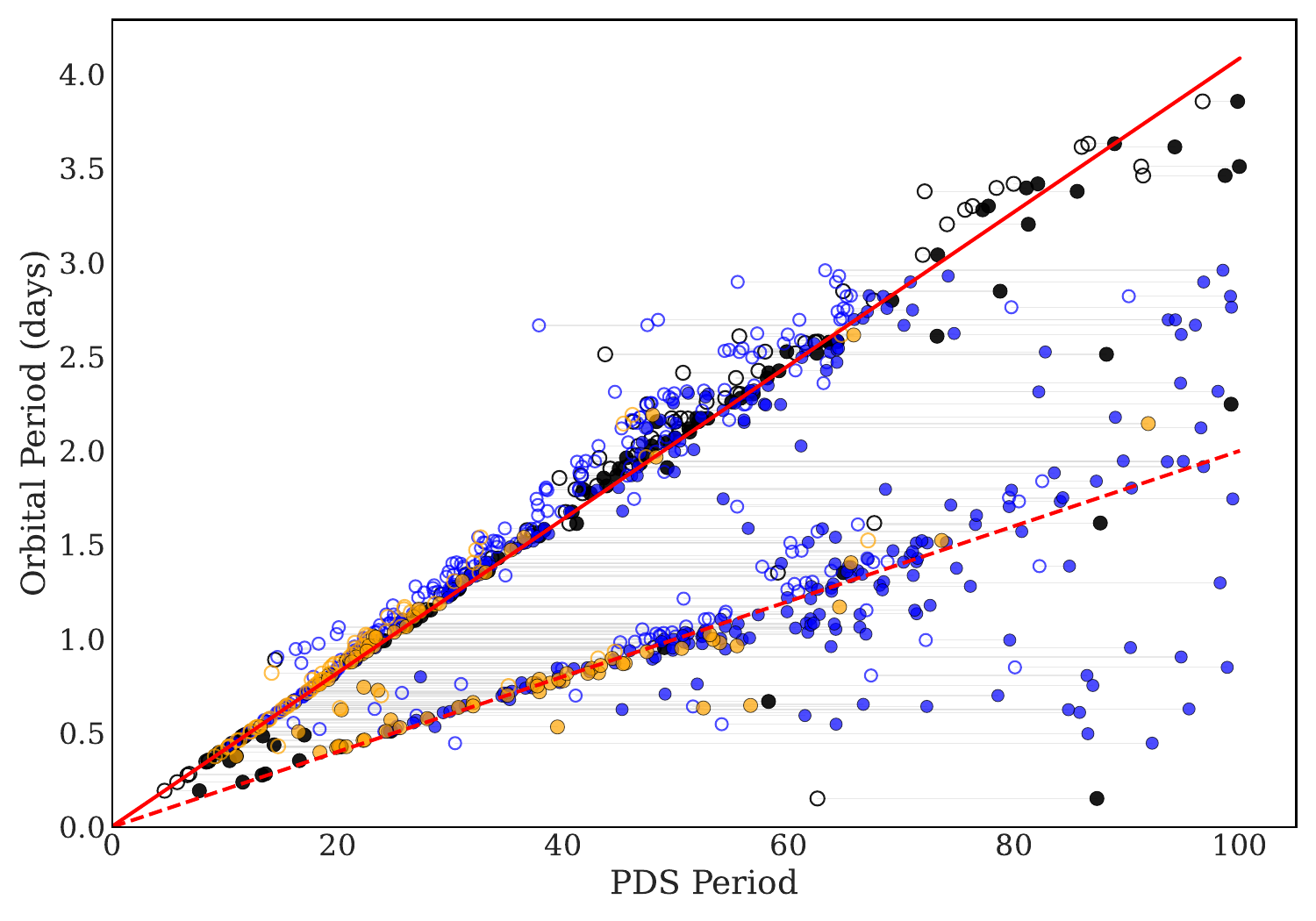}
  \end{subfigure}
  \caption{Distribution of TV systems with $P_{\rm PDS}$ on the x-axis and $P_{\rm orb}$ on the y-axis.
The solid red line shows equation~(\ref{eq:orbital_period}), while the dashed red line shows equation~(\ref{eq:manual_slope_sd}). 
For each system, quarter-wise $P_{\rm PDS}$ values were connected with gray lines to indicate the spread across quarters. 
Visually classified CEBs are shown in black, DEBs in blue, and SDEBs in orange. The minima were shown as hollow circles, while the maxima were shown as solid circles in their corresponding EB colors.}
  \label{fig:DVS}
\end{figure*}

Applying this threshold to our primary sample selected 416 TV systems. Fig.~\ref{fig:DVS} places these objects in the $P_{\rm PDS}$--$P_{\rm orb}$ plane. For each EB, the minimum and maximum quarter-wise $P_{\rm PDS}$ values are connected by a gray segment to visualize the temporal spread. Most CEBs cluster tightly along the main correlation (solid red line), whereas many DEBs, and some SDEBs, exhibit larger spreads and tend to lie closer to the secondary trend (dashed red line). A small number of systems deviate from the main locus; in such cases, the chi-square morphology departs from the sinusoidal pattern, which can shift the fitted $P_{\rm PDS}$ values while still producing large $\Delta P_{\rm PDS}/P_{\rm PDS}$. Representative examples include KIC~5302006, a short-period system with a nearly sinusoidal light curve that was visually classified as a CEB but occupies the lower-right region of Fig.~\ref{fig:DVS}. According to \citet{Matijevic}, such systems have morphology parameters in the range $0.8 \leq c \leq 1.0$, and in the KEBC \citep{2016AJ....151...68K} KIC~5302006 has $c=0.97$, consistent with an ellipsoidal EB. Another example is KIC~8432040, which has $P_{\rm orb}=5.3$~days and $c=0.97$ in the KEBC, but exhibits polynomial-like chi-square trends and was classified as a DEB by the PDS function. We found that approximately 85 percent of unmatched CEBs from the PDS classification are ellipsoidal systems, with most having $c \geq 0.9$. While these cases represent edge effects of the chi-square morphology, they do not undermine the overall effectiveness of the $\Delta P_{\rm PDS}/P_{\rm PDS}$ metric in identifying TV systems.

Finally, Table~\ref{tab:TVStable} lists a representative subset of our TV classifications, including four EBs that showed no literature references to any interesting physical characteristics such as star spots, flares, or the O'Connell effect, which could lead to TV behaviour. These systems might represent newly identified candidates that require further investigation in future studies.

\begin{table*}
    \centering
    \caption{In this table, we present a small subset of our TV classifications. For each target, we listed the {\sl Kepler} ID, visual classification, \( P_{\rm PDS} \), \( P_{\rm orb} \), and \(\Delta P_{\rm PDS}/P_{\rm PDS}\) values, along with references from the literature survey and the physical characteristics of each EB in the Notes column. The last four systems, for which no references are provided, are new TV systems identified using our method. The full version of this table is available online.}
    \label{tab:TVStable}
    \begin{tabular}{lcccccc}
        \hline
        {\sl Kepler} ID & Visual Classification & $P_{\rm PDS}$ & $P_{\rm orb}$ & $\Delta P_{\rm PDS}/ P_{\rm PDS}$ & Reference$^{a}$ & Notes$^{b}$ \\
        \hline
        KIC 2305372 & S.D & 32.274 & 1.405 & 1.038 & [2,3,5,6] & F,OE \\
        KIC 2305543 & D   & 64.273 & 1.362 & 0.037 & [2] & SM \\
        KIC 2309587 & D   & 87.269 & 1.839 & 0.055 & [2,3,5] & SM,F\\
        KIC 2447893 & S.D & 16.068 & 0.662 & 0.994 & [1,2,3,4] & SM,F\\
        KIC 2557430 & D   & 78.001 & 1.298 & 0.471 & [1,2,3,5] & SM,F \\
        KIC 2558370 & D   & 58.891 & 0.546 & 0.172 & [2] & SM \\
        KIC 2569494 & S.D & 67.025 & 1.523 & 0.097 & [2,6] & SM,AM,OE \\
        KIC 2577756 & S.D & 20.478 & 0.870 & 1.234 & [2,3,5,6] & SM,F,AM,OE \\
        KIC 2695740 & C   & 85.968 & 3.616 & 0.096 & [1,2,3,4,6] & SM,F,S,OE \\
        KIC 2720354 & D   & 66.030 & 2.821 & 0.049 & [1,2,3] & SM,F\\
        KIC 5428668 & D   & 30.532 & 1.406 & 0.095 & ... & ...\\
        KIC 8800998 & C   & 20.9 & 0.883 & 0.013 & ... & ...\\
        KIC 9077796 & C   & 11.566 & 0.238 & 0.503 & ... & ...\\
        KIC 10991769 & C   & 76.405 & 3.281 & 0.02 & ... & ...\\
        \hline
    \end{tabular}
    
    \vspace{2mm}
    \raggedright
    \footnotesize
    $^{a}$~Reference codes: 
    1: \citet{2016ApJ...829...23D}, 
    2: \citet{2017AJ....154..250L}, 
    3: \citet{2011A&A...529A..89D}, 
    4: \citet{2015MNRAS.448..429B}, 
    5: \citet{2016ApJS..224...37G}, 
    6: \citet{knote2022characteristics}. \\
    $^{b}$~Notes: 
    SM : literature indicates the presence of Starspot Modulation; 
    F : literature indicates the presence of Flare; 
    AM : literature indicates the presence of Asymmetric Minima ; 
    OE : literature indicates the presence of O’Connell Effect; 
    S : literature indicates the presence of Spots 
\end{table*}

\subsection{The TV nature}\label{sect53}
\begin{figure*}
  \centering
  \begin{subfigure}{0.45\textwidth}
\includegraphics[width=\textwidth]{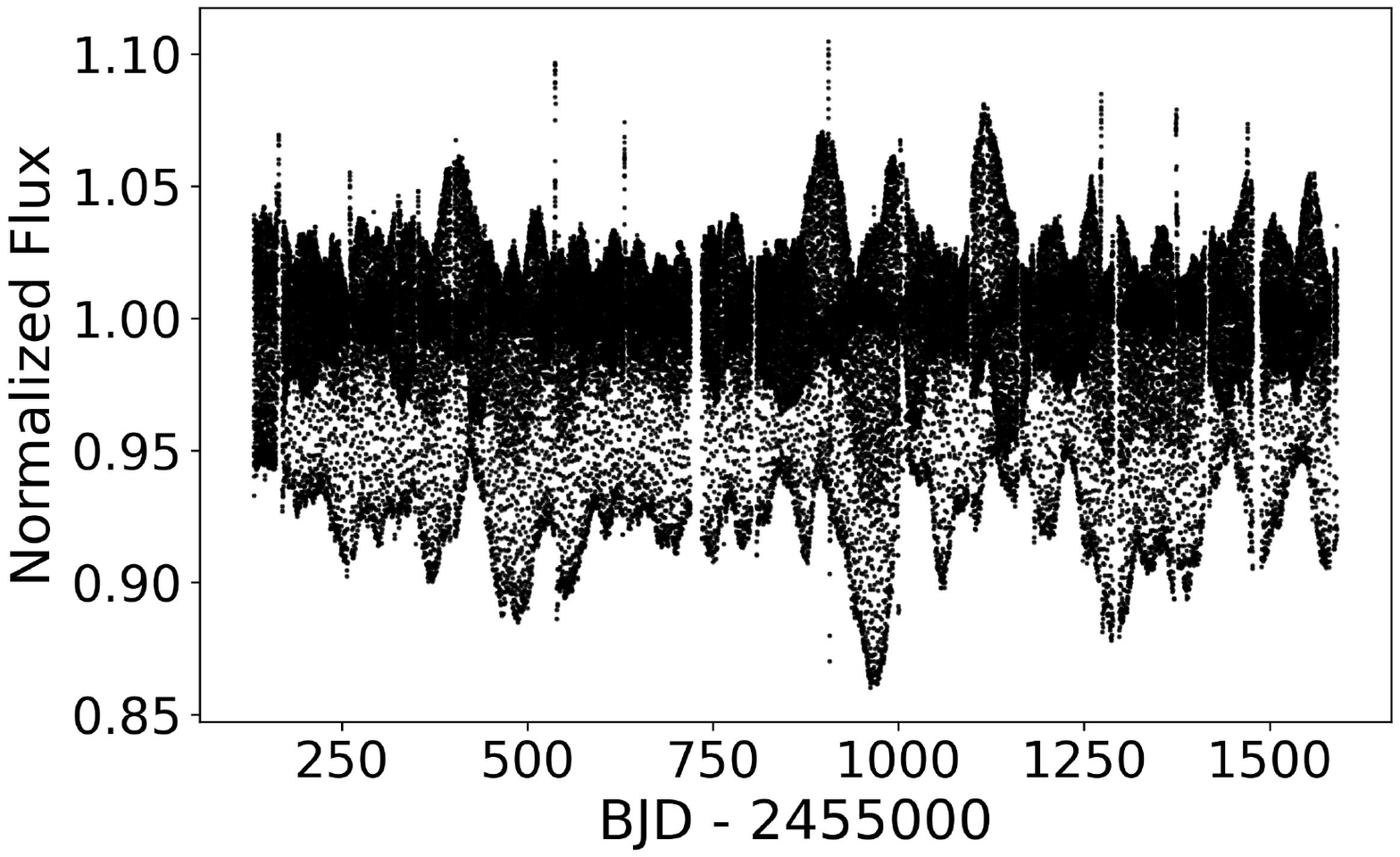}
    \caption{}
  \end{subfigure}
  \begin{subfigure}{0.45\textwidth}
    \includegraphics[width=\textwidth]{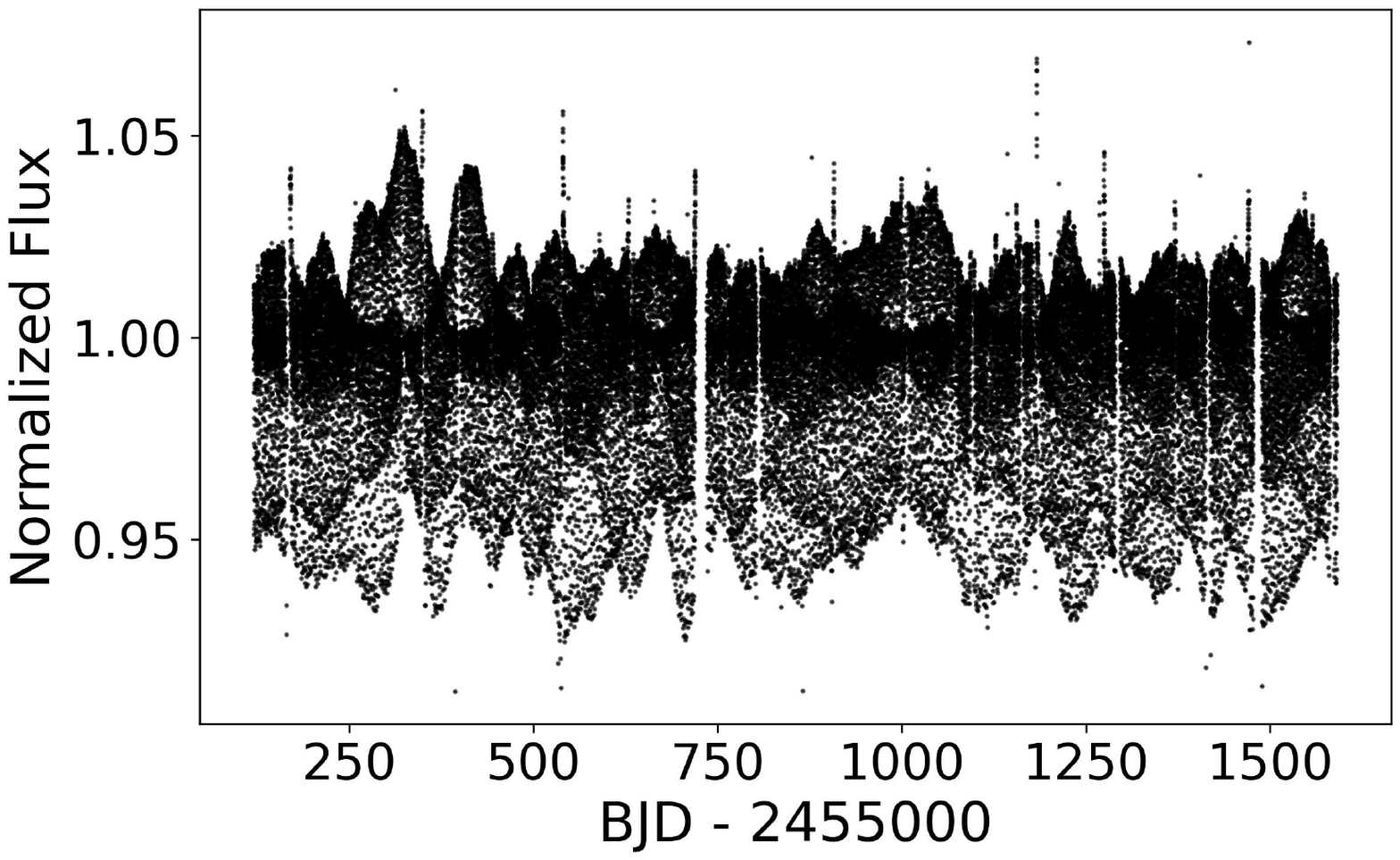}
    \caption{}
  \end{subfigure}
  \caption{(a) Time-domain {\it Kepler} light curve of KIC~2577756, plotted as normalized flux versus BJD, showing strong variability and clear flare signatures consistent with its high $\Delta P_{\rm PDS}/P_{\rm PDS}$ value. (b) Time-domain {\it Kepler} light curve of KIC~7284688, plotted as normalized flux versus BJD, showing strong variability and varying O'Connell effect due to asynchronous rotation of starspots consistent with its high $\Delta P_{\rm PDS}/P_{\rm PDS}$ value.}
  \label{fig:TV EXAMPLES}
\end{figure*}

To investigate the origin of the TV behaviour, we cross-matched our TV sample with 416 systems reported in relevant {\sl Kepler} EB studies using {\sl Kepler} IDs and system coordinates. Using the catalogue of stellar temperatures for {\sl Kepler} EB \citep{10.1093/mnras/stt2146}, we found that cooler stars (late-F, G, K, and M types) exhibited systematically larger $\Delta P_{\rm PDS}/P_{\rm PDS}$ values than hotter stars. This trend is consistent with enhanced magnetic activity in stars with convective envelopes, driven by rotational dynamos, leading to increased starspot coverage \citep{2009A&ARv..17..251S}, higher flare rates—particularly in K and M dwarfs \citep{2016ApJ...829...23D}—and coronal X-ray emission from magnetically confined plasma \citep{1981ApJ...245..163V,Wright_2011}.


To further constrain the physical origin of the observed variability, we cross-matched the 416 TV systems with several catalogues related to starspot modulation, flares, and O'Connell-effect variability \citep{2016ApJ...829...23D,2017AJ....154..250L,2011A&A...529A..89D,2015MNRAS.448..429B,2016ApJS..224...37G,knote2022characteristics}. The cross-matched systems and their reported properties are summarized in Table~\ref{tab:TVStable}, where SM denotes starspot modulation, F and S denote flares and starspots, OE denotes the O'Connell effect, and AM denotes asymmetric minima.

From this cross-match, we found a median $\Delta P_{\rm PDS}/P_{\rm PDS}$ value of 0.5 for flare-active close binaries \citep{2016ApJS..224...37G}, which is significantly higher than the adopted threshold of 0.0133. As an illustrative example, Fig.~\ref{fig:TV EXAMPLES}(a) shows KIC~2577756, which has $\Delta P_{\rm PDS}/P_{\rm PDS}=1.234$ and exhibits strong flare activity in its time-domain light curve, consistent with previous studies. This system is also highlighted in Fig.~\ref{fig:f_WBULL} by a blue vertical line.\\
Figure~\ref{fig:TV EXAMPLES}(b) presents KIC~7284688, a solar-type EB displaying a rapidly varying O'Connell effect attributed to asynchronous starspot rotation \citep{Pan_2023}. This system shows $\Delta P_{\rm PDS}/P_{\rm PDS}=2.612$ and is marked by a green vertical line in Fig.~\ref{fig:f_WBULL}.

Overall, these results reinforce the strong connection between magnetic activity and temporal variability in our TV sample, and demonstrate that flare- and starspot-active systems systematically exhibit higher $\Delta P_{\rm PDS}/P_{\rm PDS}$ values, making them particularly important targets for detailed follow-up studies.


\section{Summary and Conclusions}\label{sec:con}

In this study, we classified 2806 EBs into CEBs, DEBs, and SDEBs using long-cadence {\sl Kepler} data. We created a new classification scheme that was different from the traditional LC shape analysis, which was based on an indirect measure of the time-series data and that did not require any LC modelling.
We generated LCs from MAST FITS files cross-matched with the KEBC and normalised them via quarter-wise chi-square versus box-size analysis, which revealed characteristic trends for each EB type. We used box sizes between 10 and 101, which showed the most distinctive chi-square versus box-size shapes when tested on a few targets and were also chosen for practical computational reasons. Because the chi-square patterns were stable across different quarters for most EBs, we generated plots for all available quarters to provide sufficient training data for the network, while reserving the highest-epoch quarter of each system exclusively for the testing set.

In this analysis, we first fitted a PDS function to the quarter with the largest number of epochs, extracted the $P_{\rm PDS}$ values, and, using the conditions on $R^2$ and $P_{\rm PDS}$, achieved a classification accuracy of 86.5 percent.
In addition, for CEBs, we found a strong linear correlation between $P_{\rm PDS}$ and $P_{\rm orb}$, with a Pearson coefficient of 0.99, confirming that 
$P_{\rm PDS}$ is intrinsically linked to the $P_{\rm orb}$ of the EBs.

To improve the PDS methodology, we built a 1D-CNN for the morphological classification and conducted two different analysis to identify the best-performing model and dataset. We found that chi-square plots remained consistent across quarters for most CEBs and DEBs, while some SDEBs and low-$P_{\rm orb}$ DEBs showed inconsistencies. By using data from all available quarters and removing targets with inconsistent chi-square patterns, we improved the model’s ability to learn general features and achieved an accuracy of 90 percent. In a separate experiment using only CEBs and DEBs, our method distinguished these two classes with an overall accuracy of 95 percent. 

To further enhance performance, we combined observed chi-square plots with synthetic data generated using PHOEBE. This approach improved classification, particularly for CEBs and DEBs. However, PHOEBE models still struggled to reproduce the full variability observed in SDEBs from {\sl Kepler} LC data. This resulted in more homogeneous chi-square plots and contributed to the improved accuracy for SDEBs. When we tested the PHOEBE model only on CEBs and DEBs, the accuracy increased to 99 percent, since chi-square variations in these classes were primarily due to changes in $P_{\rm orb}$, which PHOEBE replicated effectively by adjusting the period.  


The CNN1 provided the best accuracy across the three EB classes together, correctly representing the observed data. In contrast, the CNN2 achieved the highest accuracy when classifying only CEBs and DEBs but was unable to classify the SDEBs properly, as we believed the variation in chi-square plots for SDEBs was linked to various physical phenomena.

To investigate the misclassified systems, we identified EBs whose chi-square plots and light curves varied from quarter to quarter. We termed these systems as TV systems and developed an automated method to detect them. TV systems exhibited quarterly chi-square variations that led to changes in $P_{\rm PDS}$ values. By measuring the difference between the maximum and minimum $P_{\rm PDS}$ values across quarters, we quantified the extent of variability, which was also reflected in their LCs.

The chi-square plots were very sensitive to variations in the binary LC, especially those generated due to magnetic activity, which was one of the prime reasons for temporal variation in EBs. Our method detected even small changes in chi-square plots, as their shapes varied from quarter to quarter. After normalizing these variations, we obtained the $\Delta P_{\rm PDS}/P_{\rm PDS}$ parameter, which detected systems with enhanced magnetic activity. This sensitivity enabled us to identify four EBs without any literature references, which require further investigation.

The chi-square plots not only helped in classifying the EBs but were also highly sensitive to physical changes in the binaries. We observed that DEBs with shorter $P_{\rm orb}$ tended to show damping-like chi-square patterns, though not as strongly as CEBs. In contrast, CEBs exhibited more stable, damping-like plots, possibly because these systems had reached equilibrium through mass transfer or other processes. SDEBs did not show  very distinctive chi-square patterns, as most of the SD sample was dominated by Temporally Varying systems, showing variable and distorted chi-square shapes. Leveraging these variations, we classified the targets using the PDS function and CNN. In future work, we plan to explore these unique chi-square shapes further and relate them to specific physical phenomena, which may lead to a more accurate classification of SDEBs and help unravel the magnetic effects influencing the chi-square plots.

\section{Data and Code Availability}\label{sec:data-code}

All code developed for this work is available at the following GitHub repository:
\url{https://github.com/mousam100}

The full machine-readable classification table and the complete temporally varying systems (TVS) table have been submitted to CDS/VizieR for ingestion. Following the VizieR procedure, the catalogue will be released publicly after publication of the corresponding paper.

The authors declare that there are no conflicts of interest regarding the publication of this paper.

\section*{Acknowledgements}
We are very grateful to the anonymous referee for their illuminating and insightful suggestions, which led to substantial 
improvements in this manuscript.

P.C., J.A. acknowledges support from the Spanish Virtual Observatory (\url{https://svo.cab.inta-csic.es}) project funded by the Spanish Ministry of Science, Innovation and Universities /State Agency of Research MCIU/AEI/10.13039/501100011033 through grant PID2023-146210NB-I00. HJ acknowledges STFC support for this research provided through infrastructure grant ST/R000905/1. MCGO acknowledges financial support from the Agencia Estatal de Investigación (AEI/10.13039/501100011033) of the Ministerio de Ciencia e Innovación and the ERDF 'A way of making Europe' through project PID2022-137241NB-C42.  J.A.'s contribution to this work is part of his academic activities as a professor at Universidad Militar Nueva Granada, Bogotá, Colombia.

\bibliographystyle{mnras}
\bibliography{mybib}

\label{lastpage} 
\end{document}